%% file: frontiers_vector.tex
\documentclass[letterpaper]{article} 
\usepackage{graphicx}
\usepackage[margin=1in]{geometry}

\usepackage{url,hyperref,lineno,microtype,subcaption}
\usepackage[onehalfspacing]{setspace}

\usepackage{booktabs} % For formal tables

\usepackage{xparse}% http://ctan.org/pkg/xparse
\usepackage{mathtools}

\newcommand{\Ceil}[1]{\left\lceil #1 \right\rceil}
\newcommand{\Floor}[1]{\left\lfloor #1 \right\rfloor}
\graphicspath{ {Figures/} {RasterFigures/} }
\usepackage{url}

\usepackage{algorithm}
\usepackage[noend]{algpseudocode}
\usepackage{graphicx,caption}
\usepackage[normalem]{ulem}
\usepackage{authblk}
\usepackage{xcolor}

\usepackage{amsthm}
\newtheorem{definition}{Definition}
\newtheorem{lemma}{Lemma}

\def\firstAuthorLast{T. Vu and A. Eldawy} %use et al only if is more than 1 author

\def\Address{Department of Computer Science and Engineering, University of California, Riverside, California}
\def\corrAuthor{Corresponding Author}
\def\corrEmail{eldawy@ucr.edu}

\begin{document}

% \title[R*-Grove for Big Spatial Data Partitioning]{From Trees to Groves: An in-depth Study of Big Spatial Data Partitioning} 
% \title[R*-Grove for Big Spatial Data Partitioning]{R*-Grove: Toward Balanced Spatial Partitioning for Large-scale Datasets with Variable-size Entities} 
\title{R*-Grove: Balanced Spatial Partitioning for Large-scale Datasets}
\date{July, 21, 2020}

% \author[*]{Author A}
% \author[*]{Author B}
% \author[*]{Author C}
% \author[**]{Author D}
% \author[**]{Author E}
% \affil[*]{Department of Computer Science, \LaTeX\ University}
% \affil[**]{Department of Mechanical Engineering, \LaTeX\ University}

% \renewcommand\Authands{ and }

\author{Tin Vu and Ahmed Eldawy\\
Department of Computer Science \& Engineering\\
University of California, Riverside, CA 92521, USA\\
\{tvu032,eldawy\}@ucr.edu 
}

% \author{Ahmed Eldawy\\
% Computer Science and Engineering\\
% University of California, Riverside\\
% eldawy@ucr.edu
% }

\maketitle

% -------------- Abstract -------------
\begin{abstract}
The rapid growth of big spatial data urged the research community to develop several big spatial data systems. Regardless of their architecture, one of the fundamental requirements of all these systems is to spatially partition the data efficiently across machines. The core challenges of big spatial partitioning are building high spatial quality partitions while simultaneously taking advantages of distributed processing models by providing load balanced partitions. Previous works on big spatial partitioning are to reuse existing index search trees as-is, e.g., the R-tree family, STR, Kd-tree, and Quad-tree, by building a temporary tree for a sample of the input and use its leaf nodes as partition boundaries. However, we show in this paper that none of those techniques has addressed the mentioned challenges completely. This paper proposes a novel partitioning method, termed R*-Grove, which can partition very large spatial datasets into high quality partitions with excellent load balance and block utilization. This appealing property allows R*-Grove to outperform existing techniques in spatial query processing. R*-Grove can be easily integrated into any big data platforms such as Apache Spark or Apache Hadoop. Our experiments show that R*-Grove outperforms the existing partitioning techniques for big spatial data systems. With all the proposed work publicly available as open source, we envision that R*-Grove will be adopted by the community to better serve big spatial data research.

\small
 \paragraph{Keywords:} big spatial data, partitioning, R*-Grove, index optimization, query processing
\end{abstract}

% -------------- Introduction -------------
\section{Introduction}\label{sec:introduction}

The recent few years witnessed a rapid growth of big spatial data collected by different applications such as satellite imagery \cite{EMA+15}, social networks \cite{MAA+14}, smart phones \cite{HBC+16}, and VGI \cite{G07}. Traditional Spatial DBMS technology could not scale up to these petabytes of data which led to the birth of many big spatial data management systems such as SpatialHadoop \cite{EM15}, GeoSpark \cite{YWS15}, Simba \cite{XLY+16}, LocationSpark \cite{TYM+16}, and Sphinx \cite{ESE+17}, to name a few.

\begin{figure}[t]
	\centering
	\includegraphics[width=\columnwidth]{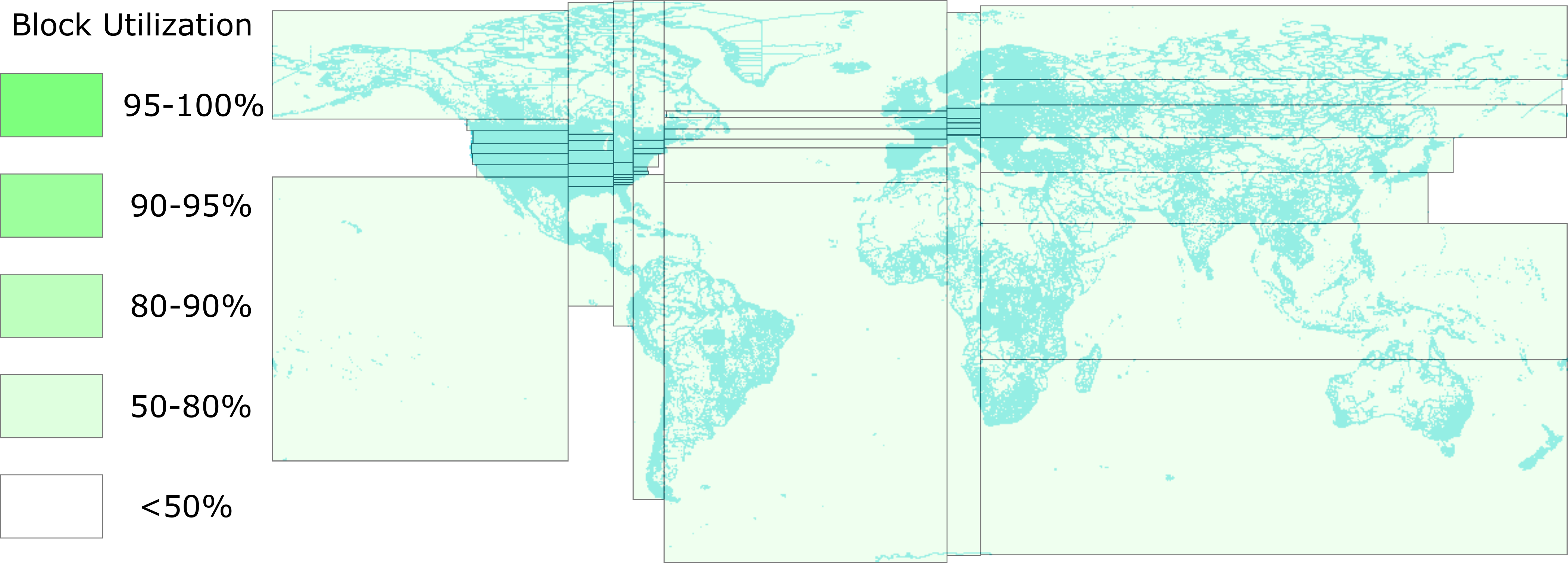}\\
	(a)~STR-based partitioning \cite{EM15}. All the thin and wide partitions reduce the query efficiency.\\
	\includegraphics[width=\columnwidth]{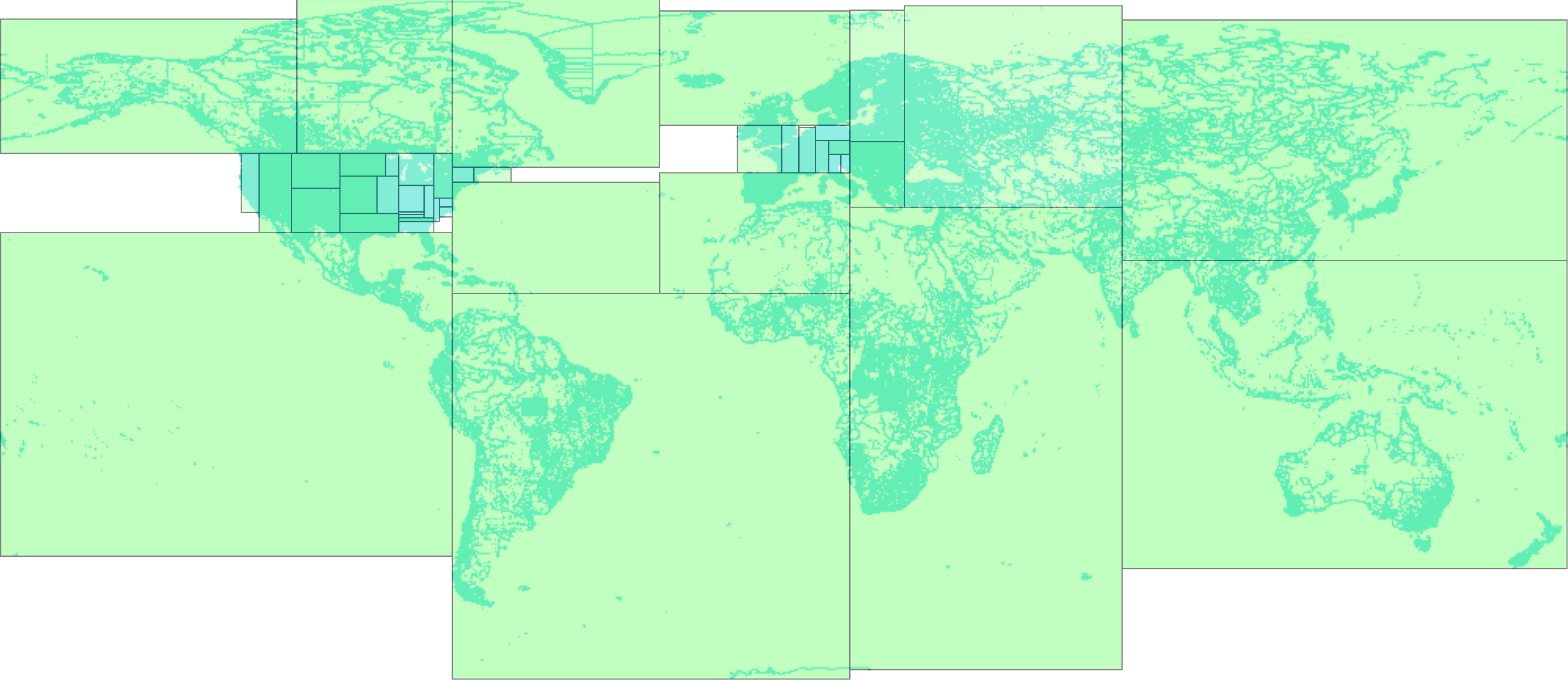}
	(b)~The proposed R*-Grove method with square-like and balanced partitions.\\
	\caption{Comparison between STR and R*-Grove}
	\label{fig:str_rsgrove}
\end{figure}

Regardless of their architecture, all these systems need an essential preliminary step that partitions the data across machines before the execution can be parallelized. This is also known as {\em global indexing} \cite{EM16}. A common method that was first introduced in SpatialHadoop \cite{EM15}, is the sample-based STR partitioner. This method picks a small sample of the input to determine its distribution, packs this sample using the STR packing algorithm \cite{leutenegger1997str}, and then uses the boundaries of the leaf nodes to partition the entire data. Figure~\ref{fig:str_rsgrove}(a) shows an example of an STR-based partitioning where each data partition is depicted by a rectangle. The method was later generalized by replacing the STR bulk loading algorithm with other spatial indexes such as Quad-tree \cite{S84}, Kd-Tree, and Hilbert R-trees \cite{KF94,EAM15}. That STR-based partitioning was very attractive due to its simplicity and good load balancing which is very important for distributed applications.
Its simplicity urged many other researchers to adopt it in their systems such as GeoSpark \cite{YWS15} and Simba \cite{XLY+16} for in-memory distributed processing; Sphinx \cite{ESE+17} for SQL big spatial data processing; HadoopViz \cite{EMJ16,GEJ19} for scalable visualization of big spatial data; and in distributed spatial join \cite{SM17}.

Despite their wide use, the existing partitioning techniques all suffer from one or more of the following three limitations.
First, some partitioning techniques (STR, Kd-tree) prioritize load balance over spatial quality which results in suboptimal partitions. This is apparent in Figure~\ref{fig:str_rsgrove}(a) where the thin and wide partitions result in low overall quality for the partitions since square-like partitions are preferred for most spatial queries. Square-like partitions are preferred in indexing because they indicate that the index is not biased towards one dimension. Also, since most queries are shaped like a square or a circle, square-like partitions would minimize the overlap with the queries \cite{BKS+90}.
Second, they could produce partitions that do not fill the HDFS blocks in which they are stored. Big data systems are optimized to process full blocks, i.e., 128~MB, to offset the fixed overhead in processing each block. However, the index structures used in existing partitioning techniques, e.g., R-trees, Kd-tree, Quad-tree, produce nodes with number of records in the range $[m, M]$, where $m\le M/2$. In practice, $m$ can be as low as $0.2M$ \cite{BKS+90,BS09}. While those underutilized index nodes were desirable for disk indexing as they can accommodate future inserts, they result in underutilized blocks as depicted in Figure~\ref{fig:str_rsgrove}(a) where all blocks are less than 80\% full. %This is a drawback in big spatial data systems where fully utilized partitions are preferred since further insertions are not supported.
Moreover, this design might also produces poor load balance among partitions due to the wide range of partition sizes.
Third, all existing partitioning techniques rely on a sample and try to balance the number of records per partition. This resembles traditional indexes where the index contains record IDs. However, in big spatial data partitioning, the entire record is written in each partition, not just its ID. When records are highly variant in size, all existing techniques end up with extremely unbalanced partitions.
%Third, all existing partitioning techniques rely on a sample and use the number of sample points as a proxy for the data size. This becomes very inaccurate for variable-size records since the entire record is written to the partition rather than just the key as in traditional disk indexes. In other words, the load balance in terms of number of records does not guarantee the load balance in terms of storage size.

This paper proposes a novel spatial partitioning technique for big data, termed R*-Grove, which completely addresses all of three aforementioned limitations.
First, it produces high quality partitions by utilizing the R*-tree optimization techniques \cite{BKS+90} which aim at minimizing the total area, overlap area, and margins. The key idea of the R*-Grove partitioning technique is to start with one partition that contains all sample points and then use the node split algorithm of the R*-tree to split it into smaller partitions. This results in compact square-like partitions as shown in Figure~\ref{fig:str_rsgrove}(b).
Second, in order to ensure that we produce full blocks and balanced partitions, R*-Grove introduces a new constraint that puts a lower bound on the ratio between the smallest and the largest block, e.g., 95\%. This property is theoretically proven and practically validated by our experiments.
Third, when the input records have variable sizes, R*-Grove combines a data size histogram with the sample points to assign a weight for each sample point. These weights are utilized to guarantee that the size of each partition falls in a user-defined range.

Given the wide adoption of the previous STR-based partitioner, we believe the proposed R*-Grove will be widely used in big spatial data systems. This impacts a wide range of spatial analytics and processing algorithms including indexing \cite{EAM15,VAW14}, range queries \cite{EM15,YWS15}, kNN queries \cite{EM15}, visualization \cite{EMJ16,GEJ19}, spatial join \cite{jacox2007spatial}, and computational geometry \cite{ELM+13,LEX+19}. All the work proposed in this paper is publicly available as open source and supports both Apache Spark and Apache Hadoop. We run an extensive experimental evaluation with up-to 500~GB and 7~billion record datasets and up-to nine dimensions. The experiments show that R*-Grove consistently outperforms existing STR-based, Z-curve-based, Hilbert-Curve-based, and Kd-tree-based techniques in both partitions quality and query efficiency.

The rest of this paper is organized as follow.
Section~\ref{sec:related_work} describes the related works.
Section~\ref{sec:background} gives a background about big spatial data partitioning.
Section~\ref{sec:rsgrove} describes the proposed R*-Grove technique.
Section~\ref{sec:case-studies} describes the advantages of R*-Grove in popular case studies of big spatial data systems.
Section~\ref{sec:experiments} gives a comprehensive experimental evaluation of the proposed work.
Finally, Section~\ref{sec:conclusion} concludes the paper.

% -------------- Related Work -------------
\section{Related Work}\label{sec:related_work}

This section discusses the related work in big spatial data partitioning. In general, distributed indexes for big spatial data are constructed in two levels, one global index that partitions the data across machines, and several local indexes that organize records in each partition. Previous work \cite{EM15,LCC+14,EM16} showed that the global index provides far much improvement than local indexes. Therefore, in this paper we focus on global indexing and it can be easily combined with any of the existing local indexes. The work in global indexing can be broadly categorized into three approaches, namely, sampling-based methods, space-filling-curve (SFC)-based methods, and quad-tree-based methods.

The {\em sampling-based} method picks a small sample from the input data to infer its distribution. The sample is loaded into an in-memory index structure while adjusting the data page capacity, e.g., leaf node capacity, such that the number of data pages is roughly equal to the desired number of partitions. The order of sample objects does not affect the partition quality, since the sample is uniformly taken from the entire input dataset. Furthermore, most of algorithms sort the data as part of the partitioning process so the original order is completely lost. Some R-tree bulk-loading algorithms (STR \cite{leutenegger1997str} or OMT \cite{lee2003omt}) can also be used to speed up the tree construction time. Then, the minimum bounding rectangles (MBRs) of the data pages are used to partition the entire dataset. This method was originally proposed for spatial join and denoted the seeded-tree \cite{LR94}. It was then used for big spatial indexing in many systems including SpatialHadoop \cite{EM15,EAM15}, Scala-GiST \cite{LCC+14}, GeoSpark \cite{YWS15}, Sphinx \cite{ESE+17}, Simba \cite{XLY+16}, and many other systems. This technique can be used with existing R-tree indexes but it suffers from two limitations, load imbalance and low quality of spatial partitions. Additionally, when there is a big variance in record sizes, the load imbalance is further amplified due to the use of the sample. We will further discuss these limitations in Section~\ref{sec:rsgrove}.

The {\em SFC-based} method builds a spatial index on top of an existing one-dimensional index by applying any space-filling curve, e.g., Z-curve or Hilbert curve. MD-HBase \cite{nishimura2013mathcal} builds Kd-tree-like and Quad-tree-like indexes on top of HBase by applying the Z-curve on the input data and customizing the region split method in HBase to respect the structure of both indexes. GeoMesa \cite{fox2013spatio} uses geo-hashing which is also based on the Z-curve to build spatio-temporal indexes on top of Accumulo. Unlike MD-HBase which only supports point data, GeoMesa can support rectangular of polygonal geometries by replicating a record to all overlapping buckets in the geohash. While this method can ensure a near-perfect load balance, it produces an even bigger spatial overlap between partitions as compared to the sampling-based approach described above. This drawback leads to the inefficient performance of spatial queries.

The {\em quad-tree-based} method relies heavily on the Quad-tree structure to build efficient and scalable Quad-tree index in Hadoop \cite{WPA+14}. It starts by splitting the input data into equi-sized chunks and building a partial Quad-tree for each split. Then, it combines the leaf nodes of the partial trees based on the Quad-tree structure to merge them into the final tree. While highly efficient, this method cannot generalize to other spatial indexes and is tightly tied to the Quad-tree structure. In addition, this Quad-tree-based partitioning tends to produce much more than the desired number of partitions which also leads to load imbalance.

Although there are several partitioning techniques for large-scale spatial data as mentioned above, sampling-based method is the most ubiquitous option, which is integrated in most of existing spatial data systems. Sampling-based methods are preferred as they are simple to implement and provide very good results. In this paper, we follow the sampling-based approach, and propose a method which utilizes R*-tree's advantages that were never used before for big spatial data partitioning. The proposed R*-Grove index has three advantages over the existing work. First, it inherits and improves the R*-tree index structure to produce high-quality partitions that are tailored to big spatial data. Second, the improved algorithm produces balanced partitions by employing a user-defined parameter, termed {\em balance factor}, $\alpha$, e.g., 95\%. In addition, it can produce spatially disjoint partitions which are necessary for some spatial analysis algorithms. Third, R*-Grove can couple a sample with a data size histogram to guarantee the desired load balance even when the input record sizes are highly variant.
While R*-Grove is not the only framework for big spatial partitioning, it is the first one that is tailored for large-scale spatial datasets while existing techniques reuse traditional index structures, such as R-tree, STR, or Quad-tree, as black boxes.
% -------------- Background -------------
\section{Background}
\label{sec:background}

\subsection{R*-tree}
The R*-tree \cite{BKS+90} belongs to the R-tree family \cite{G84} and it improves the insertion algorithm to provide high quality index. In R-tree, the number of children in each nodes has to be in the range $[m,M]$. By design, $m$ can be at most $\Floor{M/2}$ to ensure that splitting a node of size $M+1$ is feasible. In this paper, we utilize and enhance two main functions of the R*-tree index, namely, {\sc ChooseSubtree} and {\sc SplitNode} which are both used in the insertion process. For the {\sc ChooseSubtree} method, given the MBR of a record and a tree node, it chooses the best subtree to assign this record to. The {\sc SplitNode} method takes an overflow node with $M+1$ records and splits it into two nodes.

\subsection{Sample-based Partitioning Workflow}\label{sec:sample_based_partitioning}

\begin{figure}[t]
	\centering
	\includegraphics[width=0.5\textwidth]{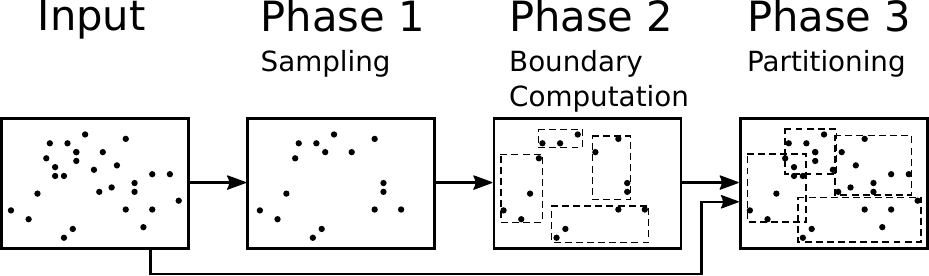}
	\caption{The sampling-based partitioning process}
	\label{fig:indexing-process}
\end{figure}

This section gives a background on the sampling-based partitioning technique \cite{EM15,EAM15,VAW14}, just partitioning hereafter, that this paper relies on. Figure~\ref{fig:indexing-process} shows the workflow for the partitioning algorithm which consists of three phases, namely, sampling, boundary computation, and partitioning. The sampling phase (Phase~1) draws a random sample of the input records and converts each one to a point. Notice that sample points are picked from the entire file at no particular order so the order of points does not affect the next steps. The boundary computation phase (Phase~2) runs on a single machine and processes the sample to produce partition boundaries as a set of rectangles. Given a sample $S$, the input size $D$, and the desired partition size $B$, this phase adjusts the capacity of each partition to contain $M=\Ceil{|S|\cdot B/D}$ sample points which is expected to produce final partitions with the size of one block each. The final partitioning phase (Phase~3) scans the entire input in parallel and assigns each record to these partitions based on the MBR of the record and the partition boundaries. If each record is assigned to exactly one partition, the partitions will be spatially overlapping with no data replication. If each record is assigned to all overlapping partitions, the partitions will be spatially disjoint but some records can be replicated and duplicate handling will be needed in the query processing \cite{DS00}. Some algorithms can only work if the partitions are spatially disjoint such as visualization \cite{EMJ16} and some computational geometry functions \cite{LEX+19}.

The proposed R*-Grove method expands Phase~1 by optionally building a histogram of storage size that assists in the partitioning algorithm at Phase~2. In Phase~2, it adapts R*-tree-based algorithms to produce the partition boundaries with desired level of load balance. In Phase~3, we propose a new data structure that improves the performance of that phase and allows us to produce spatially disjoint partitions if needed.

\subsection{Quality Metrics}
This paper uses the quality metrics defined in \cite{eldawy2015spatial}. Below, we redefine these metrics while accounting for the case of partitions that span multiple HDFS blocks. A single partition $\pi_i$ is defined by two parameters, minimum bounding box $mbb_i$ and size in bytes $size_i$. Given the HDFS block size $B$, e.g., 128~MB, we define the number of blocks for a partition $\pi_i$ as $b_i=\Ceil{size_i/B}$. Given a dataset that is partitioned into a set of $l$ partitions, $\mathcal P=\{\pi_i\}$, we define the five quality metrics as follows.

\begin{definition}[Total Volume - $Q_1$]
The total volume is the sum of the volume of all partitions where the volume of a partition is the product of its side lengths.
\[Q_1(\mathcal P)=\sum_{\pi_i \in \mathcal P}{b_i\cdot volume(mbb_i)}\]
We multiply by the number of blocks $b_i$ because big spatial data systems process each block separately. Lowering the total volume is preferred to minimize the overlap with a query. Given the popularity of the two-dimensional case, this is usually used under the term {\em total area}.
\end{definition}

\begin{definition}[Total Volume Overlap - $Q_2$]
This quality metric measures the sum of the overlap between pairs of partitions.
\[Q_2(\mathcal P)=\sum_{\pi_i,\pi_j\in \mathcal P, i\neq j}{b_i\cdot b_j\cdot volume(mbb_i \cap mbb_j)} + \sum_{\pi_i \in \mathcal P}{\frac{b_i(b_i-1)}{2} \cdot volume(mbb_i)}\]
where $mbb_i\cap mbb_j$ is the intersection region between the two boxes. The first term calculates the overlaps between pairs of partitions and the second term accounts for {\em self-overlap} which treats a partition with multiple blocks as overlapping partitions. Lowering the volume overlap is preferred to keep the partitions apart.
\end{definition}

\begin{definition}[Total Margin - $Q_3$]
The margin of a block is the sum of its side lengths. The total margin is the sum of all margins as given below.
\[Q_3(\mathcal P)=\sum_{\pi_i \in \mathcal P}{b_i\cdot margin(mbb_i)}\]
Similar to $Q_1$, multiplying by the number of blocks $b_i$ treats each block as a separate partition. Lowering the total margin is preferred to produce square-like partitions.
\end{definition}

\begin{definition}[Block Utilization - $Q_4$]
Block utilization measures how {\em full} the HDFS blocks are.
\[Q_4(\mathcal P)=\frac{\sum_{\pi_i \in \mathcal P}{size_i}}{B\cdot \sum_{\pi_i \in \mathcal P}{b_i}} \]
The numerator $\sum size_i$ represents the total size of all partitions and denominator $B\sum b_i$ is the maximum amount of data that can be stored in all blocks used by these partitions. In big data applications, each block is processed in a separate task which has a setup time of a few seconds. Having full or near-full blocks minimize the overhead of the setup. The maximum value of block utilization is $1.0$, or $100\%$.
\end{definition}

\begin{definition}[Standard Deviation of Sizes]
\[Q_5(\mathcal P)=\sqrt{\frac{\sum_{\pi_i \in \mathcal P}{(size_i-\overline{size})^2}}{l}}\]
Where $\overline{size}=\sum{size_i}/l$ is the average partition size. Lowering this value is preferred to balance the load across partitions.
\end{definition}

% -------------- R*-Grove -------------
\section{R*-Grove Partitioning}
\label{sec:rsgrove}

This section describes the details of the proposed R*-Grove partitioning algorithm. R*-Grove employs three techniques that overcome the limitations of existing works. The first technique adapts the R*-tree index structure for spatial partitioning by utilizing the {\sc ChooseSubTree} and {\sc SplitNode} functions in the sample-based approach described in Section~\ref{sec:background}. This technique ensures a high spatial quality of partitions.
The second technique addresses the problem of load balancing by introducing a new constraint that guarantees a user-defined ratio between smallest and largest partitions.
The third technique combines the sample points with its storage histogram to balance the sizes of the partitions rather than the number of records. This combination allows R*-Grove to precisely produce partitions with a desired block utilization, which cannot be achieved by any other partitioning techniques.

\subsection{R*-tree-based Partitioning}
\label{sec:rsgrove:basic}

This part describes how R*-Grove utilizes the R*-tree index structure to produce high quality partitions. It utilizes the {\sc SplitNode} and {\sc ChooseSubtree} functions from the R*-tree algorithm in Phases~2 and~3 as described shortly. A na\"ive method \cite{VE18} is to use the R*-tree as a blackbox in Phase~2 in Figure~\ref{fig:indexing-process} and insert all the sample points into an R*-tree. Then it emits the MBRs of the leaf nodes as the output partition boundaries. However, this technique was shown to be inefficient as it processes the sample points one-by-one and does not integrate the R*-tree index well in the partitioning algorithm. Therefore, we propose an efficient approach that runs much faster and produces higher quality partitions. It extends Phases~2 and~3 as follows.

Phase~2 computes partition boundaries by only using the {\sc SplitNode} algorithm from the R*-tree index which splits a node with $M+1$ records into two nodes with the size of each one in the range $[m,M]$. This algorithm starts by choosing the split axis, e.g., $x$ or $y$, that minimizes the total margin. Then, all the points are sorted along the chosen axis and the split point is chosen as depicted in Algorithm~\ref{alg:choosesplitpoint}. The {\sc ChooseSplitPoint} algorithm simply considers all the split points and chooses the one that minimizes some cost function which is typically the total area of the two resulting partitions.

\begin{algorithm}[t]
\begin{algorithmic}[1]
\Function{ChooseSplitPoint}{$P$,$m$}
\State chosenK = -1; minCost = $\infty$
\For{$k$ in $[m,|P|-m]$}
    \State $P_1=P[1..k]$ \Comment{$P_1$ is the first $k$ records of $P$} \label{alg:choosesplitpoint:split1}
    \State $P_2=P[k+1..|P|]$ \Comment{$P_2$ is all the remaining records $P-P_1$}
    \State Calculate the cost of the partitions $P_1$ and $P_2$
    \label{alg:choosesplitpoint:calc}
    \If{the cost is smaller than minCost}
        \State Set chosenK = $k$ and update minCost
    \EndIf
    \label{alg:choosesplitpoint:updatemin}
\EndFor
\State \Return chosenK
\EndFunction
\end{algorithmic}
\caption{A simplified version of the traditional R*-tree splitting mechanism.\\
Inputs: $P$ is the all sample records; $m$ is the minimum size of a node.\\
Output: the optimal splitting position.}
\label{alg:choosesplitpoint}
\end{algorithm}

We set $M=\Ceil{|S|\cdot B/|D|}$ as explained in Section~\ref{sec:background} and $m=0.3M$ as recommended in the R*-tree paper. In particular, this phase starts by creating a single big tree node that has all the sample points $S$. Then, it recursively calls the {\sc SplitNode} algorithm as long as the resulting node has more than $M$ elements. This top-down approach has a key advantage over building the tree record-by-record as it allows the algorithm to look at all the records at the beginning and optimize for all of them. Furthermore, it avoids the {\sc ForcedReinsert} algorithm which is known to slow down the R*-tree insertion process. Notice that this is different than the bulk loading algorithms as it does not produce a full tree. Rather, it just produces a set of boundaries that are produced as partitions. Phase~3 treats all the MBRs as leaf nodes in an R-tree and uses the {\sc ChooseLeaf} method from the R*-tree to assign an input record to a partition.

\textbf{Run-time analysis:} The {\sc SplitNode} algorithm can be modeled as a recursive algorithm where each iteration sorts all the points and runs the linear-time splitting algorithm to produce two smaller partitions. The run-time can be expressed as $T(n)=T(k)+T(n-k)+O(n\log n)$, where $k$ is the size of one group resulting from the partitioning, $n$ is the number of records in the input partition. In particular, $T(k)$ and $T(n-k)$ are the running times to partition two partitions from splitting process. The term $O(n\log n)$ is the running time for the splitting part which requires sorting all the points. This recurrence relation has a worst case of $n^2 \log n$ if $k$ is always $n-1$. In order to guarantee a run-time of $O(n\log^2 n)$, we define a parameter $\rho\in[0, 0.5]$ which defines the minimum splitting ratio $k/n$. Setting this parameter to any non-zero fraction guarantees an $O(n\log^2 n)$ run-time. However, the restriction of $k/n$ also limits the range of possible value of $k$. For example, if $n=100$ and $\rho=0.3$, $k$ must be a number in the range $[30, 70]$. As this parameter gets closer to $0.5$, the two partitions become closer in size and the run-time decreases but the quality of the index might also deteriorate due to the limited search space imposed by this parameter. To incorporate this parameter in the node-splitting algorithm, we call the {\sc ChooseSplitPoint} function with the parameters ($P$,$\max\{m,\rho \cdot |P|\}$), where $|P|$ is the number of points in the list $P$.

\subsection{Load Balancing for Partitions with Equal-size records}
\label{sec:rsgrove:balance}

\begin{algorithm}[t]
\begin{algorithmic}[1]
\Function{ChooseValidSplitPoint}{$P$,$m$}
\For{$k$ in $[m,|P|-m]$} \label{alg:choosesvalidsplitpoint:for}
    \If{either $k$ or $|P|-k$ is invalid} \Comment{ Lemma~\ref{lemma:check_valid_partition_size}}
    \label{alg:choosesvalidsplitpoint:test}
      \State Skip this iteration and {\bf continue}
    \EndIf
    \State Similar to Lines~\ref{alg:choosesplitpoint:split1}-\ref{alg:choosesplitpoint:updatemin} in Algorithm~\ref{alg:choosesplitpoint}
\EndFor
\State \Return chosenK
\EndFunction
\end{algorithmic}
\caption{R*-tree-based split while ensuring valid partitions\\
Inputs: $P$ is the all sample records; $[m,M]$ is the target range of sizes for final partitions.\\
Output: the optimal splitting position.}
\label{alg:choosesvalidsplitpoint}
\end{algorithm}

In this section, we focus on balancing the number of records in partitions assuming equal-size records. We further extend this in the next section to support variable-size records. The method in Section~\ref{sec:rsgrove:basic} does a good job in producing high-quality partitions similar to what the R*-tree provides. However, it does not address the second limitation, that is, balancing the sizes of the partitions. Recall that the R-tree index family requires the leaf nodes to have sizes in the range $[m,M]$, where $m\le M/2$. With the R*-tree algorithm explained earlier, some partitions might be 30\% full which reduces block utilization and load balance.
We would like to be able to set $m$ to a larger value, say, $m=0.95M$. Unfortunately, if we do so, the {\sc SplitNode} algorithm would just fail because it will face a situation where there is no valid partitioning.

To illustrate the limitation of the {\sc SplitNode} mechanism, consider the following simple example. Let us assume we choose $m=9$ and $M=10$ while the list $P$ contains $28$ points. If we call the {\sc SplitNode} algorithm on the 28 points, it might produce two partitions with 14 records each. Since both contain more than $M=10$ points, the splitting method will be called again on each of them which will produce an incorrect answer since there is no way to split 14 records into two groups while each of them contain between 9 and 10 records. A correct splitting algorithm would produce three partitions with sizes 9, 9, and 10. Therefore, we need to introduce a new constraint to the splitting algorithm so that it always produces partitions with sizes in the range $[m,M]$.

\textbf{The Final Finding:} The {\sc SplitNode} algorithm can be minimally modified to guarantee final leaf partitions in the range $[m,M]$ by satisfying the following validity constraint:
\[\Ceil{S_i/M}\le \Floor{S_i/m}, i\in\{1,2\}\], where $S_1$ and $S_2$ are the sizes of the two resulting partitions of the split. Algorithm~\ref{alg:choosesvalidsplitpoint} depicts the main changes to the algorithm that introduces a new constraint test in Line~\ref{alg:choosesvalidsplitpoint:test} that skips over invalid partitioning.
The rest of this section provides the theoretical proof that this simple constraint guarantees the algorithm termination with leaf partitions in the range $[m,M]$. We start with the following definition.

\begin{definition}\textbf{Valid Partition Size:}
An integer number $S$ is said to be a valid partition size with respect to a range $[m,M]$ if there exists a set of integers $X=\{x_1,\cdots,x_n\}$ such that $\sum{x_i}=S$ and $x_i\in[m,M]$ $\forall\ i \in [1,n]$. In words, if we have $S$ records, there is at least one way to split them such that each split has between $m$ and $M$ records.
\label{def:valid}
\end{definition}

For example, if $m=9$ and $M=10$, the sizes 14, 31, and 62, are all invalid while the sizes 9, 27, and 63, are valid. Therefore, to produce balanced partitions, the {\sc SplitNode} algorithm should keep the invariant that the partition sizes are always valid according to the above definition. Going back to the earlier example, if $S=28$, the answer $S_1=S_2=14$ will be rejected because $S_1=14$ is invalid. Rather, the result of the first call to the {\em SplitNode} algorithm will result in two partitions with sizes $\{10,18\}$ or $\{9,19\}$. The following lemma shows how to test a size for validity in constant time.

\begin{lemma}\textbf{Validity Test:}
An integer $S$ is a valid partition size w.r.t a range $[m,M]$ iff $L \leq U$ in which L (lower bound) and U (upper bound) are computed as:
\begin{align}
\nonumber L&= \lceil S/M \rceil \\
\nonumber U&= \lfloor S/m \rfloor 
\end{align}
\label{lemma:check_valid_partition_size}
\end{lemma}

\begin{proof}
First, if $S$ is valid then, by definition, there is a partitioning of $S$ into $n$ partitions such that each partition is in the range $[m,M]$. It is easy to show that $L\le U$ and we omit this part for brevity.
The second part is to show that if the inequality $L\leq U$ holds, then there is at least one valid partitioning.
Based on the definition of $L$ and $U$, we have:

\begin{align}
U&= \lfloor S/m \rfloor \Rightarrow U \leq S/m \Rightarrow S \geq m \cdot U \Rightarrow S \geq m \cdot L \Rightarrow S - m \cdot L \geq 0
\label{eqn:inequality_1} \\
L&= \lceil S/M \rceil \Rightarrow L \geq S/M \Rightarrow S \leq M \cdot L \Rightarrow S - m \cdot L \leq (M - m) \cdot L
\label{eqn:inequality_2}
\end{align}

% \begin{align}
% U&= \lfloor S/m \rfloor \Rightarrow U \leq S/m \Rightarrow S \geq m \cdot U \Rightarrow S \geq m \cdot L \Rightarrow S - m \cdot L \geq 0
% \label{eqn:inequality_1}
% \end{align}

% \begin{align}
% \centering
% L&= \lceil S/M \rceil \Rightarrow L \geq S/M \Rightarrow S \leq M \cdot L \Rightarrow S - m \cdot L \leq (M - m) \cdot L
% \label{eqn:inequality_2}
% \end{align}

% Finally, from \ref{eqn:inequality_1} and \ref{eqn:inequality_2}, we have:
% \begin{align}
% 0 \leq S - m \cdot L& \leq (M - m) \cdot L
% \label{eqn:inequality}
% \end{align}

Based on Inequalities~\ref{eqn:inequality_1} and \ref{eqn:inequality_2}, we can make a valid partitioning as follows:

\begin{enumerate}
    \item Start with $L$ empty partitions. Assign $m$ records to each partition. The remaining number of records is $S - m \cdot L \geq 0$. This is satisfied due to Inequality~\ref{eqn:inequality_1}.

    \item Since each partition now has $m$ records, it can receive up-to $M-m$ additional records in order to keep its validity. Overall, $L$ partitions of size $m$ can accommodate up-to $(M-m) \cdot L$ records to keep a valid partitioning. But the remaining number of records $S - m \cdot L$ is not larger than the upper limit of what the partitions can accommodate, $(M-m) \cdot L$ as shown in Inequality~\ref{eqn:inequality_2}.
Therefore, this condition is satisfied as well.
\end{enumerate}

In conclusion, it follows that if the condition $L \leq U$ holds, we can always find a valid partitioning scheme for $S$ records which completes the proof.
\end{proof}

If we apply this test for the example above, we find that 28 is valid because $L = \lceil 28/10 \rceil=3 \le U = \lfloor 28/9 \rfloor=3$ while 62 is invalid because $L = \lceil 62/10 \rceil=7 > U = \lfloor 62/9 \rfloor=6$. This approach works fine as long as the initial sample size $S$ is valid but how do we guarantee the validity of $S$? We show that this is easily guaranteed if the size $S$ is above some threshold $S^*$ as shown in the following lemma.

\begin{lemma}
Given a range $[m,M]$, any partition of size $S \ge S^*$ is valid where $S^*$ is defined by the following formula:
\begin{equation}
S^*=\Ceil{\frac{m}{M-m}} \cdot m
\label{eqn:S*}
\end{equation}
\label{lemma:min_valid_size}
\end{lemma}

\begin{proof}
Following Definition~\ref{def:valid}, we will prove that for any partition size $S\ge S^*$, there exists a way to split it into $k$ groups such that the size of each group is in the range $[m, M]$. 

First, let $i = \Ceil{\frac{m}{M-m}}$, we have:

\begin{align}
S& \ge S^* = \Ceil{\frac{m}{M-m}} \cdot m = i\cdot m \\
\Rightarrow S&= i \cdot m + X, X \ge 0.\ Let\ X = a \cdot m + b&, a\ge 0, 0\le b<m \\
\Rightarrow S&=i \cdot m + (a \cdot m + b) = (i + a)\cdot m + b&, a\ge 0, 0\le b<m 
\label{eqn:inequality_lemma2_1}
\end{align}

Second, since $b < m$, we have:

\begin{align}
\frac{b}{M-m} < \frac{m}{M-m} \le i \Rightarrow \frac{b}{i} < M-m
\label{eqn:inequality_lemma2_2}
\end{align}

From Equation~\ref{eqn:inequality_lemma2_1} and \ref{eqn:inequality_lemma2_2}, we can make a valid partitioning for a partition size $S$ as follows: 
\begin{enumerate}
    \item Start with $i+a$ empty partitions. Assign $m$ records to each partition. The remaining number of records is $b$. This step is based on Equation~\ref{eqn:inequality_lemma2_1}.
    
    \item Equation~\ref{eqn:inequality_lemma2_2} means that we can split $b$ records over $i$ groups such that each group receives at most $M-m$ records. Since we already have $i+a$ groups each of size $m$, adding $M-m$ to $i$ groups out of them will increase their sizes to $M$ which still keeps them in the valid range $[m,M]$. The remaining groups will still have $m$ records making them valid too.
\end{enumerate}

This completes the proof of Lemma~\ref{lemma:min_valid_size}.
\end{proof}

Based on Lemma~\ref{lemma:min_valid_size}, a question is raised as how large the size of sample points $S$ should be to ensure that a good block utilization is achievable. As we mentioned from beginning, R*-Grove allows us to configure a parameter $\alpha = m/M$, that called \emph{balance factor}, is computed as the ratio between minimum and maximum number of records of a leaf node in the tree. $\alpha$ should be close to $1$ to guarantee a good block utilization. Let's assume that $0<r\le 1$ is the sampling ratio and $p$ is the storage size of a single point. The maximum number of records $M$ is computed in the Section~\ref{sec:rsgrove:basic} as:

\begin{align}
    M=\Ceil{\frac{|S|\cdot B}{D}} \Rightarrow M = \Ceil{\frac{|S|\cdot p}{D} \cdot \frac{B}{p}} \Rightarrow M = \Ceil{\frac{r \cdot B}{p}}
    \label{eqn:m_b}
\end{align}

From Equation~\ref{eqn:m_b}, we can rewrite Lemma~\ref{lemma:min_valid_size} as:
\begin{equation}
|S| \geq S^*=\Ceil{\frac{m}{M-m}} \cdot m \Rightarrow |S| \geq \Ceil{\frac{\alpha}{1-\alpha}} \cdot \alpha \cdot \Ceil{\frac{r \cdot B}{p}} \Rightarrow |S| \cdot p \geq \Ceil{\frac{\alpha}{1-\alpha}} \cdot \alpha \cdot \Ceil{r \cdot B}
\label{eqn:S*2}
\end{equation}

Therefore, assume that we want to configure the \emph{balance factor} as $\alpha=0.95$, sample ratio $r=1\%$ and block size $B=128$ MB, then the term $|S|\cdot p$ in Equation~\ref{eqn:S*2} would be computed as $23$ MB. In other words, if the storage size of sample points  $|S| \cdot p \geq 23$ MB, it will be guaranteed to produce a valid partitioning. This is a reasonable size that can be stored in main memory and processed in a single machine.

\subsection{Load Balancing for Datasets with Variable-size Records}
\label{sec:rsgrove:histogram}

The above two approaches can be combined to produce high-quality and balanced partitions in terms of number of records. However, the partitioning technique needs to write the actual records in each partition and often these records are of variable sizes. For example, the sizes of records in the {\tt OSM-Objects} dataset~\cite{allobjects} range from 12 bytes to 10 MB per record. Therefore, balancing the number of records can result in a huge variance in the partition sizes in terms of number of bytes.

\begin{algorithm}[t]
\begin{algorithmic}[1]
\Function{ChooseWeightedSplitPoint}{$P$,$w$,$m$}
\State $W=\sum_{1\le i\le |P|}{w_i}$
\For{$k$ in $[m,|P|-m]$} \label{alg:choosesweightedsplitpoint:for}
    \State $W_1=\sum_{1 \le i\le k}{w_i}$ \label{alg:choosesweightedsplitpoint:calcw}
    \If{either $W_1$ or $W-W_1$ is invalid} \Comment{ Lemma~\ref{lemma:check_valid_partition_size}}
    \label{alg:choosesweightedsplitpoint:test}
      \State Skip this iteration and {\bf continue}
    \EndIf
    \State Similar to Lines~\ref{alg:choosesplitpoint:split1}-\ref{alg:choosesplitpoint:updatemin} in Algorithm~\ref{alg:choosesplitpoint}
\EndFor
\State \Return chosenK
\EndFunction
\end{algorithmic}
\caption{Choose the split point with weights\\
Inputs: $P$ is the all sample records; $w$ is an array of weights of corresponding records in $P$; $[m,M]$ is the target range of sizes for final partitions.\\
Output: the optimal splitting position.}
\label{alg:choosesweightedsplitpoint}
\end{algorithm}

To overcome this limitation, we combine the sample points with a {\em storage size histogram} of the input as follows. The storage size histogram is used to assign a weight to each sample point that represents the total size of all records in its vicinity. To find these weights, Phase~1 computes, in addition to the sample, a storage size histogram of the input. This histogram is created by overlaying a uniform grid on the input space and computing the total size of all records that lie in each grid cell \cite{CE17, siddique2019comparing}. This histogram is computed on the full dataset not the sample, therefore, it catches the actual size of the input. After that, we count the number of {\em sample} points in each grid cell. Finally, we divide the total weight of each cell among all sample points in this cell. For example, if a cell has a weight of 1,000 bytes and contains five sample points, the weight of each point in this cell becomes 200 bytes.

In Phase~2, the {\sc SplitNode} function is further improved to balance the {\em total weight} of the points in each partition rather than the number of points. This also requires modifying the value of $M$ to be $M=\Ceil{\sum w_i / N}$, where $w_i$ is the assigned weight to the sample point $p_i$, and $N$ is the desired number of partitions. Algorithm~\ref{alg:choosesweightedsplitpoint} shows how the algorithm is modified to take the weights into account. Line~\ref{alg:choosesweightedsplitpoint:calcw} calculates the weight of each partitioning point which is used to test the validity of this split point as shown in Line~\ref{alg:choosesweightedsplitpoint:test}.

Unfortunately, if we apply this change, the algorithm is no longer guaranteed to produce balanced partitions. The reason is that the proof of Lemma~\ref{lemma:check_valid_partition_size} is no longer valid. That proof assumed that the partition sizes are defined in terms of number of records which makes all possible partition sizes part of the search space in the {\bf for-loop} in Line~\ref{alg:choosesvalidsplitpoint:for} of Algorithm~\ref{alg:choosesvalidsplitpoint}. However, when the size of each partition is the sum of the weights, the possible sizes are limited to the weights of the points. For example, let us assume a partition with five points all of the same weight $w_i=200$ while $m=450$ and $M=550$. The condition in Definition~\ref{def:valid} suggests that the total weight $1,000$ is valid because $L=\Ceil{1000/550}=2\le U=\lfloor1000/450\rfloor=2$. However, given the weights $w_i=200$ for $i\in[1,5]$, there is no valid partitioning, i.e., there is no way to make two partitions each with a total weight in the range $[450,550]$.

To overcome this issue, this part further improves the {\sc SplitNode} algorithm so that it still guarantees a valid partitioning even for the case described above. The key idea is to make minimal changes to the weights to ensure that the algorithm will terminate with a valid partitioning; we call this process {\em weight correction}. For example, the case described earlier will be resolved by changing the weights of two points from $200$ and $200$ to $100$ and $300$. This will result in the valid partitioning $\{200, 200, 100\}$ and $\{300, 200\}$ which is valid. Keep in mind that these weights are approximate anyway as they are based on the sample and histogram so these minimal changes would not hugely affect the overall quality, yet, they ensure that the algorithm will terminate correctly. The following part describes how these weight changes are applied while ensuring a valid answer.

First of all, we assume that the points are already sorted along the chosen axis as explained in Section~\ref{sec:rsgrove:basic}. Further, we assume that Algorithm~\ref{alg:choosesweightedsplitpoint} failed by not finding any valid partitions, i.e., return -1. Now, we make the following definitions to use them in the weight update function.

\begin{figure}[t]
%\centering
    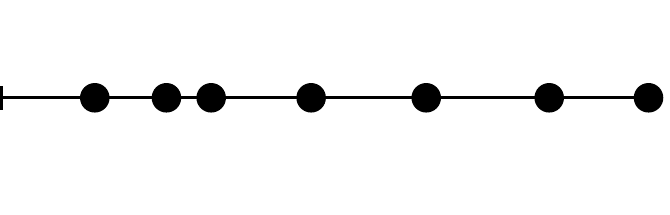 \\
    (a)~Positions of points \\~~\\~~\\
    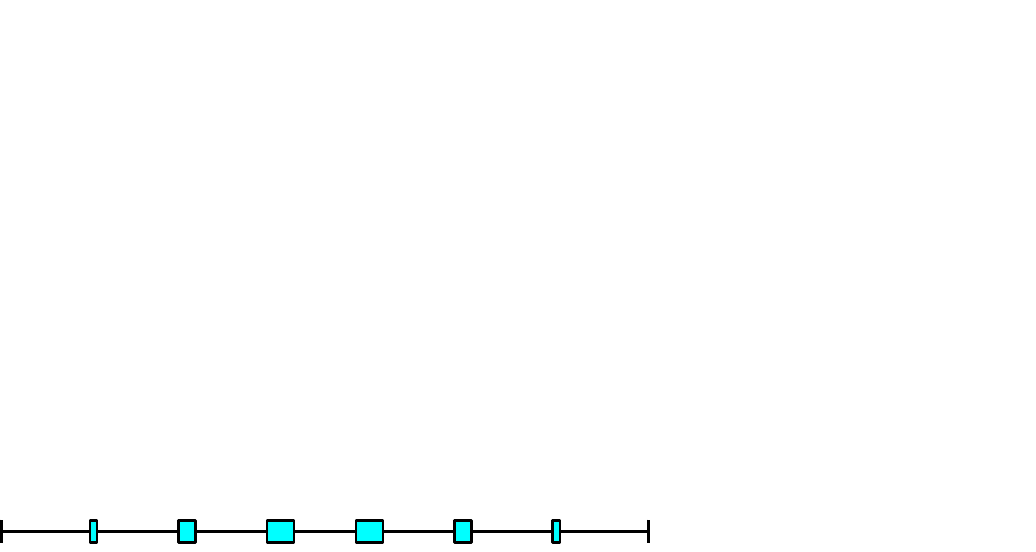 \\
    (b)~Valid ranges \\~~\\~~\\
    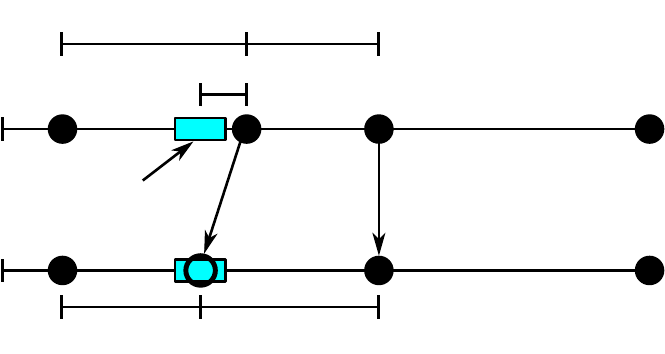 \\
    (c)~Weight correction
    \caption{Load balancing for datasets with variable-size records}
    \label{fig:load_balance}
\end{figure}

\begin{definition}\textbf{Point position:}
Let $p_i$ be point $\#i$ in the sort order and its weight is $w_i$. We define the position of the point $i$ as $pos_i=\sum_{j\leq i}{w_j}$.
\end{definition}

Based on this definition, we can place all the points on a linear scale based on their position as shown in Figure~\ref{fig:load_balance}(a).

\begin{definition}\textbf{Valid left range:}
A range of positions $VL=[vl_s,vl_e]$ is a valid left range if for all positions $vl\in VL$ the value $vl$ is valid w.r.t. $[m,M]$. All the valid left ranges can be written in the form $[im,iM]$ where $i$ is a natural number and they might overlap for large values of $i$. (See Figure~\ref{fig:load_balance}(b).)
\end{definition}

\begin{definition}\textbf{Valid right range:}
A range of positions $VR=[vr_s,vr_e]$ is a valid right range if for all positions $vr\in VR$ the value $W-vr$ is valid w.r.t. $[m,M]$. Similar to valid left ranges, all valid right ranges can be written in the form $[W-jM,W-jm]$, where $W=\sum{w_i}$. (See Figure~\ref{fig:load_balance}(b).)
\end{definition}

\begin{definition}\textbf{Valid range:}
A range of positions $V=[v_s,v_e]$ is valid if for all positions $v\in V$, $v$ belongs to at least one valid left range and at least one valid right range. In other words, the valid ranges are the intersection of the valid left ranges and valid right ranges.
\end{definition}

Figure~\ref{fig:load_balance}(b) illustrates the valid left, valid right, and valid ranges. If we split a partition around a point with a position in a valid left range, the first partition will be valid. Similarly for valid right positions the second partition (on the right) will be valid. Therefore, we would like to split a partition around a point in one of the valid ranges (intersection of left and right).

\begin{lemma}\textbf{Empty valid ranges:}
If Algorithm~\ref{alg:choosesweightedsplitpoint} fails by returning -1, then none of the point positions in $P$ falls in a valid range.
\label{lem:empty-ranges}
\end{lemma}

\begin{proof}
By contradiction, let a point $p_i$ has a position $pos_i$ that falls in a valid range. In this case, the partitions $P_1=\{p_k:k\leq i\}$ and $P_2=\{p_l:l> i\}$ are both valid partitions because the total weight of $P_1$ is equal to the position $pos_i$ which is valid because $pos_i$ falls in a valid left range. Similarly, the total weight of $P_2$ is valid because $pos_i$ falls in a valid right range. In this case, Algorithm~\ref{alg:choosesweightedsplitpoint} should have found this partitioning as a valid partitioning because it tests all the points which is a contradiction.
\end{proof}

A corollary to Lemma~\ref{lem:empty-ranges} is that when Algorithm~\ref{alg:choosesweightedsplitpoint} fails by returning -1, then all valid ranges are empty.

As a result, we would like to slightly modify the weights of some points in the sample points in order to enforce some points to fall in valid ranges. We call this the \emph{weight correction} process. This process is described in the following lemma:

\begin{lemma}\textbf{Weight correction:}
Given any empty valid range $[v_s,v_e]$, we can modify the weight of only two points such that the position of one of them will fall in the range.
\label{lem:weight_correction}
\end{lemma}

\begin{proof}
Figure~\ref{fig:load_balance}(c) illustrates the proof of this lemma.
Given an empty valid range, we modify the two points with positions that follow the empty valid range, $p_1$ and $p_2$, where $pos_1<pos_2$. We would like to move the point $p_1$ to the new position $pos_1'=(v_s+v_e)/2$ which is in the middle of the empty valid range. To do that, we reduce the weight $w_1$ by $\Delta pos=pos_1-pos_1'$. The updated weight $w_1'=w_1-\Delta pos$. To keep the position of $p_2$ and all the following points intact, we have to also increase the weight of $p_2$ by $\Delta pos$; that is, $w_2'=w_2+\Delta pos$.
\end{proof}

We do the weight correction process for {\em all} empty valid ranges to make them non-empty and then we repeat Algorithm~\ref{alg:choosesweightedsplitpoint} to choose the best one among them.

The only remaining part is how to enumerate all the valid ranges. The idea is to simply find a valid left range, an overlapping valid right range, and compute their intersection, all in constant time. Given a natural number $i$, the valid left range is in the form $[im,iM]$. Assume that this range overlaps a valid right range in the form $[W-jM,W-jm]$. Since they overlap, the following two inequalities should hold:

\[W-jm\ge im \Rightarrow j < \frac{W-im}{m}\]
\[W-jM\leq iM \Rightarrow j > \frac{W-iM}{M}\]

Therefore, the lower bound of $j$ is $j_1=\Ceil{\frac{W-i\cdot M}{M}}$ and the upper bound of $j$ is $j_2=\Floor{\frac{W-i\cdot m}{m}}$. If $j_1\leq j_2$, then there is a solution to these inequalities which we use to generate the bounds of the valid range $[v_s,v_e]$. Notice that if there is more than one valid solution to $j$, all of them should be considered to generate all the valid ranges but we omit this special case for brevity.

\subsection{Implementation Considerations}

\begin{figure}[t]
\centering
\includegraphics[width=0.5\textwidth]{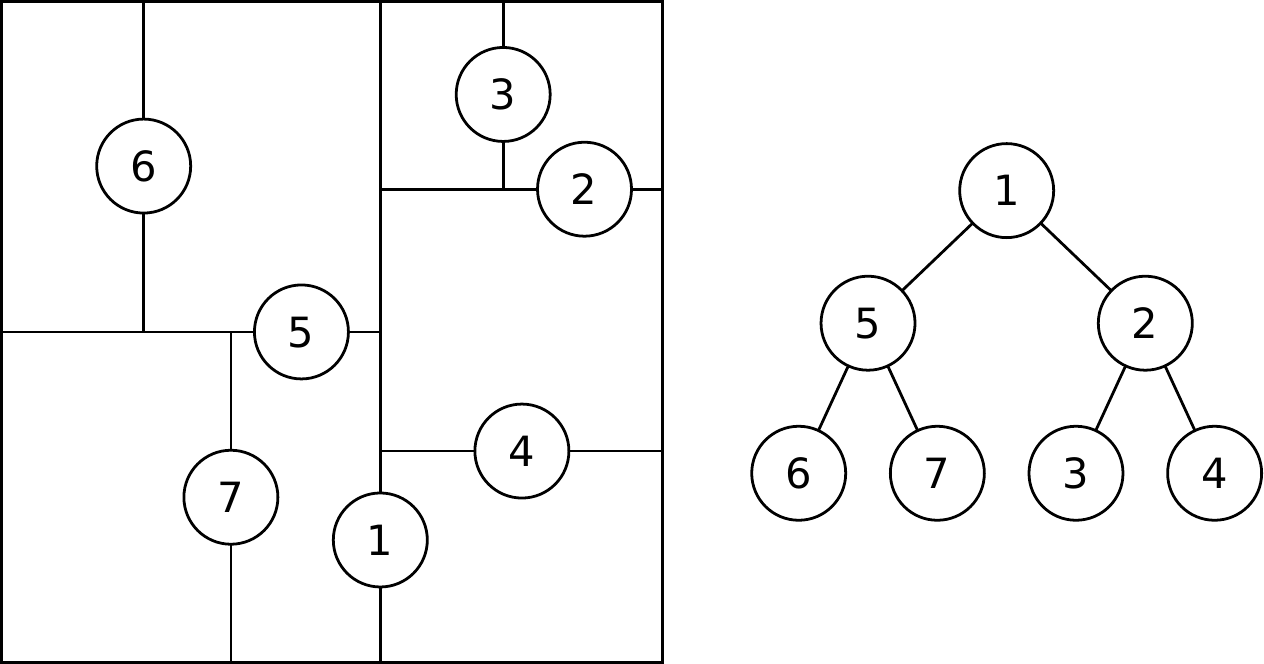}
\caption{Auxiliary search structure for R*-Grove}
\label{fig:aux}
\end{figure}

\vspace{4pt}\noindent {\bf Optimization of Phase~3:} The {\sc ChooseSubTree} operation in R*-tree chooses the node that results in the least overlap increase with its siblings~\cite{BKS+90}.
A straight-forward implementation of this method is $O(n^2)$ as it needs to compute the overlap between each candidate partition and all other partitions. In the R*-tree index, this cost is limited due to the limited size of each node. However, this step can be too slow as the number of partitions in R*-Grove can be extremely large. To speed up this step, we use a K-d-tree-like auxiliary search structure as shown in Figure~\ref{fig:aux}. This index structure is generated during Phase~2 as the partition boundaries are computed. Each time the {\sc NodeSplit} operation completes, the search structure is updated by adding a corresponding split in the direction of the chosen split axis. This auxiliary search structure is stored in memory and replicated to all nodes. It will be used in Phase 3, when we physically store the input records to the partitions. Given a spatial record, it will be assigned to the corresponding partition using a search algorithm which is similar to the K-d-tree's point search algorithm~\cite{bentley1975multidimensional}. Based on this similarity, we can estimate the running time to choose a partition to be $O(log(n))$.
Notice that this optimization is not applicable in traditional R*-trees as the partitions might be overlapping while in R*-Grove we utilize the fact that we only partition points which guarantees disjoint partitions. 

Since the partition MBRs in Phase 2 are computed from sample objects, there will be objects which do not fall in any partition in Phase 3. R*-Grove addresses this problem in two ways. First, if no disjoint partitions are desired, it chooses a single partition based on the {\sc ChooseLeaf} method in original R*-tree. In short, an object will be assigned to the partition in which the enlarged area or margin is minimal. Second, if disjoint partitions are desired, R*-Grove uses the auxiliary data structure, which covers the entire space, to assign this record to all overlapping partitions.

\vspace{4pt}\noindent {\bf Disjoint indexes:} Another advantage of using the auxiliary search structure described above, is that it allows for building a disjoint index. This search structure naturally provides disjoint partitions. To ensure that the partitions cover the entire input space, we assume that input region is infinite, that is, starts from $-\infty$ and ends at $+\infty$ in all dimensions. Then, Phase~3 replicates each record to all overlapping partitions by directly searching in this $k$-d-tree-like structure with range search algorithm, which has the $O(\sqrt{n})$ running time~\cite{lee1977worst}. This advantage was not possible with the black-box R*-tree implementation as it is not guaranteed to provide disjoint partitions.

% -------------- Case Studies -------------
\section{Case Studies}
\label{sec:case-studies}
This section describes three case studies where the R*-Grove partitioning technique can improve big spatial data processing. We consider three fundamental operations, namely, indexing, range query, and spatial join.

\subsection{Indexing}
Spatial data indexing is an essential component in most big spatial data management systems. The state-of-the-art global indexing techniques rely on reusing existing index structures with a sample which are shown to be inefficient in terms of quality and load balancing \cite{eldawy2015spatial,yu2015geospark,VAW14}.

R*-Grove partitioning can be used for the {\em global indexing} step which partitions records across machines. In big spatial data indexing, the global index is the most crucial step as it ensures load balancing and efficient pruning when the index is used. If only the number of records needs to be balanced or if the records are roughly equi-sized, then the techniques described in Sections~\ref{sec:rsgrove:basic} and~\ref{sec:rsgrove:balance} can be used. If the records are of a variable size and the total sizes of partitions need to be balanced, then the histogram-based step in Section~\ref{sec:rsgrove:histogram} can be added to ensure a higher load balance. Notice that the index would hugely benefit from the balanced partition size as it reduces the total number of blocks in the output file which improves the performance of all Spark and MapReduce queries that create one task per file block.

\subsection{Range Query}
Range query is a popular spatial query, which is also the building block of many other complex spatial queries. Previous studies found a strong correlation between the performance of range queries and the performance of other queries such as spatial join~\cite{eldawy2015spatial,HS94}. Therefore, the performance of range query could be considered as a good reflection about the quality of a partitioning technique. A good partitioning technique allows the query processor to make two optimization techniques. First, it can prune the partitions that are completely outside the query range. Second, it can directly write to the output the partitions that are completely contained in the query range without further processing \cite{ESE+17}. For very small ranges, most partitioning techniques will behave similarly as it is most likely that the small query overlaps one partition and no partitions are completely contained \cite{EAM15}. However, as the query range increases, the differences between the partitioning techniques become apparent. Since most range queries are expected to be square-like, the R*-Grove partitioning is expected to perform very well as it minimizes the total margin which produces square-like partitions. Furthermore, the balanced load across partitions minimizes the straggler effect where one partition takes significantly longer time than all other partitions.

\subsection{Spatial Join}
Spatial join is another important spatial query that benefits from the improved R*-Grove partitioning technique. In spatial join, two big datasets need to be combined to find all the overlapping pairs of records. To support spatial join on partitioned big spatial data, each dataset is partitioned independently. Then, a spatial join runs between the partition boundaries to find all pairs of overlapping partitions. Finally, these pairs of partitions are processed in parallel. An existing approach \cite{Zhou1998} preserves spatial locality to reduce the processing jobs. However, it still relies on traditional index like R-Tree, which also inherited its limitations. The R*-Grove partitioning has two advantages for the spatial join operation. First, it is expected to reduce the number of partitions by increasing the load balance which reduces the total number of pairs. Second, it produces square-like partitions which is expected to overlap with fewer partitions of the other dataset as compared to the very thin and wide partitions that the STR or other partitioning techniques produce. These advantages allows R*-Grove to significantly outperform other partitioning techniques in spatial join query performance. We will validate these advantages in the Section~\ref{sec:spatial_join}.

% -------------- Experiments -------------
\section{Experiments}
\label{sec:experiments}

In this section, we carry out an extensive experimental study to validate the advantages of R*-Grove over widely used partitioning techniques, such as bulk loading STR, Kd-tree, Z-Curve and Hilbert curve. We will show how R*-Grove addresses the current limitations of those techniques, leads to a better performance in big spatial data processing. In addition, we also show other capabilities of R*-Grove in the context of big spatial data, for example, how it works with large or multi-dimensional datasets. The experimental results in this section provide an evidence to the spatial community to start using R*-Grove if they would like to improve the system performance of their spatial applications.

\subsection{Experimental Setup}

\begin{table}[t]
\centering
\caption{Datasets for experiments}
\small
\label{tab:experimental_dataset}
\begin{tabular}{lllll}
\hline
Dataset & Type & Dimensions & Size & \# records \\
\hline
(1)~OSM-Nodes & Point & 2 & 500GB & 7.4 billions \\
(2)~OSM-Roads & Line segments & 2 & 20GB & 59 millions \\
(3)~OSM-Parks & Polygon & 2 & 7.2GB & 10 millions \\
(4)~OSM-Objects & Polygon & 2 & 96GB & 264 millions \\
(5)~NYC-Taxi & Point & 4,5,7 & 46GB & 173 millions \\
(6)~Diagonal points & Point & 3,4,5,9 & 100GB & 80 millions \\
\hline
\end{tabular}
\end{table}

\textbf{Datasets:} Table~\ref{tab:experimental_dataset} summarizes the datasets will be used in our experiments. We use both real world and synthetic datasets for our experiments: (1)~Semi-synthetic OpenStreetMap ({\tt OSM-Nodes}) dataset with $7.4$~billion points and a total size of $500$ GB. This is a semi-synthetic dataset which represents all the points in the world. The points in this dataset are generated within a pre-specified distance from original points from {\tt OSM-Nodes} dataset;
(2)~{\tt OSM-Roads} with size $20$ GB and (3)~{\tt OSM Parks} with size $7.2$ GB, which contain line segments and polygons for spatial join experiments. (4)~{\tt OSM-Objects} dataset with size $92$GB, which contains many variable-size records. (5)~{\tt NYC-Taxi} dataset with size $41.7$GB with up-to seven dimensions. All of those datasets are available online on UCR-STAR \cite{ucrstar} - our public repository for spatial data; (6)~Synthetic multi-dimensional {\tt diagonal} points, with the number of dimensions are $3, 4, 5,$ and $9$. This synthetic dataset is generated using our open source Spatial Data Generator~\cite{vu2019spatial}. Dataset (5) and (6) allow us to show the advantages of R*-Grove in multi-dimensional datasets.

{\bf Parameters and performance metrics:} In the following experiments, we partition the mentioned datasets with different datasets size $|D|$ in different techniques then we measure: (1)~partition quality metrics, namely, total partition area, total partition margin, total partition overlap, block utilization(maximum is $1.0$, i.e. $100\%$), standard deviation of partition size in MB (load balance). Notice that unit is not relevant for area, margin and overlap metric; (2)~total partitioning time (in seconds), (3)~for range queries, we measure the number of processed blocks and query running time, (4)~for spatial join, we measure the number of processed blocks and total running time. We fix the balance factor $\alpha=0.95$ and HDFS block size at $128$ MB. 

{\bf Machine specs:} All the experiments are executed on a cluster of one head node and 12~worker nodes, each having 12~cores, 64~GB of RAM, and a 10~TB HDD. They run CentOS~7 and Oracle Java 1.8.0\_131. The cluster is equipped with Apache Spark 2.3.0 and Apache Hadoop 2.9.0. The proposed indexes are available for running in both Spark and Hadoop. Unless otherwise mentioned, we use Spark by default. The source code is available at \url{https://bitbucket.org/tvu032/beast-tv/src/rsgrove/}. The implementation for R*-Grove ({\em RSGrovePartitioner}) is located at {\em indexing} package.

{\bf Baseline techniques:} We compare R*-Grove to K-d Tree, STR, Z-curve and Hilbert curve (denoted H-Curve thereafter) which are widely used in existing big spatial data systems \cite{EM16}. Z-Curve is adopted in some systems under the name Geohash which behaves in the same way.

\subsection{Effectiveness of the proposed improvements in R*-Grove}

\begin{figure}[t]
\centering
\begin{minipage}{\columnwidth}
\begin{tabular}{cc}
\includegraphics[width=0.45\columnwidth]{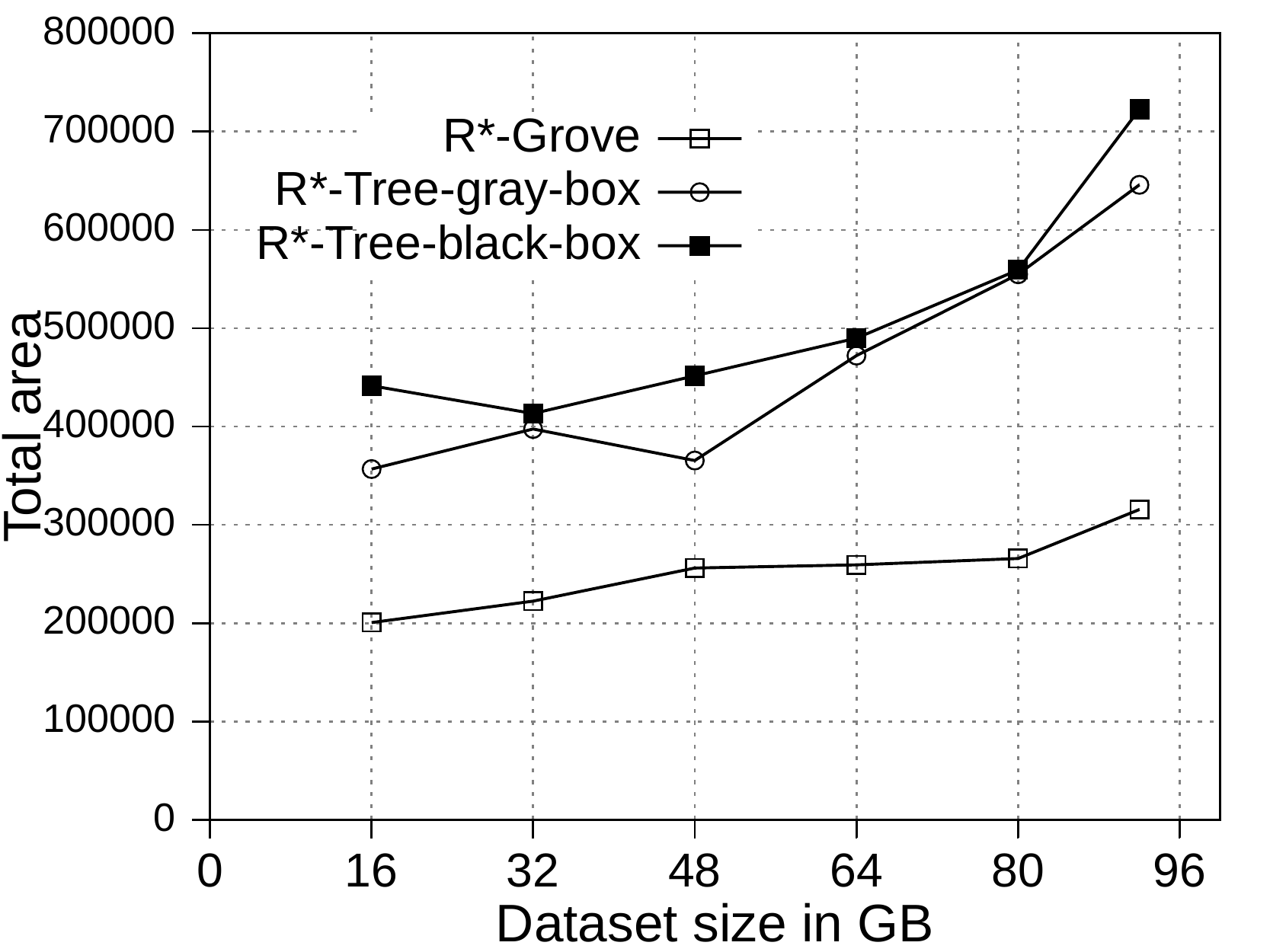}  & \includegraphics[width=0.45\columnwidth]{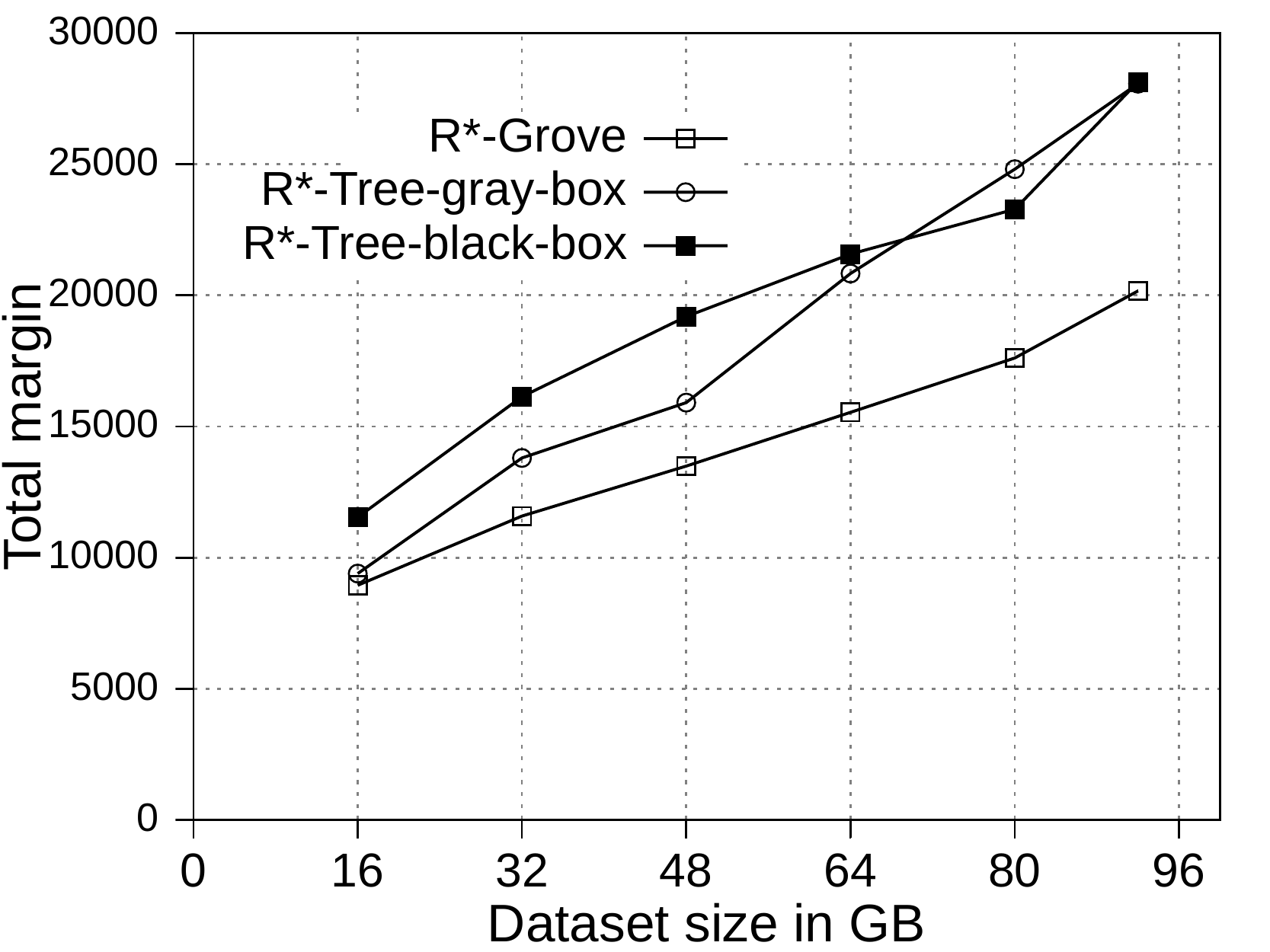}  \\
(a)~Total area & (b)~Total margin \\
\includegraphics[width=0.45\columnwidth]{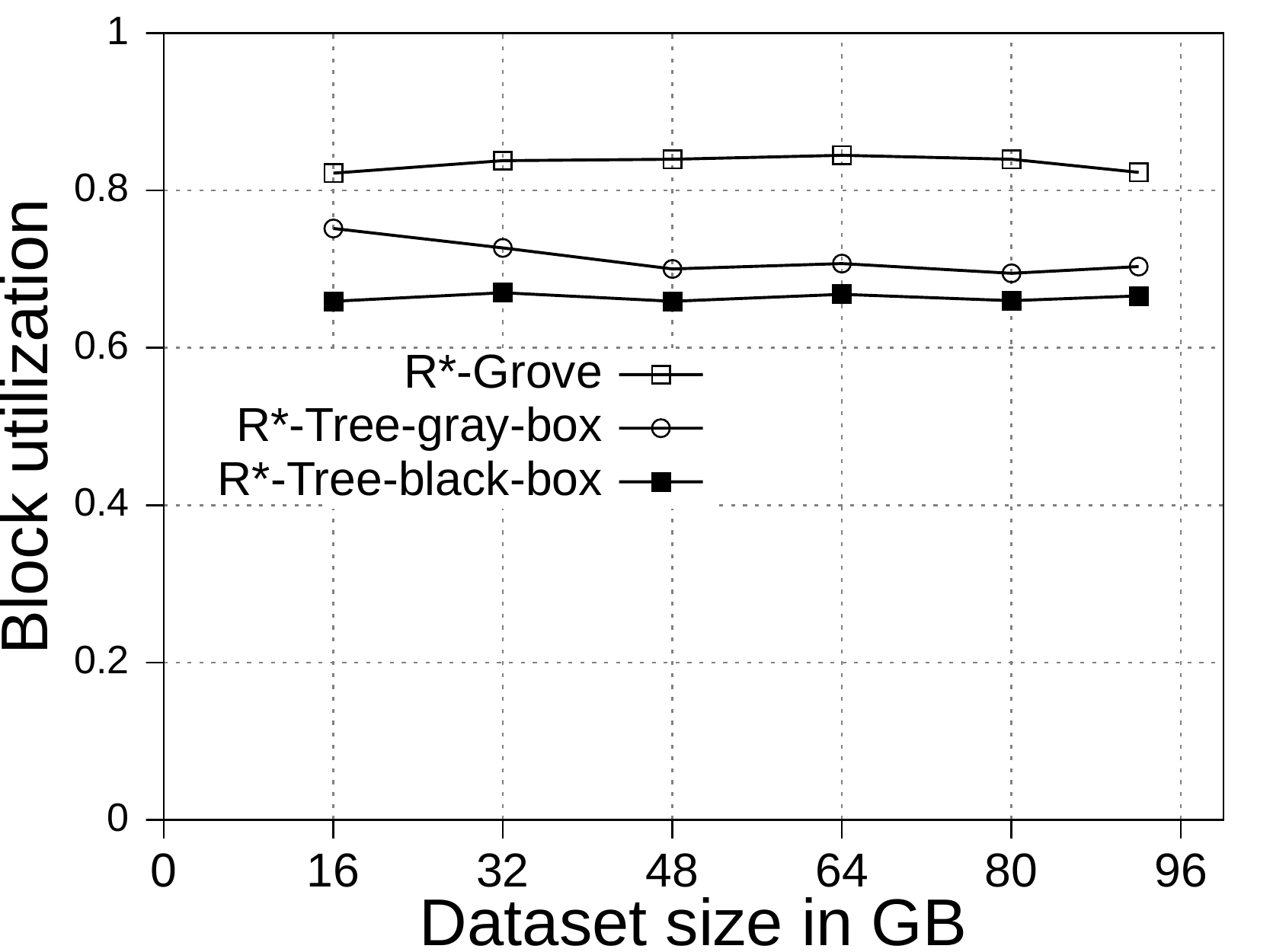}  & \includegraphics[width=0.45\columnwidth]{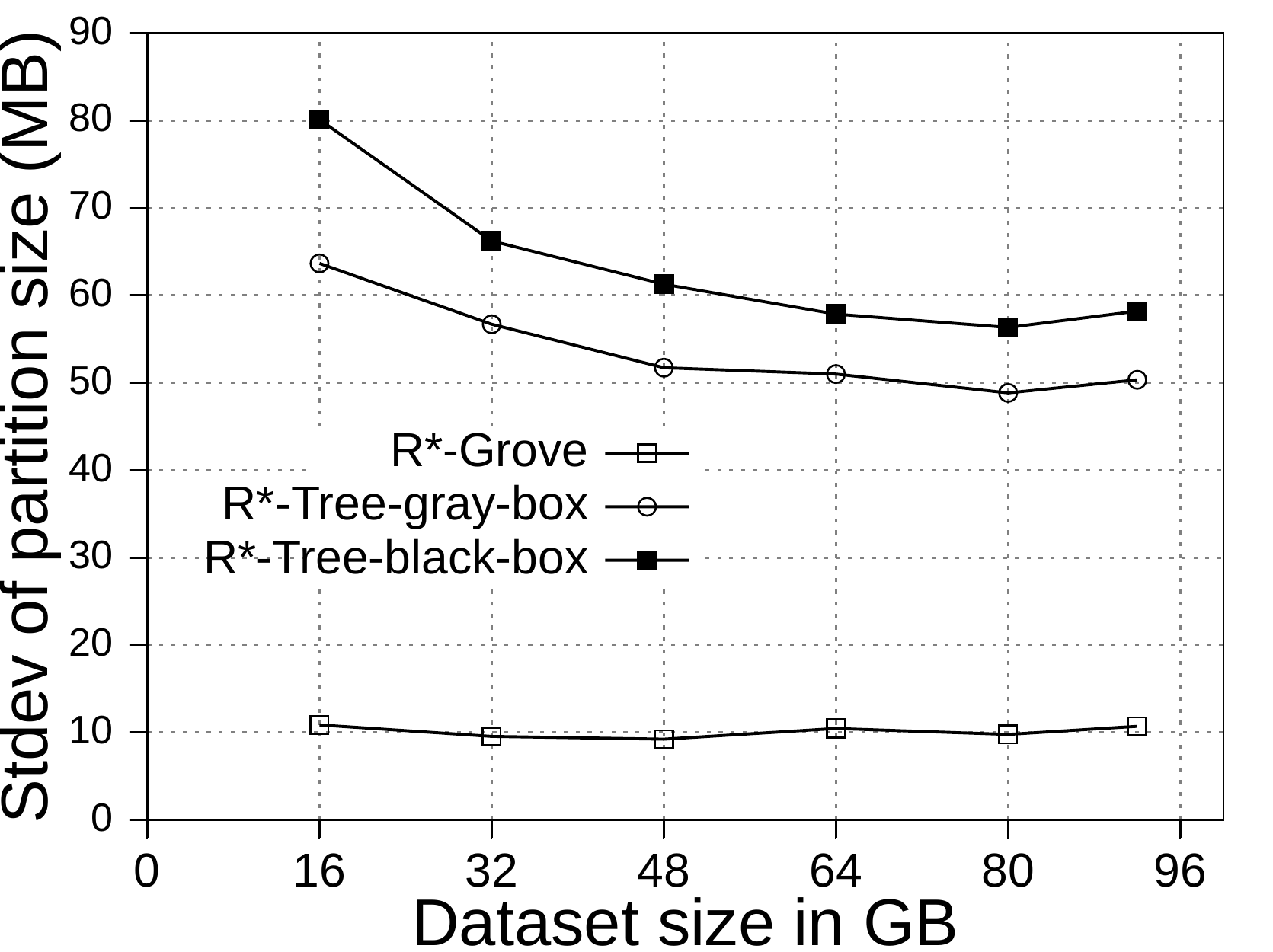} \\
(c)~Block utilization & (d)~Load balance
\end{tabular}
\end{minipage}
\caption{Partition quality with variable-record-size dataset in R*-Grove and its two variants, R*-tree-black-box and R*-tree-gray-box}
\label{fig:rstar_based_techniques}
\end{figure}

In this experiment, we compare the three following variants of R*-Grove:
(1)~\emph{R*-tree-black-box} is the application of the method in Section~\ref{sec:rsgrove:basic}. Simply, it uses the basic R*-tree algorithm to compute high-quality partition but it does not guarantee a high block utilization or load balance.
(2)~\emph{R*-tree-gray-box} applies the improvements in Sections~\ref{sec:rsgrove:basic} and~\ref{sec:rsgrove:balance}. In addition to the high-quality partition, this method can also guarantee a high block utilization in terms of number of records per partition but it does not perform well if records have highly-variable sizes since it does {\em not} include the size adjustment technique in Section~\ref{sec:rsgrove:histogram}.
(3)~\emph{R*-Grove} applies all the three improvements at Sections~\ref{sec:rsgrove:basic},~\ref{sec:rsgrove:balance} and~\ref{sec:rsgrove:histogram}. It has the advantage of producing high-quality partitions and can also guarantee a high block utilization in terms of storage size even when the record sizes are highly variable.

In Figure~\ref{fig:rstar_based_techniques}, we partition the {\tt OSM-Objects} dataset, which contains variable-size records to validate our proposed improvements. Overall, R*-Grove outperforms R*-tree-black-box and R*-tree-gray-box in all of spatial quality metrics. Especially, R*-Grove provides excellent load balance between partitions as shown in Figure~\ref{fig:rstar_based_techniques}(d), which is the standard deviation of partition size in {\tt OSM-Objects} dataset. Given the HDFS block size is $128$MB, R*-Grove has the standard deviation of partition size $5-6$ times smaller than R*-Tree-gray-box and R*-Tree-black-box. Since then, the following experiments will evaluate the performance of R*-Grove with existing widely-used spatial partitioning techniques.

\subsection{Results Overview}

\begin{figure}[t]
\centering
\includegraphics[width=1.0\columnwidth]{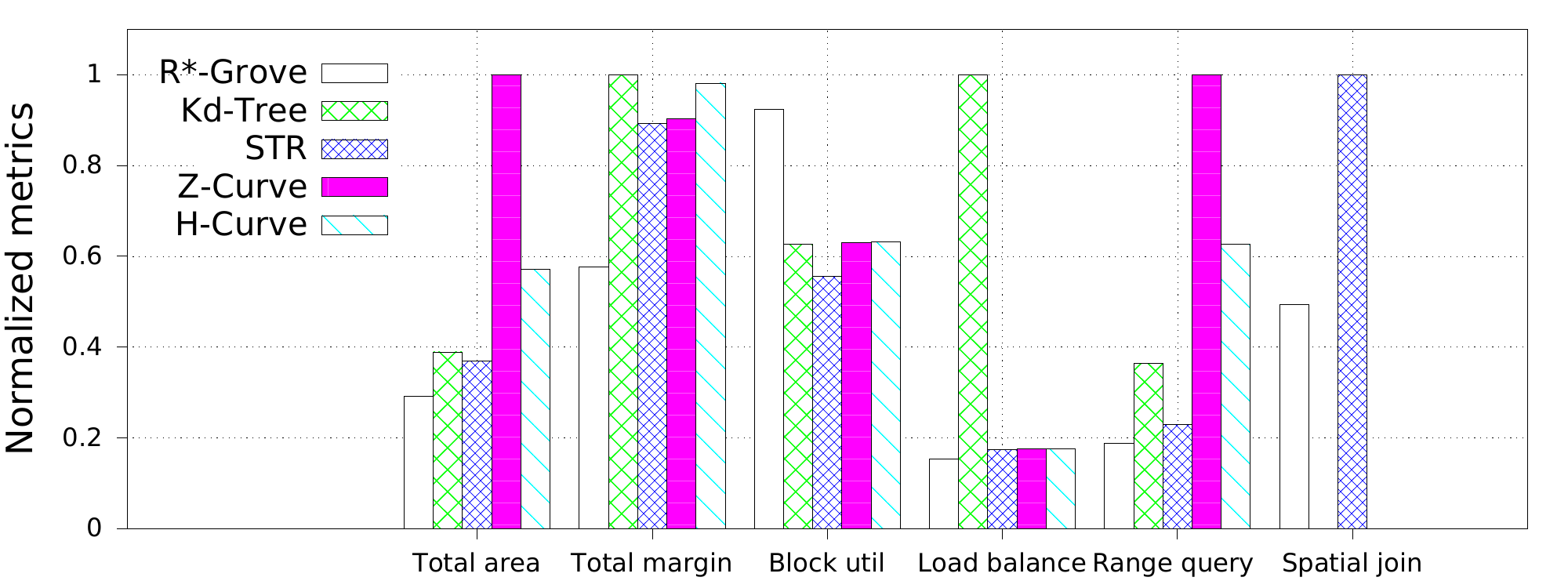}
\caption{The advantages of R*-Grove when compared to existing partitioning techniques}
\label{fig:result_overview}
\end{figure}

Figure~\ref{fig:result_overview} shows an overview of the advantages of R*-Grove over other partitioning techniques for indexing, range query, and spatial join. In this experiment, we compare to four popular baseline techniques, namely, STR, Kd-Tree, Z-Curve and H-Curve. We use {\tt OSM-Nodes} dataset \cite{ucrstar} for this experiment. The numbers on the $y-axis$ are normalized to the largest number for a better representation except for block utilization which is reported as-is. Except for block utilization, the lower the value in the chart the better it is. The first two groups, total area and total margin, show that index quality of R*-Grove is clearly better than other baselines in both measures. For block utilization, on average, a partition in R*-Grove occupy around $90\%$, while other techniques could only utilize $60-70\%$ storage capacity of an HDFS block. R*-Grove also has a better load balance when compared to other techniques. The last two groups indicate that R*-Grove significantly outperforms other partitioning techniques in terms of range query and spatial join query performance. We will go into further details in the rest of this section.

\begin{figure}[t]
\centering
\begin{minipage}{\columnwidth}
\begin{tabular}{ccc}
\includegraphics[width=0.32\columnwidth]{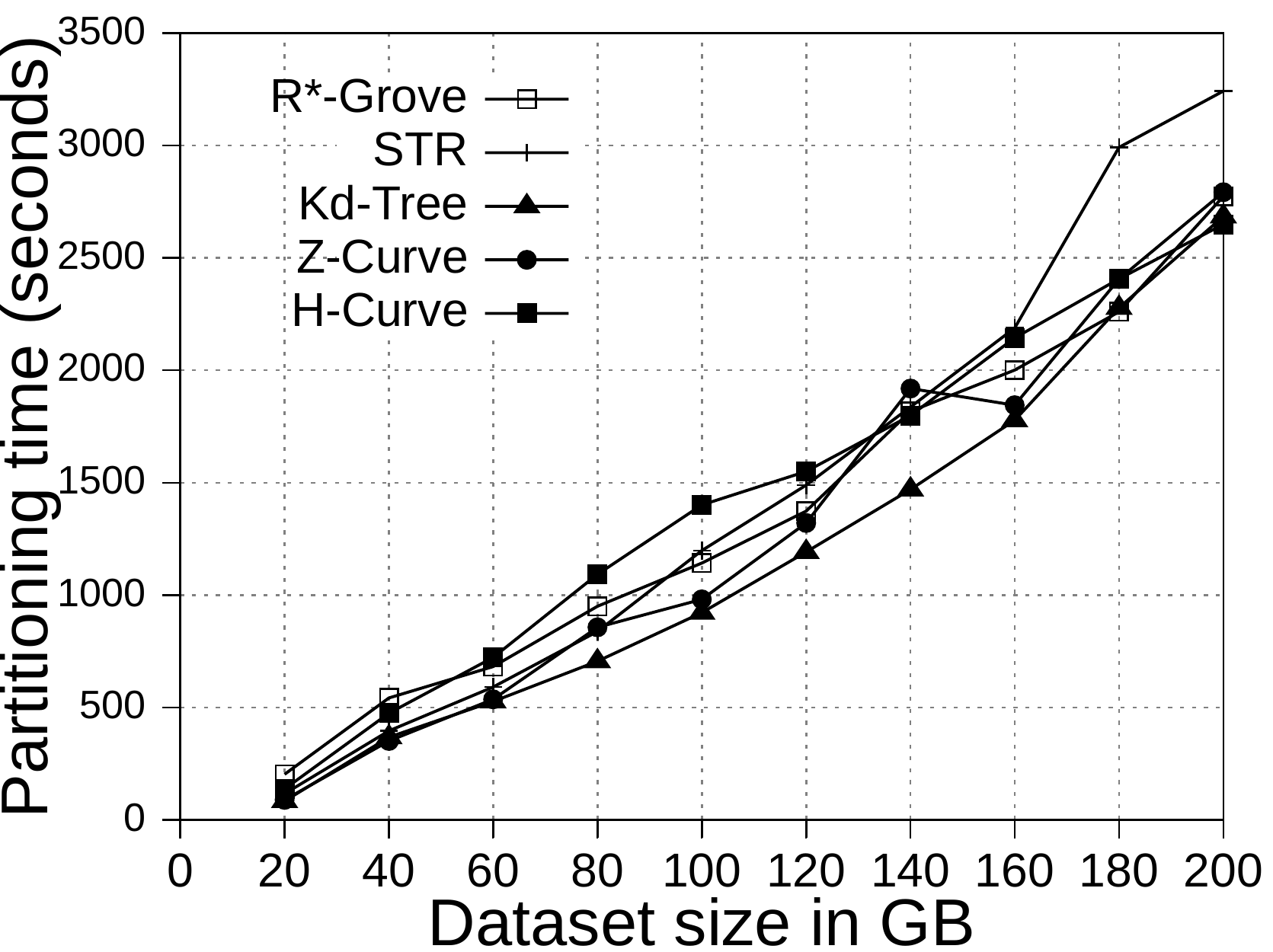}  & \includegraphics[width=0.32\columnwidth]{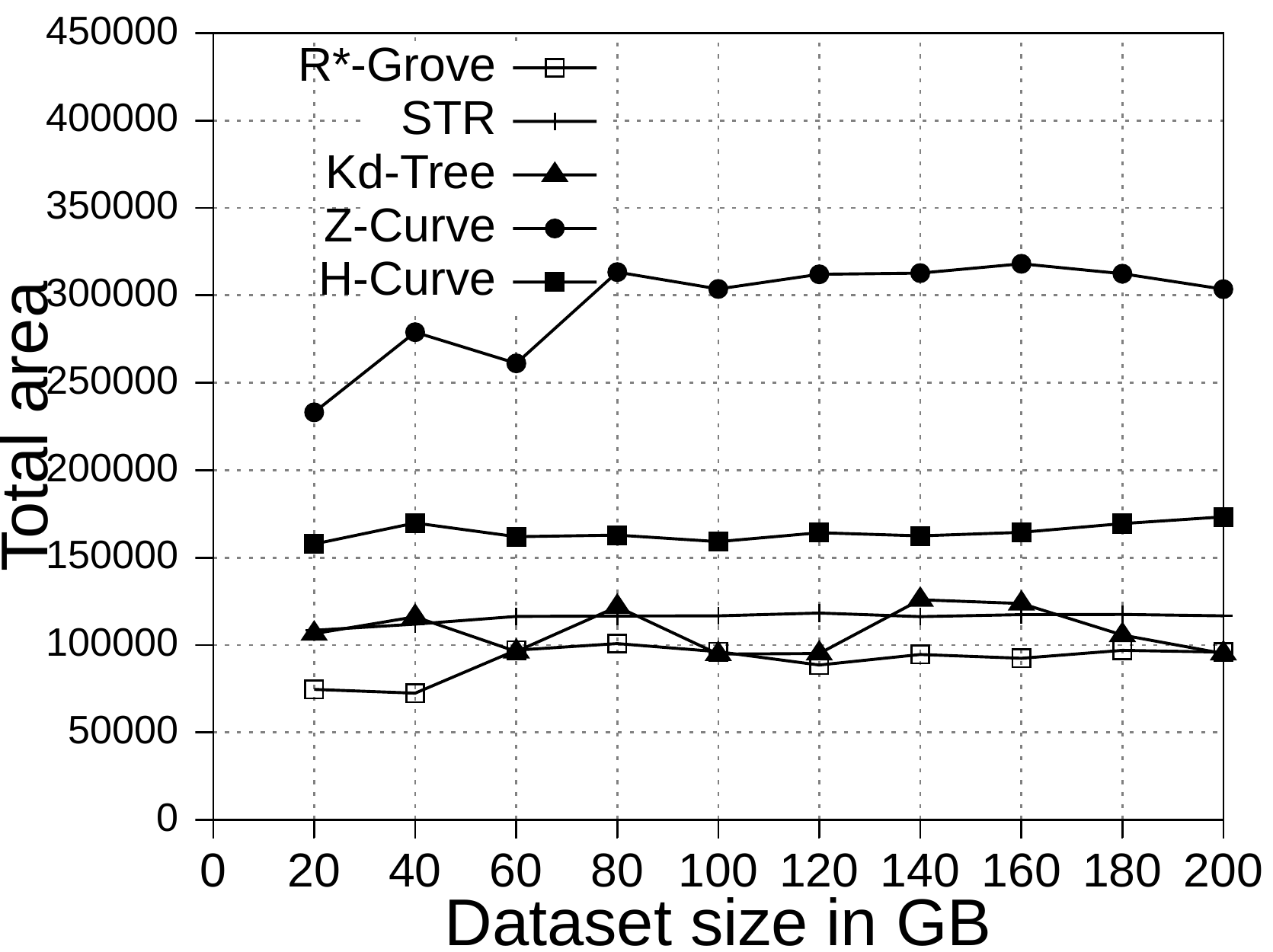} & \includegraphics[width=0.32\columnwidth]{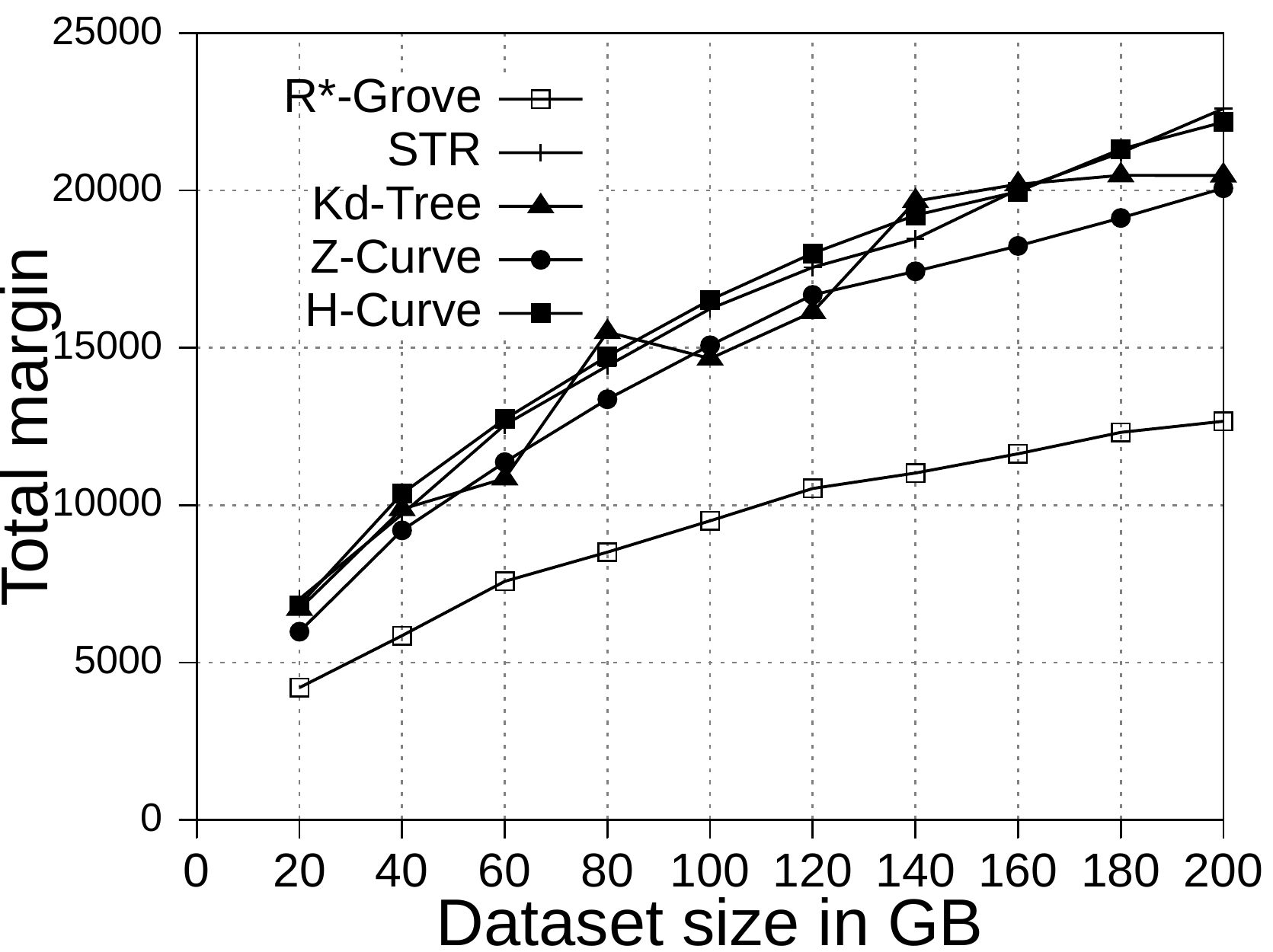} \\
(a)~Partitioning time & (b)~Total area & (c) Total margin \\
\includegraphics[width=0.32\columnwidth]{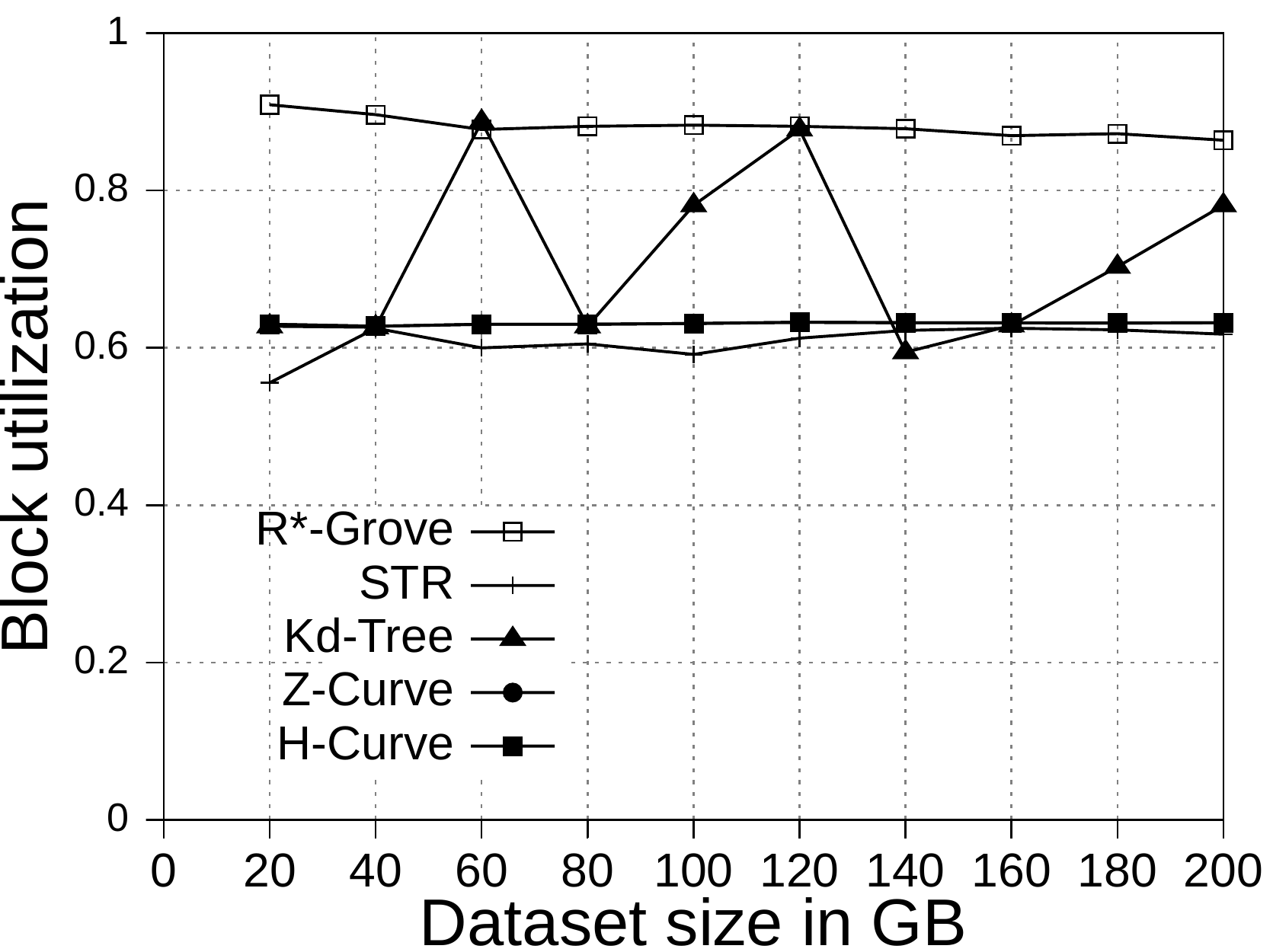}  & \includegraphics[width=0.32\columnwidth]{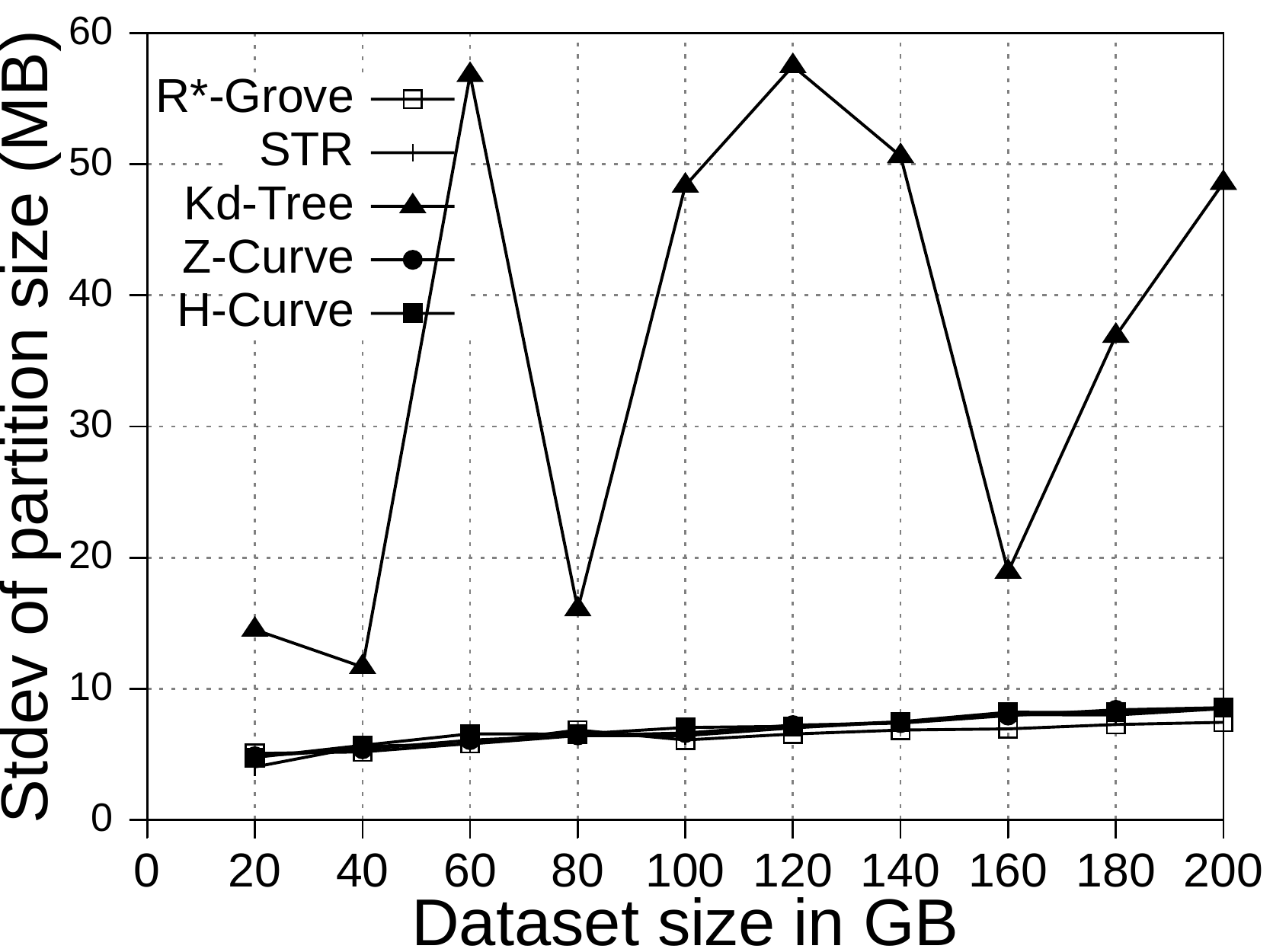} & \includegraphics[width=0.32\columnwidth]{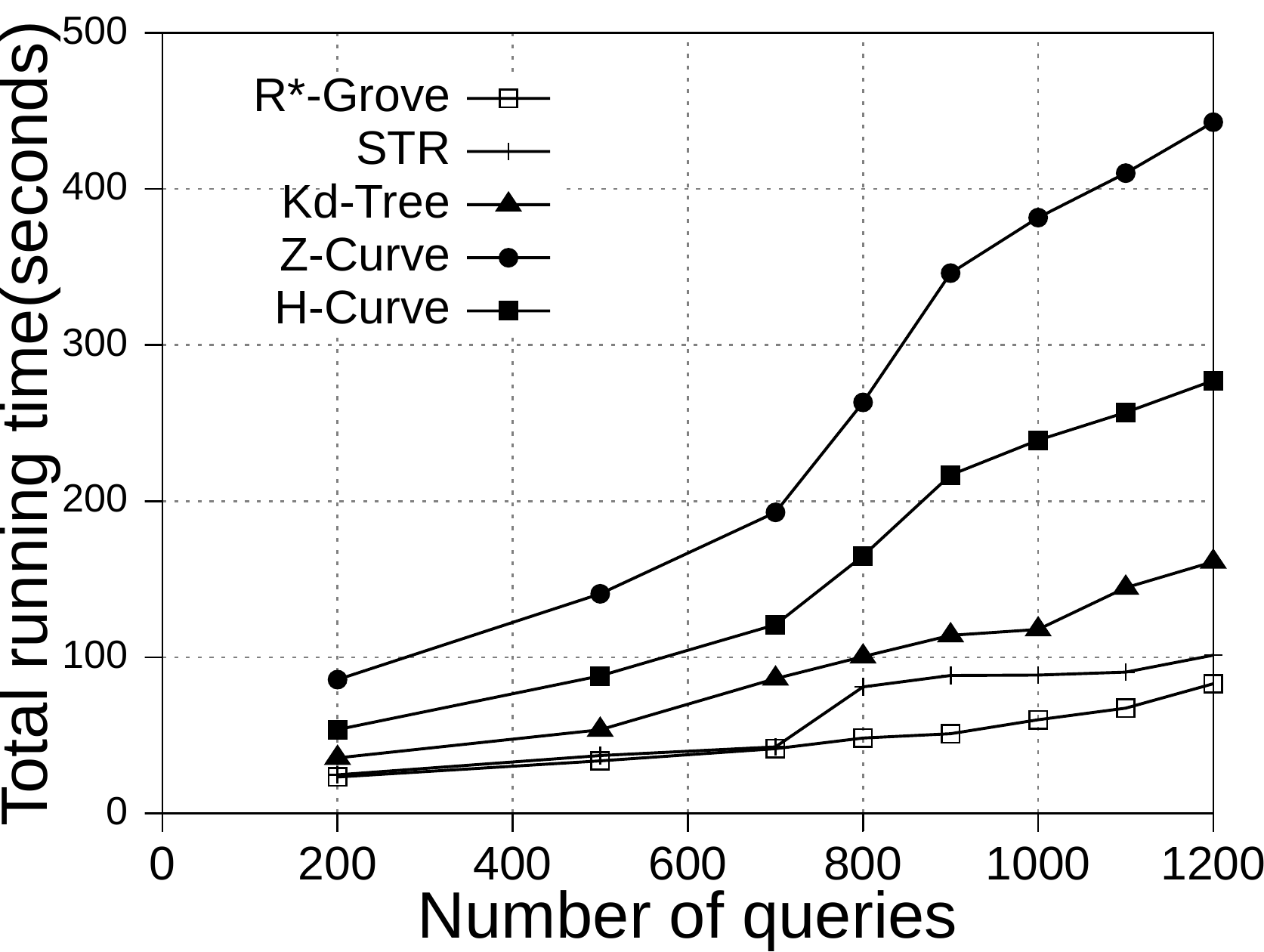} \\
(d)~Block utilization & (e)~Load balance & (f)~Range query performance
\end{tabular}
\end{minipage}
\caption{Indexing performance and partition quality of R*-Grove and other partitioning techniques in {\tt OSM-Nodes} datasets with similar-size records.}
\label{fig:comparison_all_nodes}
\end{figure}

\subsection{Partition quality}
\begin{figure}[t]
\centering
\begin{minipage}{\columnwidth}
\begin{tabular}{ccc}
\includegraphics[width=0.32\columnwidth]{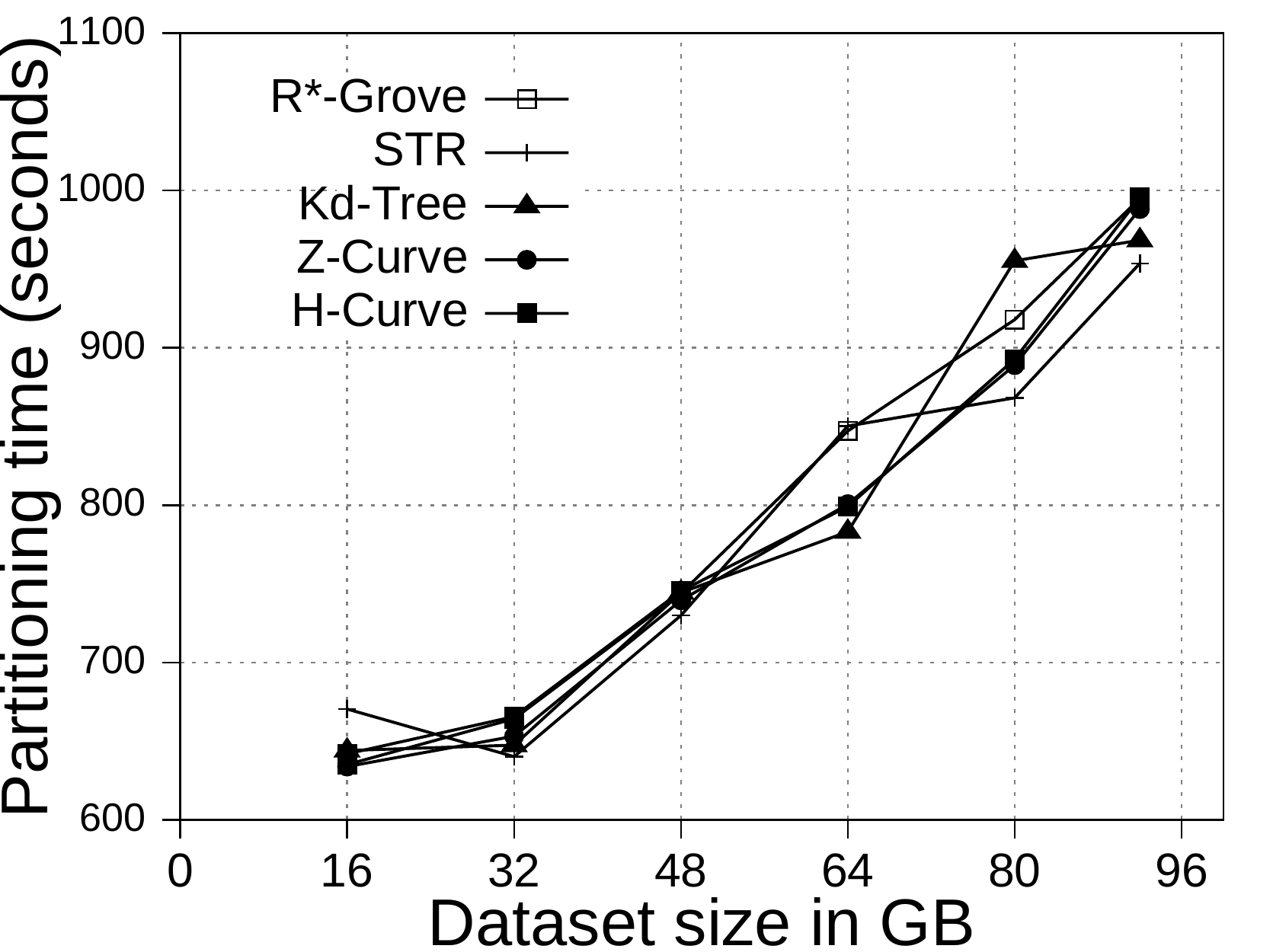}  & \includegraphics[width=0.32\columnwidth]{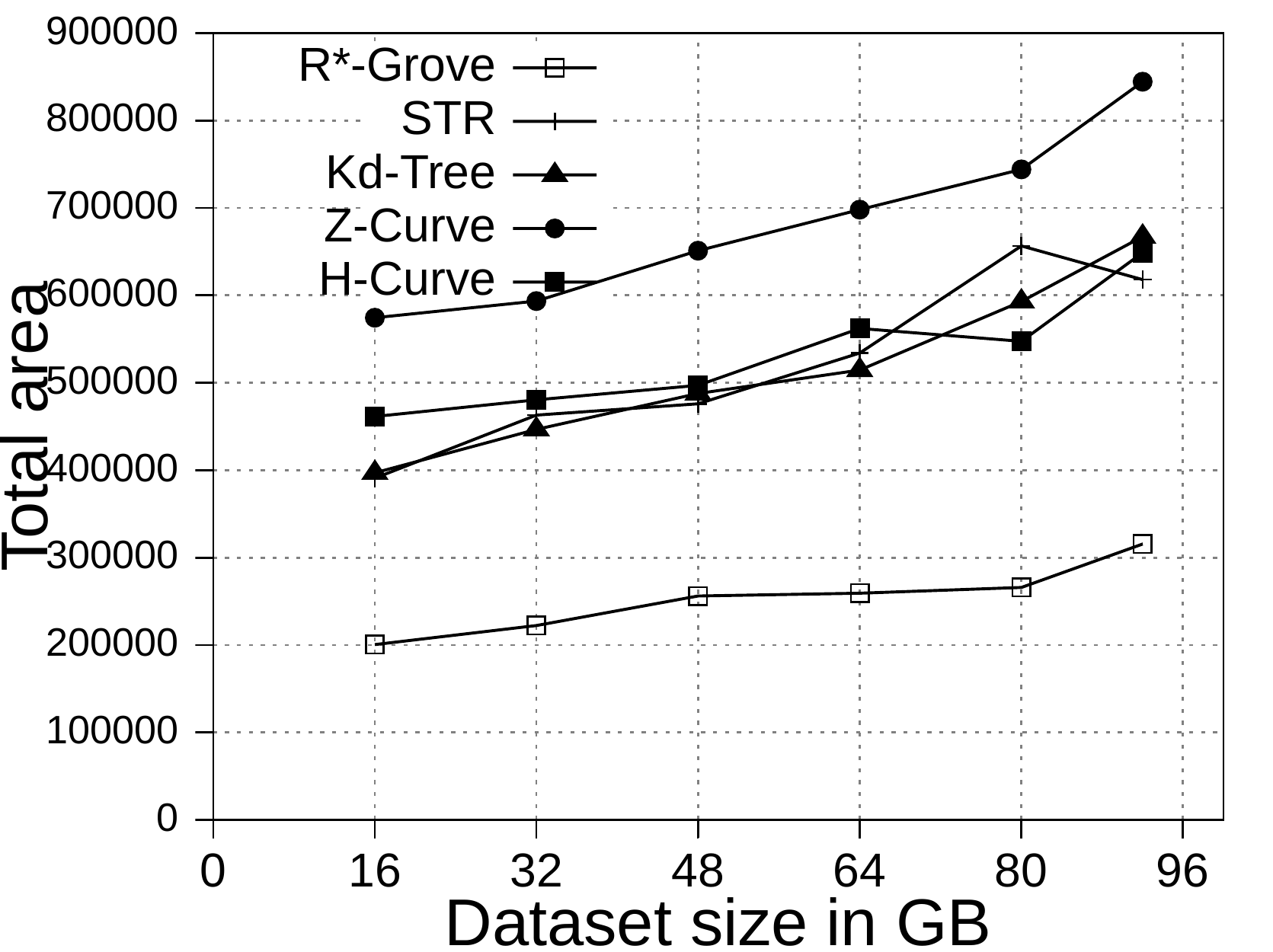} & \includegraphics[width=0.32\columnwidth]{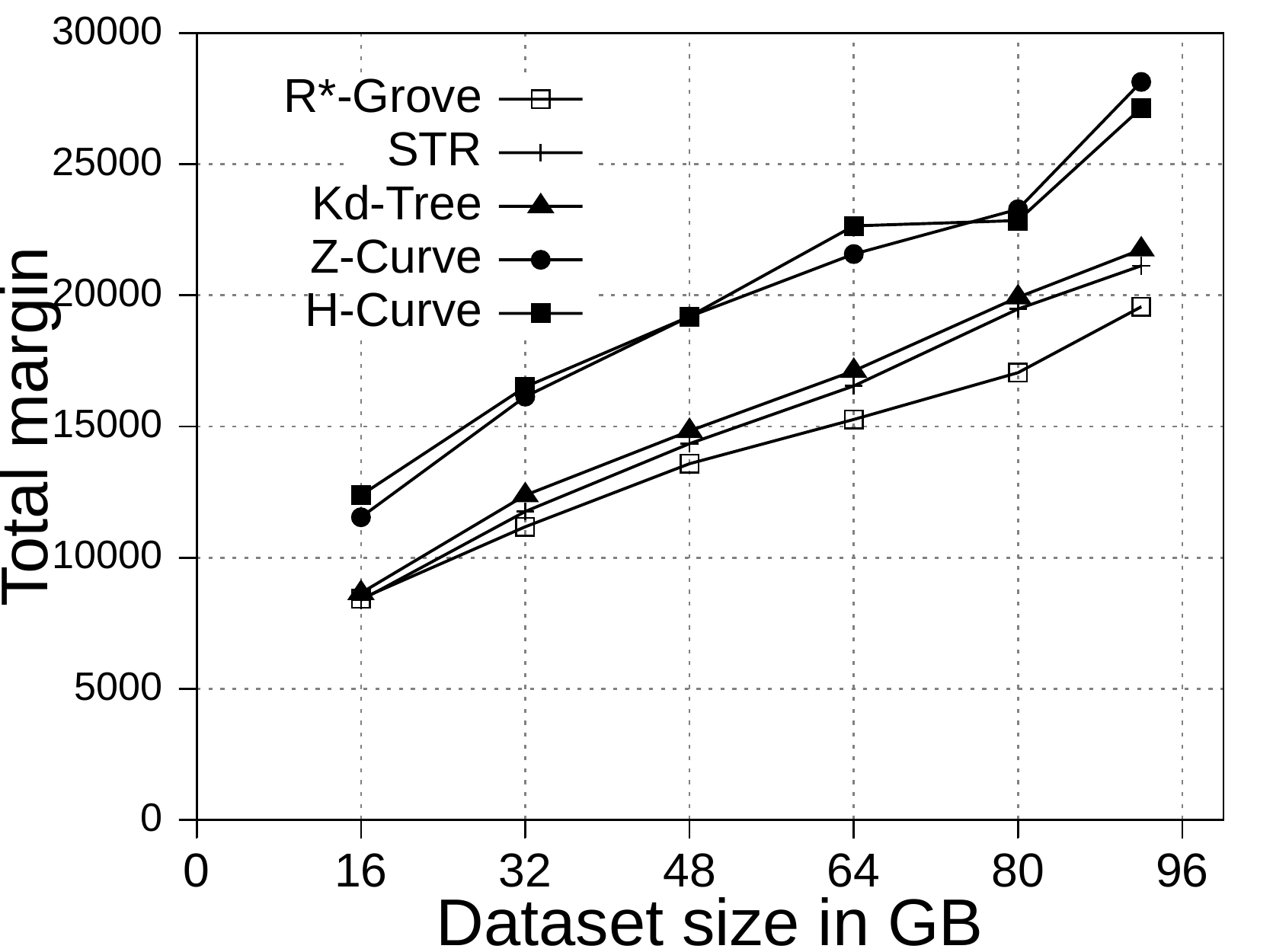} \\
(a)~Partitioning time & (b)~Total area & (c) Total margin \\
\includegraphics[width=0.32\columnwidth]{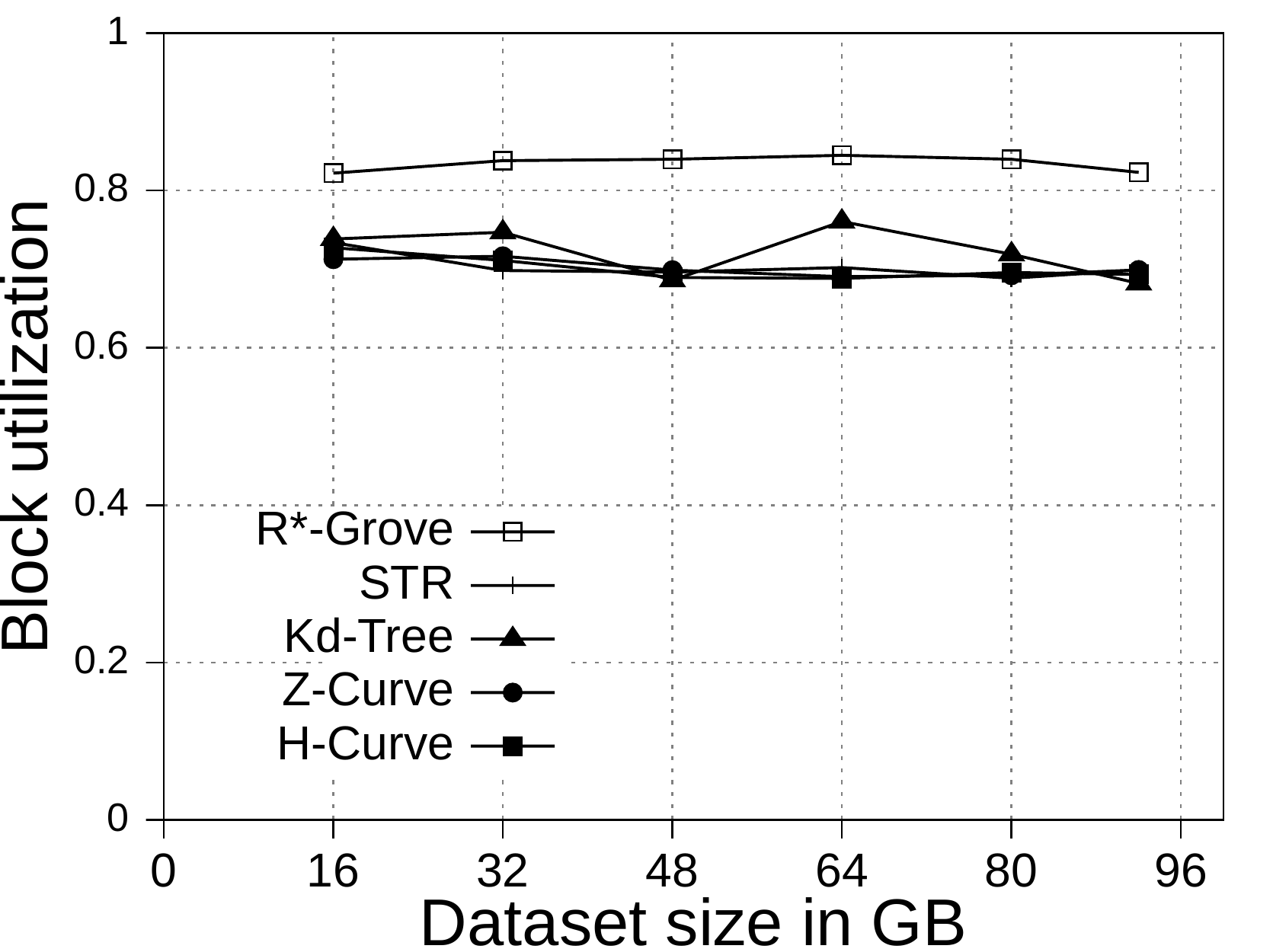}  & \includegraphics[width=0.32\columnwidth]{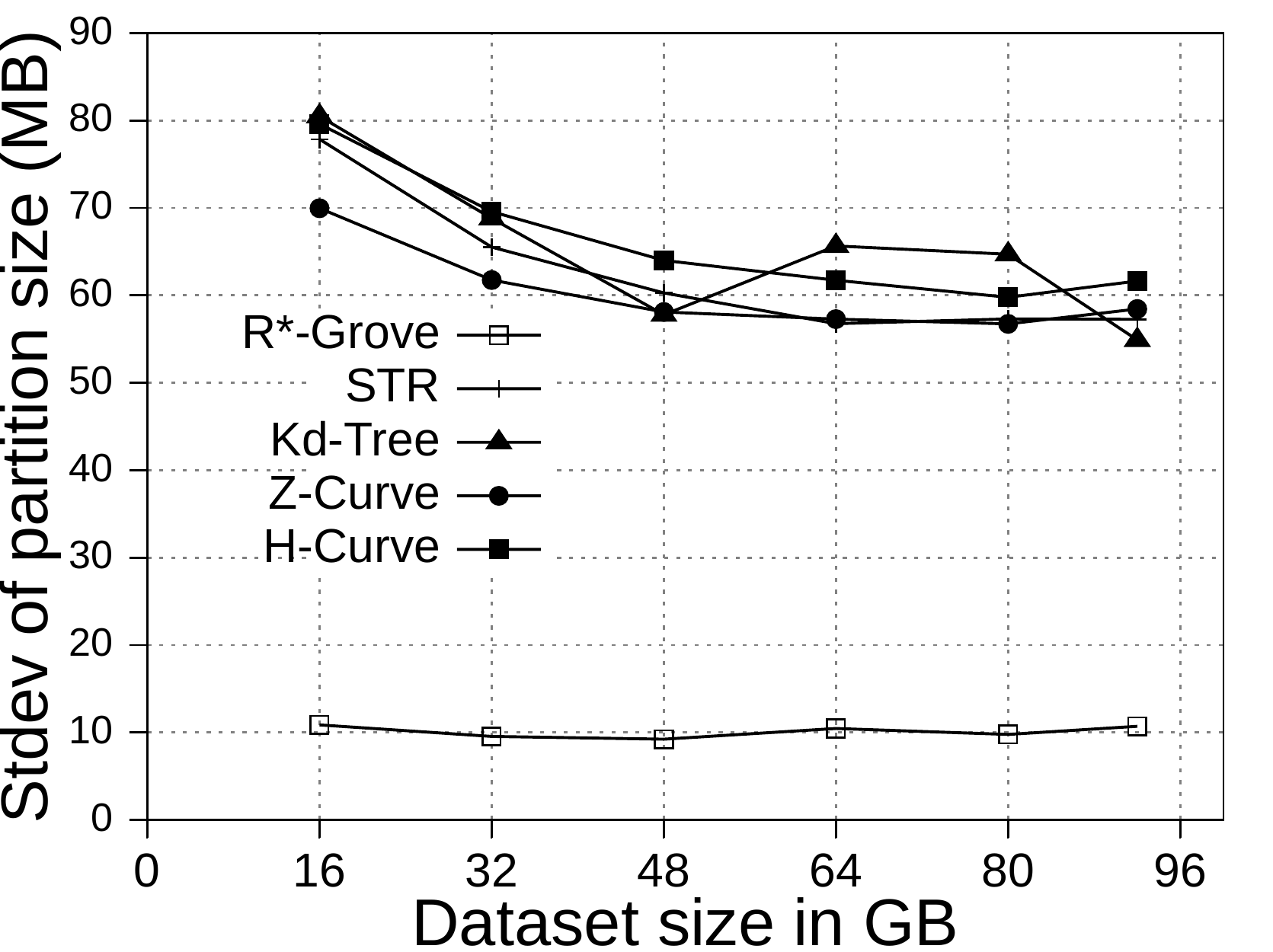} & \includegraphics[width=0.32\columnwidth]{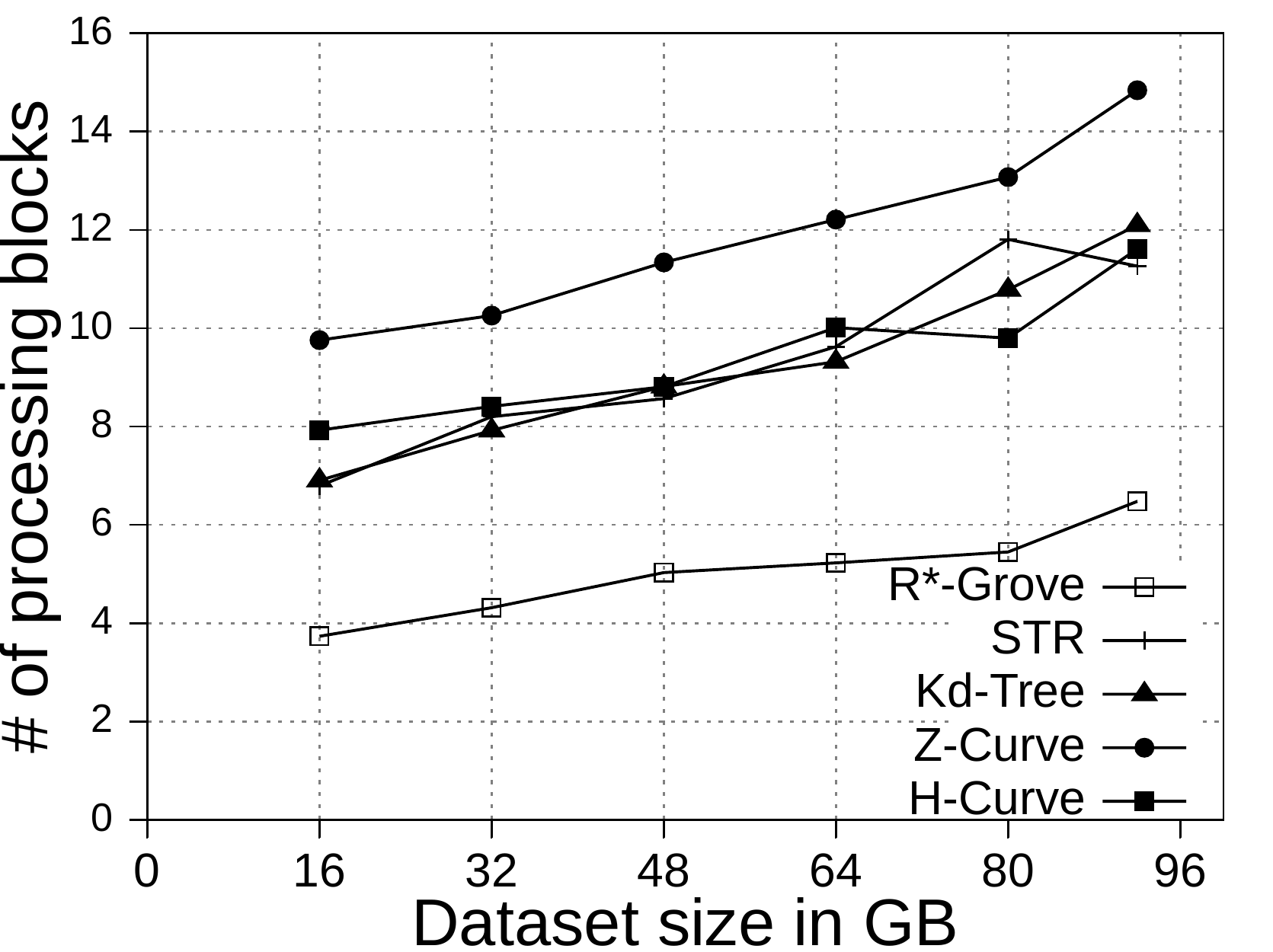} \\
(d)~Block utilization & (e)~Load balance & (f)~Average range query cost
\end{tabular}
\end{minipage}
\caption{Indexing performance and partition quality of R*-Grove and other partitioning techniques in {\tt OSM-Objects} dataset with variable-size records.}
\label{fig:comparison_all_objects}
\end{figure}

This section shows the advantages of R*-Grove for indexing big spatial data when compared to other partitioning techniques. We use {\tt OSM-Nodes} and {\tt OSM-Objects} dataset with size up to $200$ and $92$ GB, respectively. We compare five techniques, namely, R*-Grove, STR, Kd-Tree, Z-Curve and H-Curve. We implemented those techniques on Spark with sampling-based partitioning mechanism. Figures~\ref{fig:comparison_all_nodes}(a) and~\ref{fig:comparison_all_objects}(a) show that there is no significant difference of indexing performance between different techniques. This result is expected since the main difference between them is in Phase~2 which runs on a single machine on a sample of a small size (and a histogram in case of R*-Grove). Typically, Phase~2 takes only a few seconds to finish. These results suggest that the proposed R*-Grove algorithm requires the same computational resources as the baseline techniques. Meanwhile, it provides a better query performance by providing a higher partition quality as detailed next.

\subsubsection{Total area and total margin}
Figures~\ref{fig:comparison_all_nodes}(b) and~\ref{fig:comparison_all_objects}(b) show the total area of indexed datasets when we vary the {\tt OSM-Nodes} and {\tt OSM-Objects} dataset size from $20$GB to $200$GB and $16$GB to $92$GB, respectively. R*-Grove is the winner since it minimizes the total area of all partitions. While H-Curve performs generally better than Z-Curve, they are both doing bad since they do not take partition area into account in their optimization criteria. Specially, Figure~\ref{fig:comparison_all_objects}(b) strongly validates the advantages of R*-Grove in non-point datasets. Figures~\ref{fig:comparison_all_nodes}(c) and~\ref{fig:comparison_all_objects}(c) report the total margin for the same experiment. R*-Grove is the clear winner because it inherits the splitting mechanism of R*-Tree, which is the only one among all those that tries to produce square-like partitions. As the input size increases, more partitions are generated which causes the total margin to increase.

\subsubsection{Block utilization}

Figures~\ref{fig:comparison_all_nodes}(d) and~\ref{fig:comparison_all_objects}(d) show the block utilization as the input size increases. R*-Grove outperforms other partitioning techniques due to the proposed improvements in Section~\ref{sec:rsgrove:balance} and~\ref{sec:rsgrove:histogram} specifically improve block utilization. Using R*-Grove, each partition almost occupies a full block in HDFS which increases the overall block utilization. Z-Curve and H-Curve perform similarly since they produce equi-sized partition by creating split points along the curve. The high variability of the Kd-tree is due to the way it partitions the space at each iteration. Since it always partitions the space along the median, it only works perfectly if the number of partitions is a power of two; otherwise, it could be very inefficient. This occasionally results in partitions of high block utilization but they could be highly variable in size.

\subsubsection{Load balance}
\label{sec:experiments:load_balance}

Figures~\ref{fig:comparison_all_nodes}(e) and~\ref{fig:comparison_all_objects}(e) show the standard deviation of partition size in MB for the {\tt OSM-Nodes} and {\tt OSM-Objects} datasets, respectively. Note that the HDFS block size is set to $128$ MB. A smaller standard deviation indicates a better load balance. In Figure~\ref{fig:comparison_all_nodes}(e), the dataset {\tt OSM-Nodes} contains records of almost the same size so R*-Grove performs only slightly better than Z-Curve, H-Curve and STR even though these three techniques try to primarily balance the partition sizes. In Figure~\ref{fig:comparison_all_objects}(e), the {\tt OSM-Objects} dataset contains highly variable record sizes. In this case, R*-Grove is way better than all other techniques as it is the only one that employs the storage histogram to balance variable size records. In particular, we observe that the standard deviation of partition size on STR, Kd-Tree, Z-Curve and H-Curve is about $50-60\%$ of the HDFS block size. Meanwhile, R*-Grove maintains a value around $10$ MB, which is only $8\%$ of the block size.

\subsubsection{Effect of sampling ratio}
\label{sec:experiments:sampling_ratio}

\begin{figure}[t]
\centering
\begin{minipage}{\columnwidth}
\begin{tabular}{ccc}
\includegraphics[width=0.32\columnwidth]{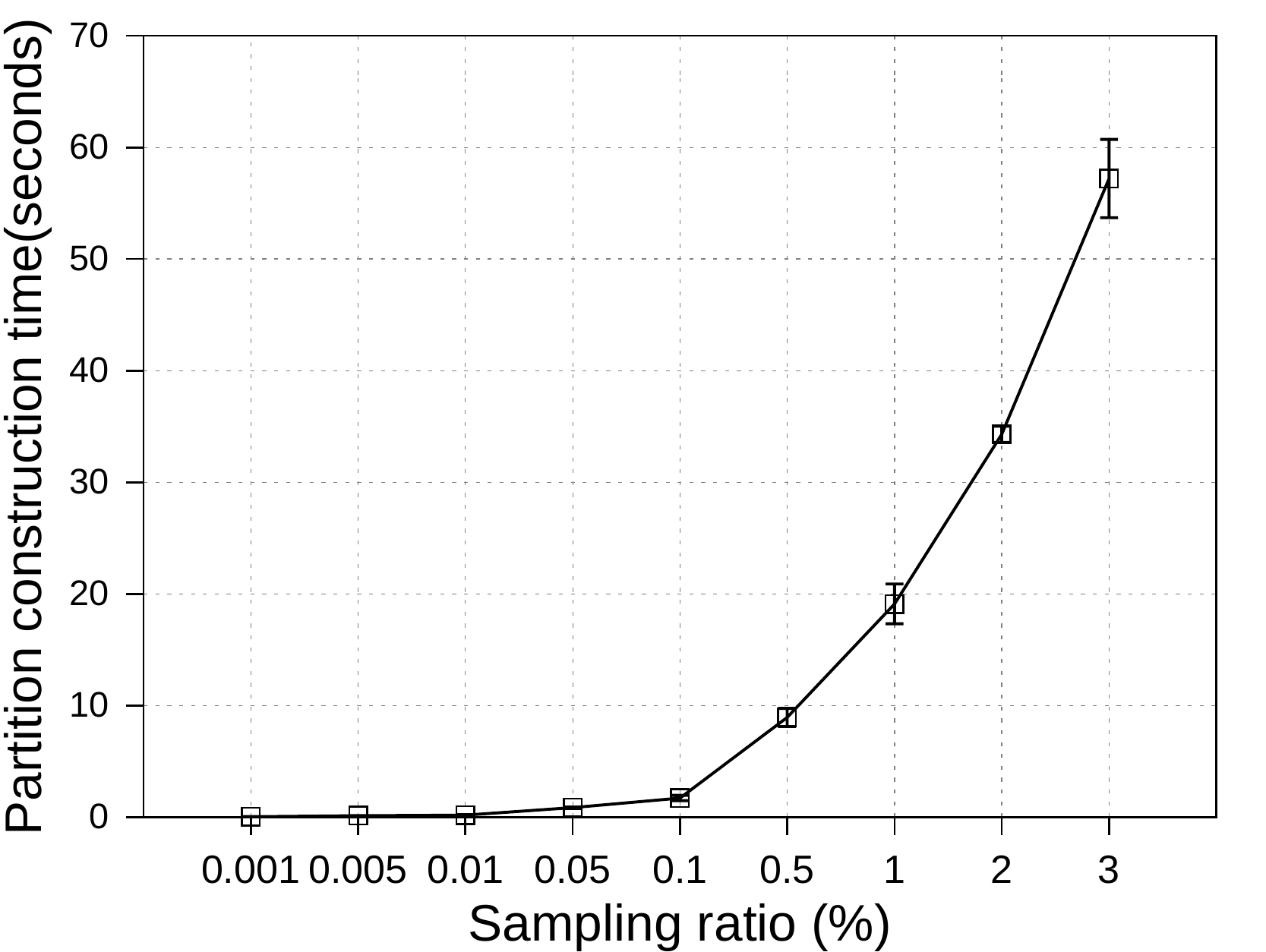}  & \includegraphics[width=0.32\columnwidth]{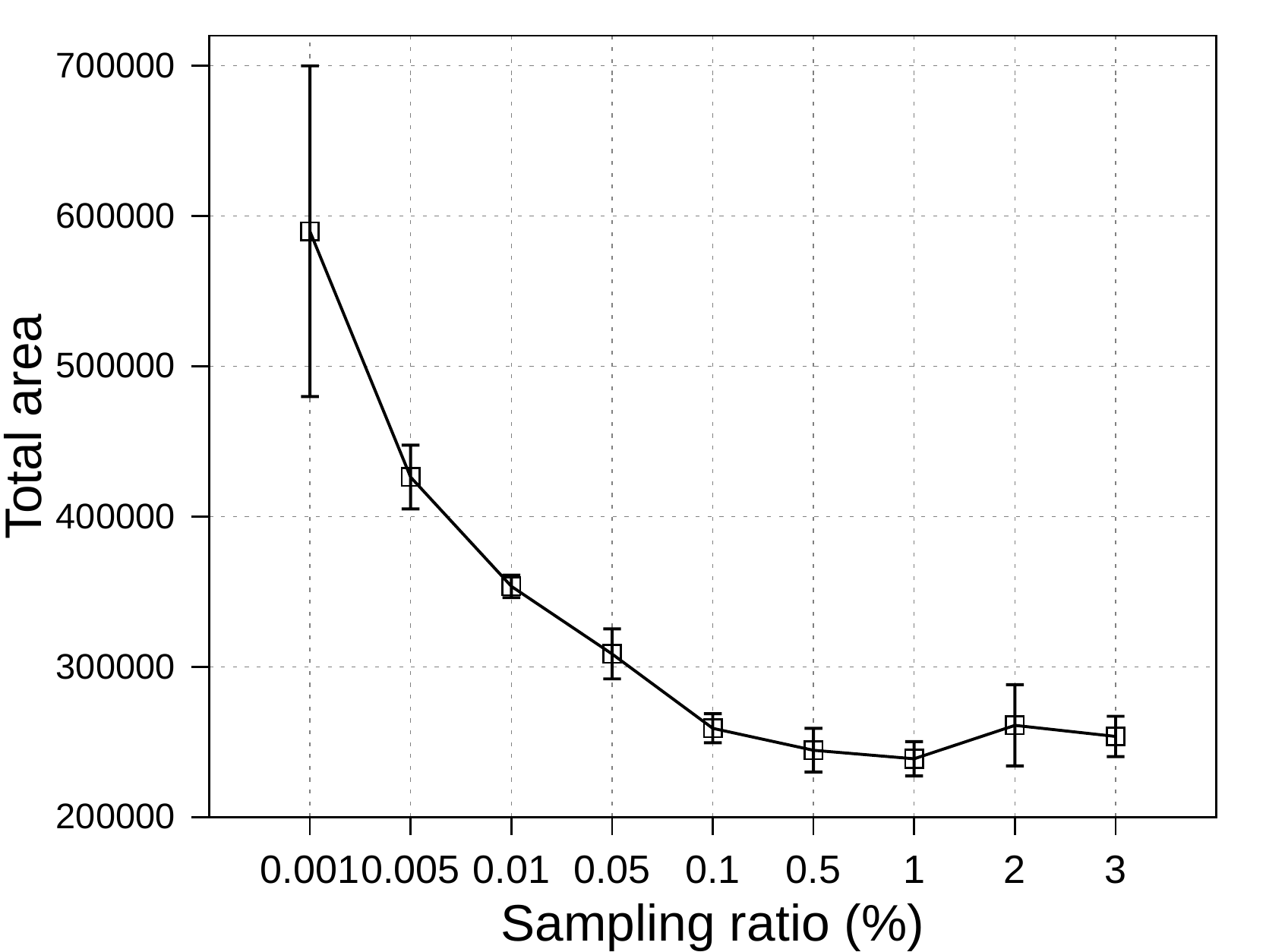} & \includegraphics[width=0.32\columnwidth]{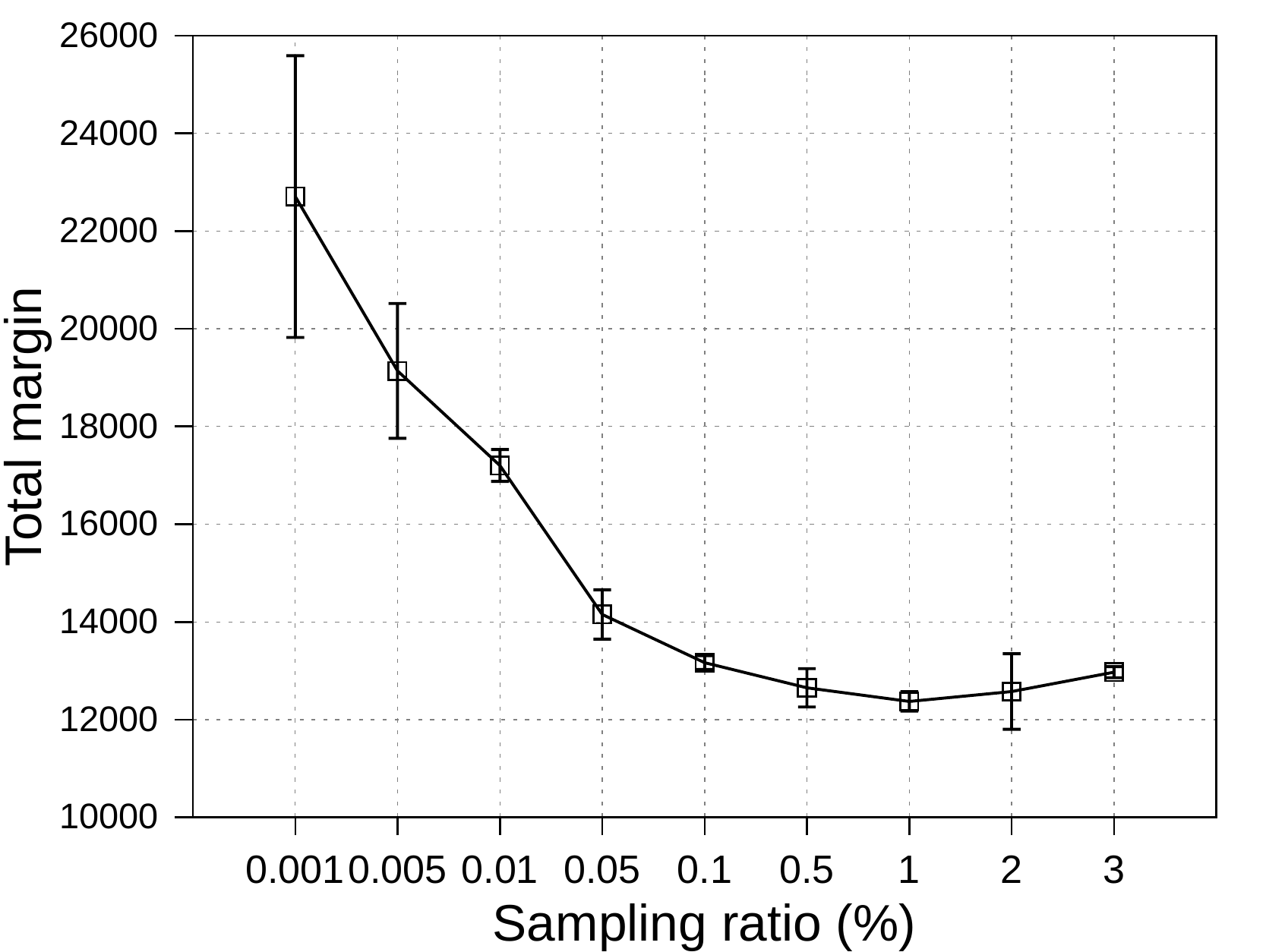} \\
(a)~Partition construction time & (b)~Total area & (c) Total margin \\
\includegraphics[width=0.32\columnwidth]{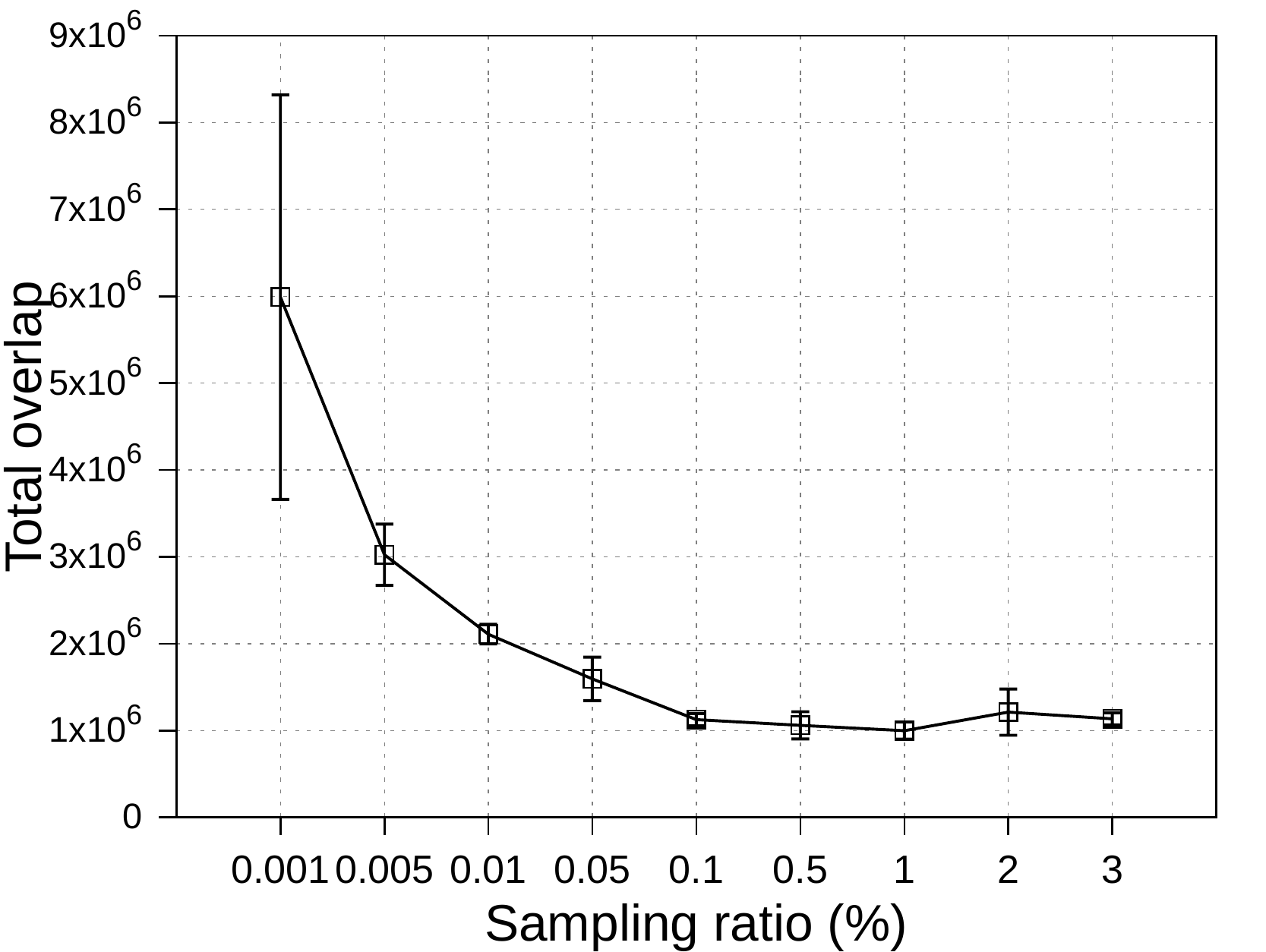}  & \includegraphics[width=0.32\columnwidth]{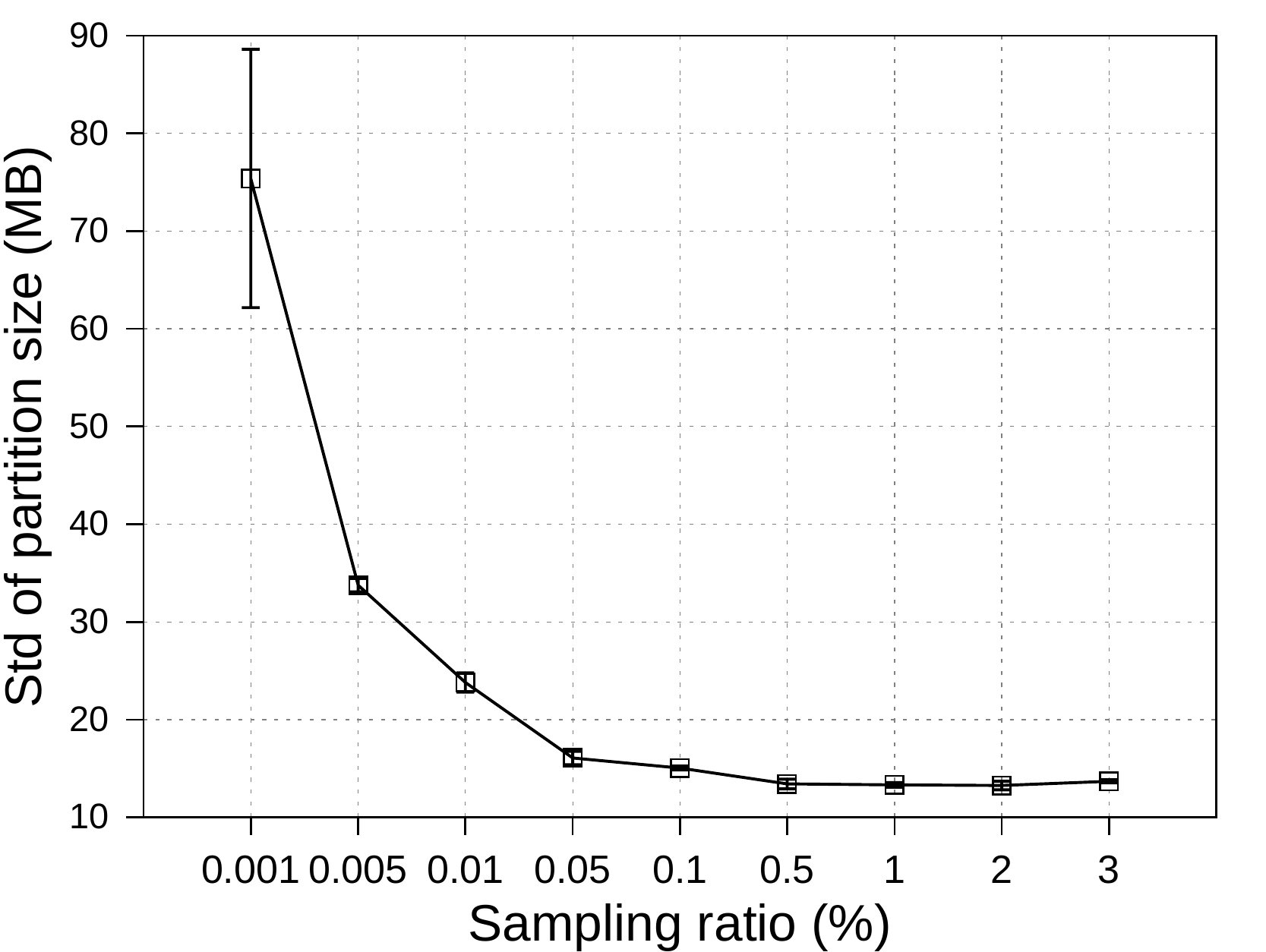} & \includegraphics[width=0.32\columnwidth]{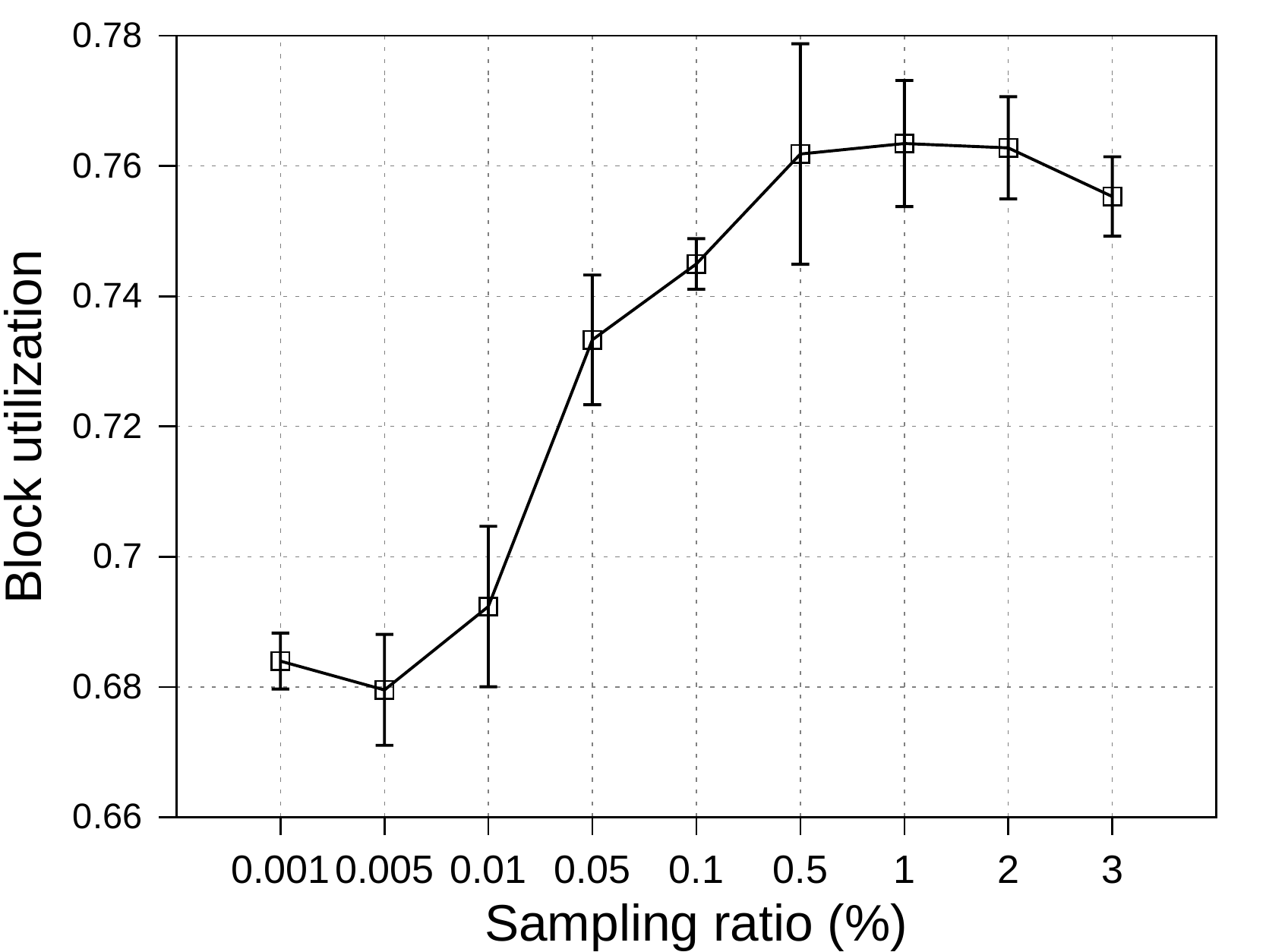} \\
(d)~Total overlap & (e)~Load balance & (f)~Block utilization
\end{tabular}
\end{minipage}
\caption{Indexing performance and partition quality of R*-Grove in {\tt OSM-Objects} datasets with different sampling ratios.}
\label{fig:effect_sampling_ratio}
\end{figure}

Since the proposed R*-Grove follows the {\em sampling-based} partitioning mechanism, a valid question is how the sampling ratio affects partition quality and performance? In this experiment, we execute several partitioning operations using R*-Grove in {\tt OSM-Objects} datasets. All the partitioning parameters are kept fixed, except the sampling ratio, which is varying from $0.001\%$ to $3\%$. For each sampling ratio value, we execute the partition operation three times, then compute the average and standard deviation of quality measures and partition construction time. Partition construction is the process that compute partition MBRs from the sample. Figure~\ref{fig:effect_sampling_ratio} uses the average values to plot the line and the standard deviation for the error bars. First, Figure~\ref{fig:effect_sampling_ratio}(a) shows that the higher sampling ratio requires higher time for the partition construction process. This is expected due to the number of sample records which the partitioner has to use to compute partition MBRs. Figures~\ref{fig:effect_sampling_ratio}(b,c,d,e) show the downward trend of total area, total margin, total overlap and standard deviation of partition size. Figure~\ref{fig:effect_sampling_ratio}(f) shows the upward trend of block utilization when the sampling ratio increases. In addition, the standard deviation of small sampling ratios is much higher than that for high sampling ratios. These results indicates that the higher sampling ratios promise better partition quality. However, an important observation is that the partition quality measures start stabilizing for sampling ratios larger than $1\%$. This behavior was also validated in a previous work~\cite{eldawy2015spatial}. In short, this work shows that a sample ratio of $1\%$ dataset is enough to achieve virtually a same partition quality as sample ratio $100\%$. In the following experiments, we choose $1\%$ as the default sampling ratio for all partitioning techniques.

\subsubsection{Effect of minimum split ratio}
\label{sec:experiments:min_split_ratio}

\begin{figure}[t]
\centering
\begin{minipage}{\columnwidth}
\begin{tabular}{ccc}
\includegraphics[width=0.32\columnwidth]{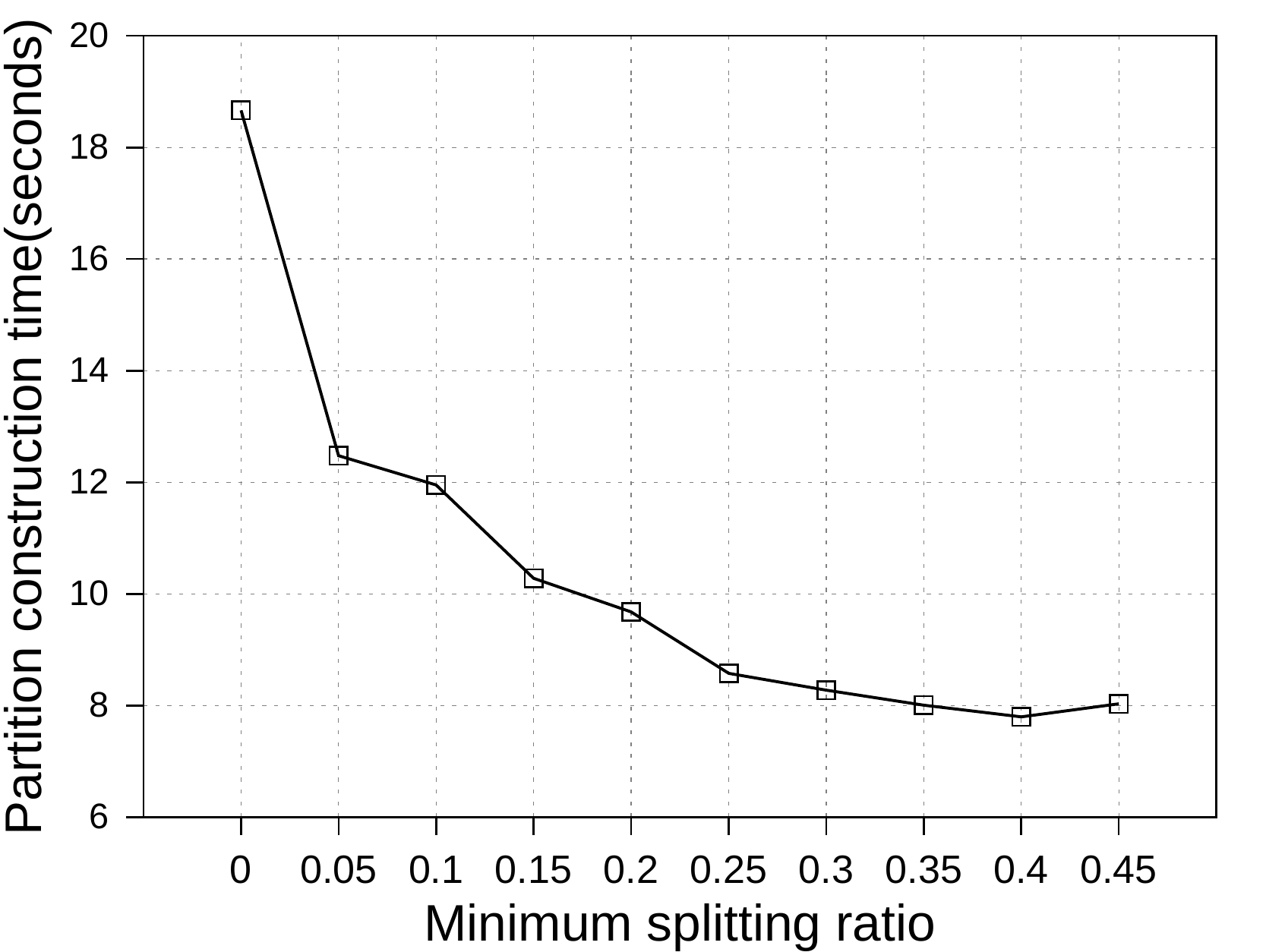}  & \includegraphics[width=0.32\columnwidth]{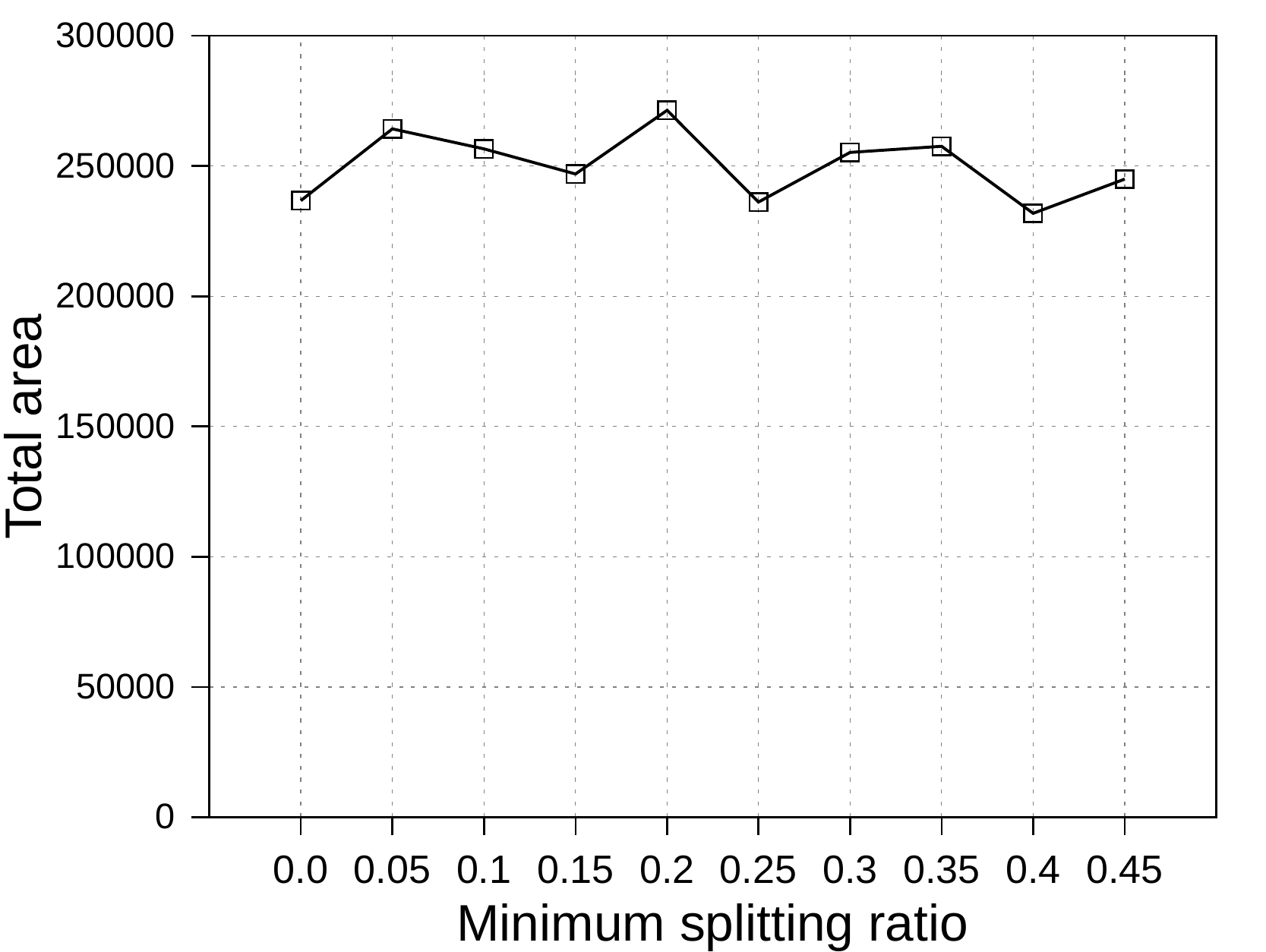} & \includegraphics[width=0.32\columnwidth]{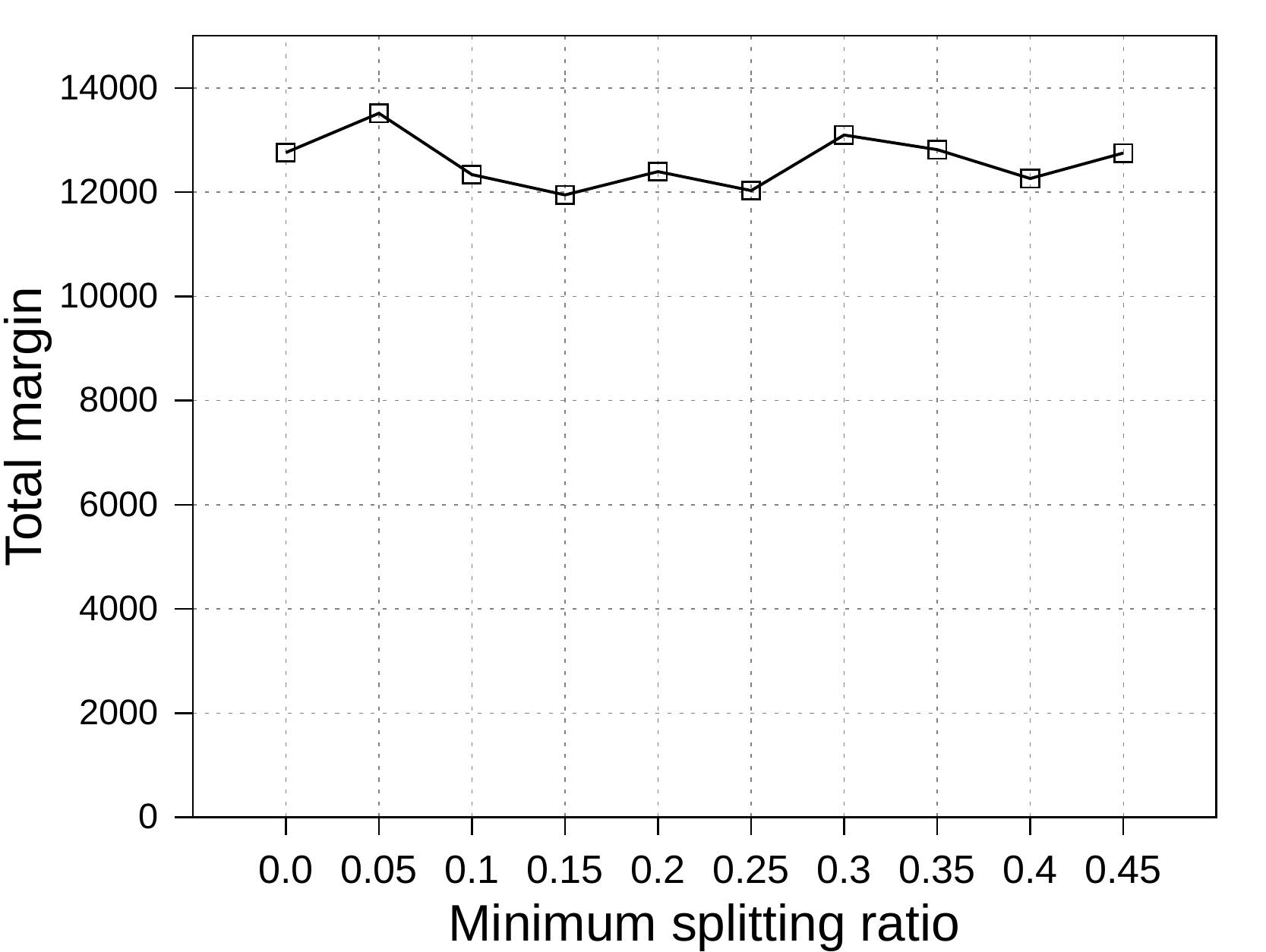} \\
(a)~Partition construction time & (b)~Total area & (c) Total margin \\
\includegraphics[width=0.32\columnwidth]{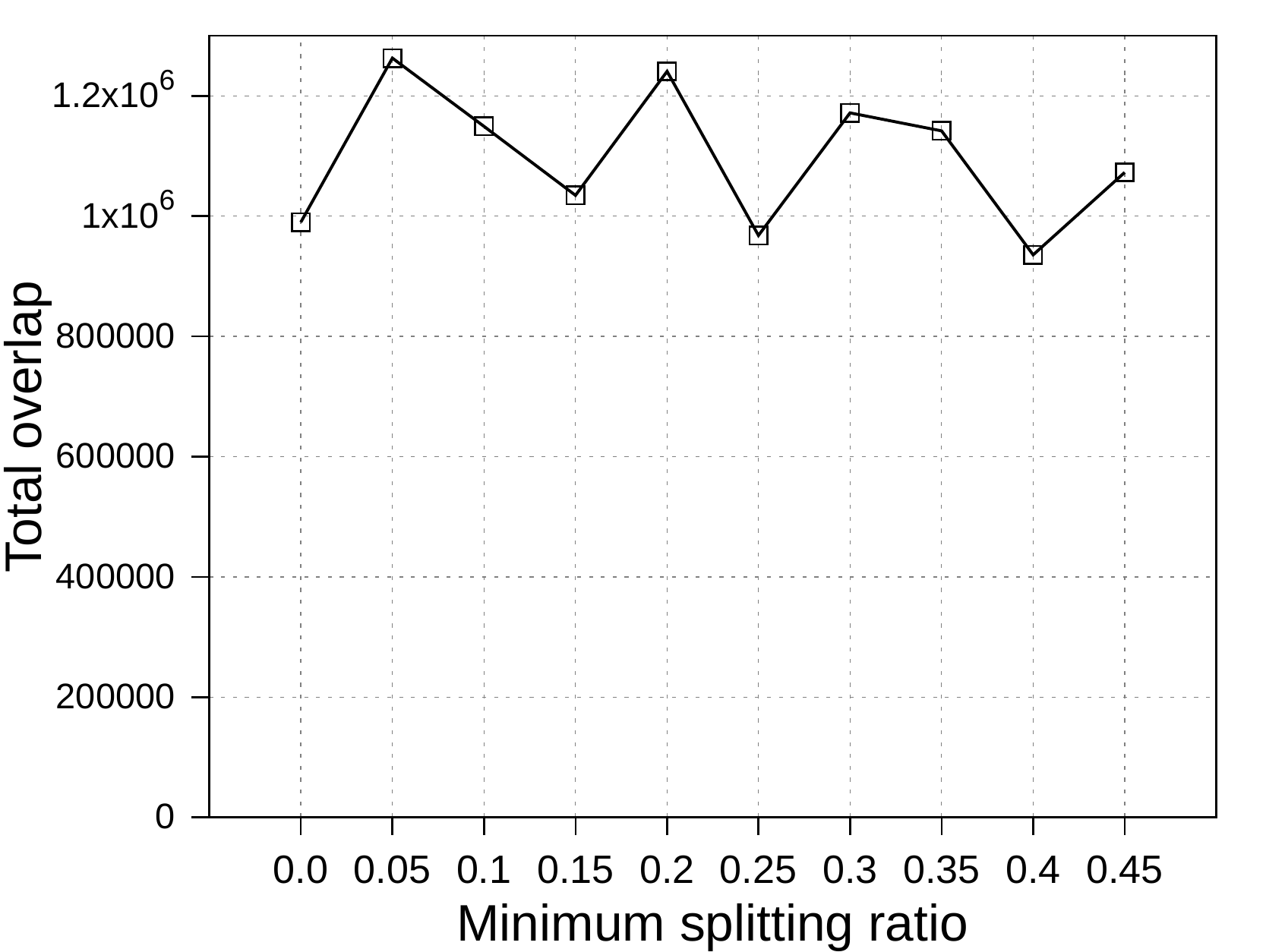}  & \includegraphics[width=0.32\columnwidth]{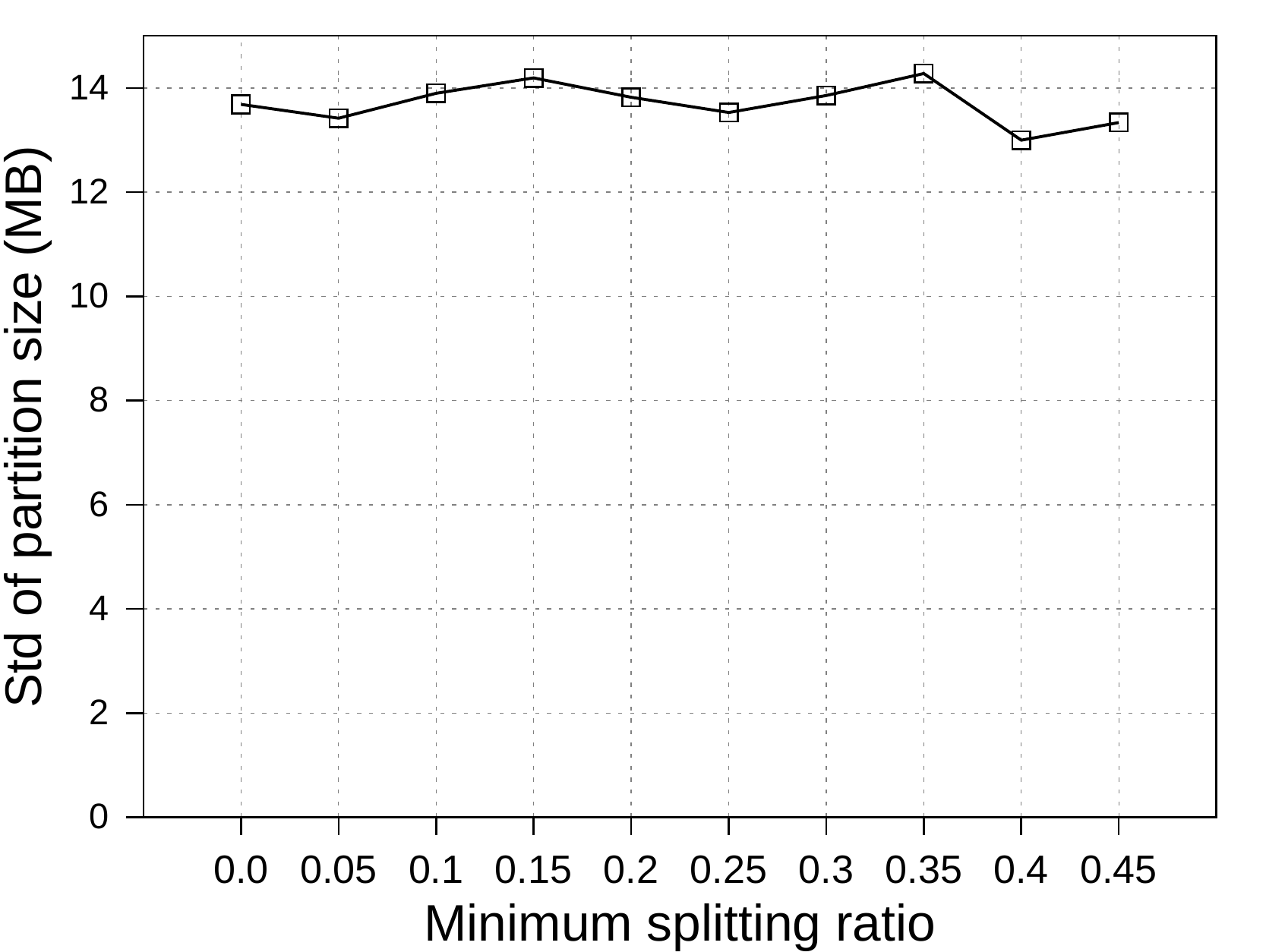} & \includegraphics[width=0.32\columnwidth]{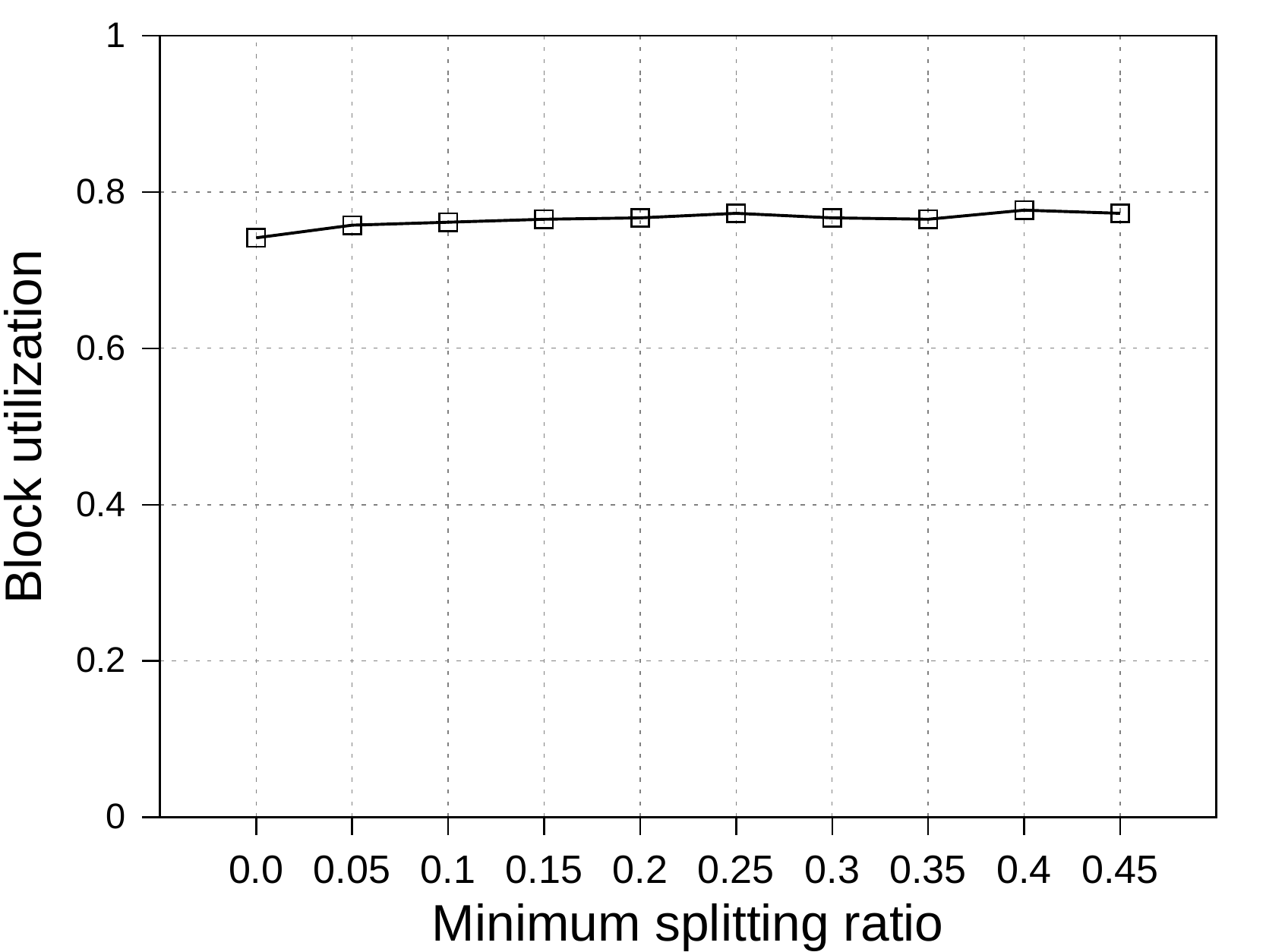} \\
(d)~Total overlap & (e)~Load balance & (f)~Block utilization
\end{tabular}
\end{minipage}
\caption{Indexing performance and partition quality of R*-Grove in {\tt OSM-Objects} datasets with different minimum splitting ratios.}
\label{fig:effect_min_split_ratio}
\end{figure}

In Section~\ref{sec:rsgrove:basic}, we introduced parameter $\rho$, namely minimum splitting ratio, to speed up the running time of {\sc SplitNode} algorithm used in Phase~2, boundary computation. In this experiment, we verify how the minimum splitting ratio impacts the partition quality and performance. We also use {\tt OSM-Objects} dataset with R*-Grove partitioning as the previous experiment in Section~\ref{sec:experiments:sampling_ratio}. We vary $\rho$ from $0$ to $0.45$. Figure~\ref{fig:effect_min_split_ratio} shows the overview of the experimental results. First, Figure~\ref{fig:effect_min_split_ratio}(a) shows that the running time of Phase~2, boundary computation, decreases as $\rho$ increases which is expected due to the balanced splitting in the recursive algorithm which causes it to terminate earlier. According to the run-time analysis in Section~\ref{sec:rsgrove:basic}, a larger value of $\rho$ reduces the depth of the recursive formula which results in a lower running time. However, this minimum splitting ratio also shrinks the search space for optimal partitioning scheme. Fortunately, the number of records in the $1\%$ sample is usually large enough such that the boundary computation algorithm could still find a good partitioning scheme even for high value of $\rho$. In the following experiments, we choose $\rho=0.4$ as the default value for R*-Grove partitioning.

\subsection{Spatial query performance}
\subsubsection{Range query}

\begin{figure}[t]
\centering
\begin{minipage}{\columnwidth}
\begin{tabular}{cc}
\includegraphics[width=0.48\columnwidth]{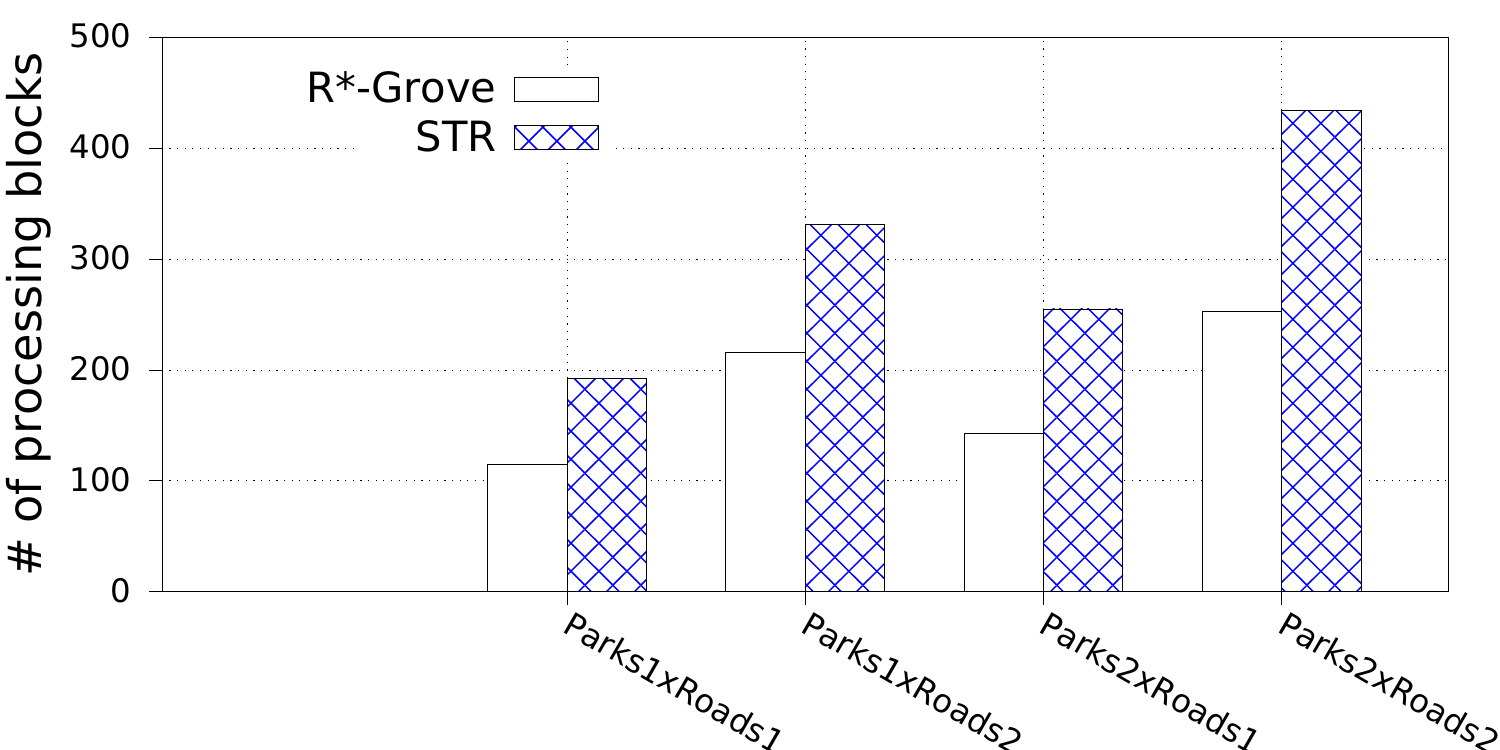} & \includegraphics[width=0.48\columnwidth]{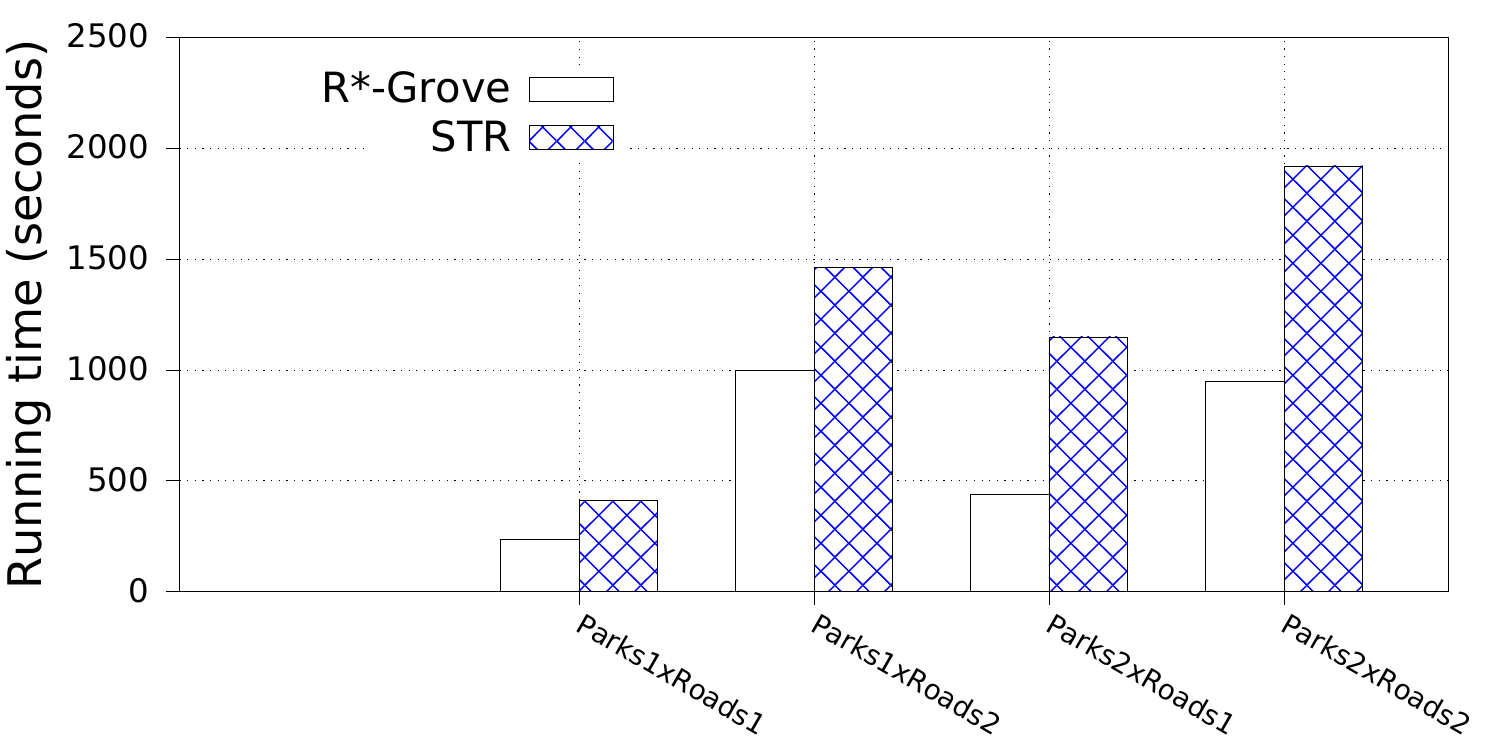} \\
(a) Number of processing blocks & (b) Running time in seconds
\end{tabular}
\end{minipage}
\caption{Spatial join performance in R*-Grove and STR partitioning}
\label{fig:spatial_join_performance}
\end{figure}

Figure~\ref{fig:comparison_all_nodes}(f) shows the performance of range query on the {\tt OSM-Nodes} dataset with size $200$GB. For partitioned {\tt OSM-Nodes} dataset, we run a number of range queries (from $200$ to $1,200$) all with the same range query size which is $0.01\%$ of the area covered by the entire input. All the queries are sent in one batch to run in parallel to put the cluster at full utilization. It is clear that R*-Grove outperforms all other techniques, especially when we run a large number of queries. This is the result of the high-quality and load-balanced partitions which minimize the number of blocks needed to process for each query. Figure~\ref{fig:comparison_all_objects}(f) shows the average cost of a range query on the {\tt OSM-Objects} dataset in terms of number of blocks that need to be processed, the lower the better. This value is also computed for a range query with size $0.01\%$ of space area. This result further confirms that R*-Grove provide a better query performance for variable-size records datasets.

\subsubsection{Spatial join}\label{sec:spatial_join}

In this experiment, we split {\tt OSM-Parks} and {\tt OSM-Roads} datsets to get multiple datasets as follows: {\tt Parks1}, {\tt Park2} with sizes $3.6$ and $7.2$ GB; {\tt Roads1} and {\tt Roads2} with sizes $10$ and $20$ GB, respectively. This allows us to study the effect of the input size on the spatial join query while keeping the input data characteristics the same, i.e., distribution and geometry size. We compare to STR since it is the best competitor of R*-Grove in previous experiments. Figure~\ref{fig:spatial_join_performance} shows the performance of the spatial join query. In general, R*-Grove significantly outperforms STR in all query instances.

Figure~\ref{fig:spatial_join_performance}(a) shows the number of accessed blocks for each spatial join query over the datasets which are partitioned by R*-Grove and STR. We can notice that R*-Grove needs to access 40\%-60\% fewer blocks than STR for two reasons. First, the better load balance in R*-Grove reduces the overall number of blocks in each dataset. Second, the higher partition quality in R*-Grove results in fewer overlapping partitions between the two datasets. The number of accessed blocks is an indicator to estimate the actual performance of spatial join queries. Indeed, this is further verified in Figure~\ref{fig:spatial_join_performance}(b), which shows actual running time for those queries. As we described, STR does not produce high quality partitions, thus the compound effect will even make it worst for spatial join query, which always relates to multiple STR partitioned datasets. On the other hand, R*-Grove addresses the limitations of STR so it can significantly improve the performance of spatial join query.

\subsection{Performance on larger datasets and multi-dimensional data}

\subsubsection{Scalability}

\begin{figure}[t]
\centering
\includegraphics[width=0.4\columnwidth]{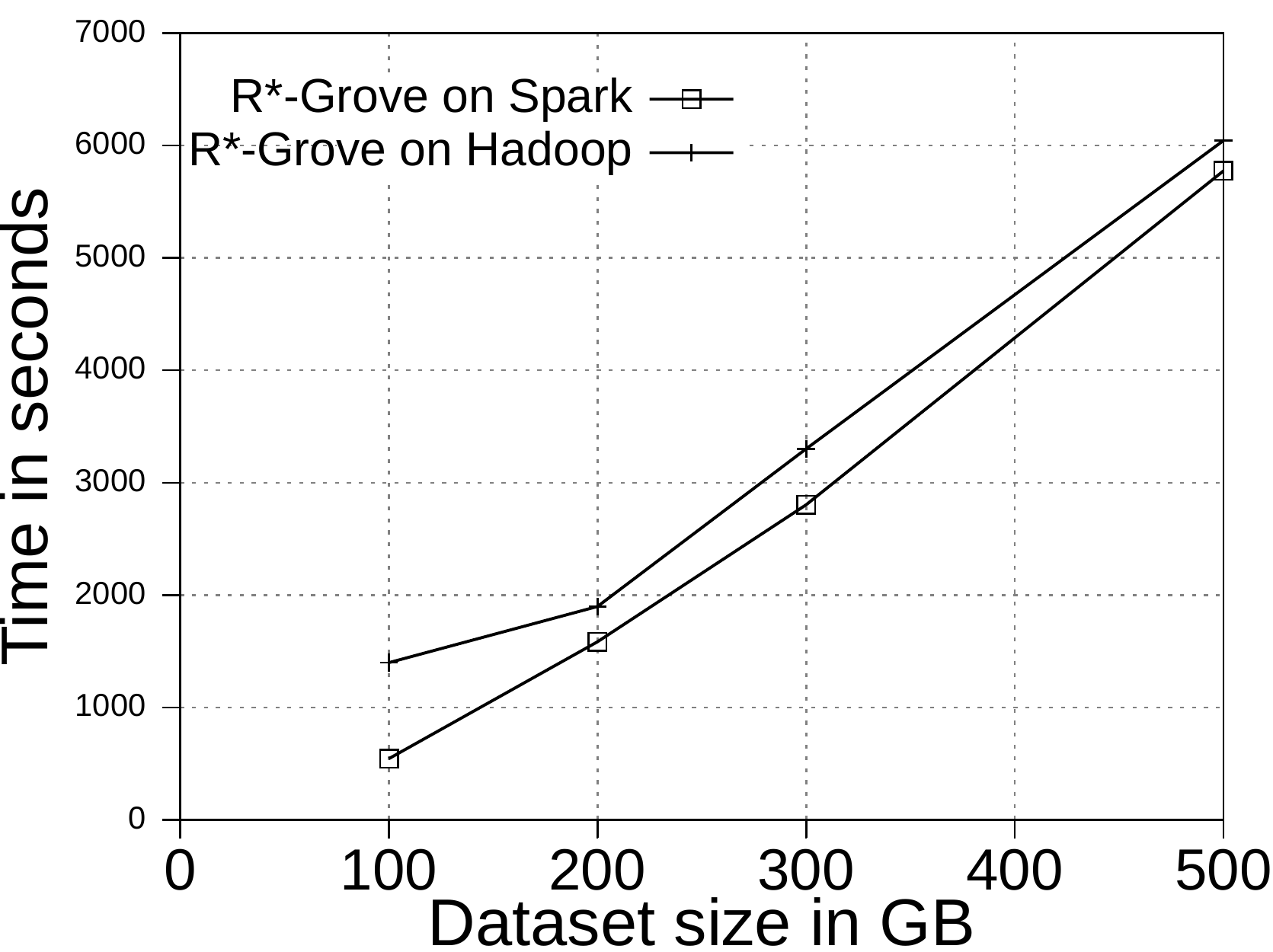}
\caption{The scalability of R*-Grove partitioning in Spark and Hadoop}
\label{fig:indexing_time_rsgrove}
\end{figure}

Figure~\ref{fig:indexing_time_rsgrove} shows the indexing time for two-dimensional {\tt OSM-Nodes} dataset with sizes $100, 200, 300$ and $500$GB. We executed the same indexing jobs in both Spark and Hadoop to see how the processing model affects the indexing performance. We observed that Spark outperforms Hadoop in terms of total indexing time. This experiment also demonstrates that R*-Grove is ready to work with large volume datasets on both Hadoop and Spark. We also observe that the gap between Hadoop and Spark decreases as the input size increases as Spark starts to spill more data to disk.

\subsubsection{Multi-dimensional datasets}

\begin{figure}[t]
\centering
\begin{minipage}{\columnwidth}
\begin{tabular}{ccc}
\includegraphics[width=0.32\columnwidth]{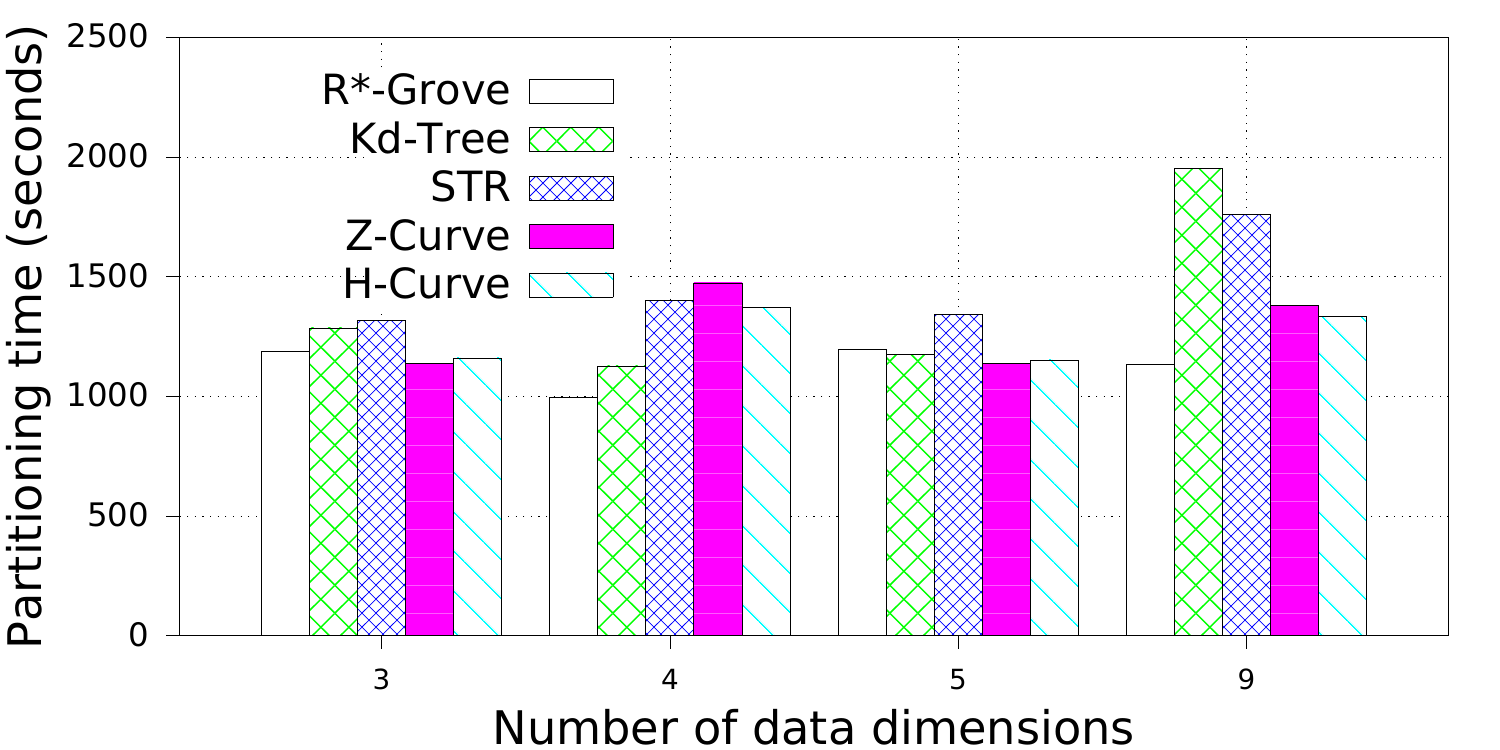}  & \includegraphics[width=0.32\columnwidth]{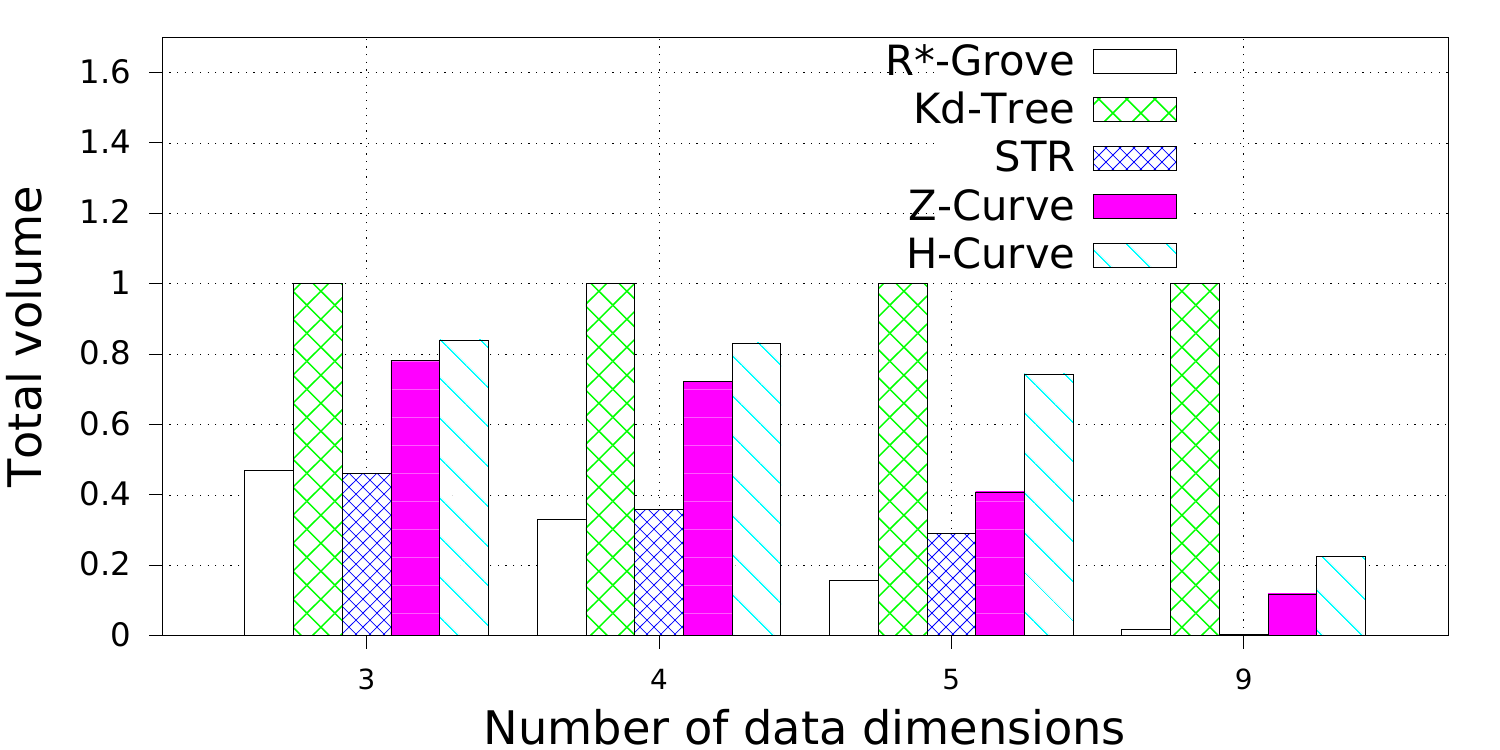} & \includegraphics[width=0.32\columnwidth]{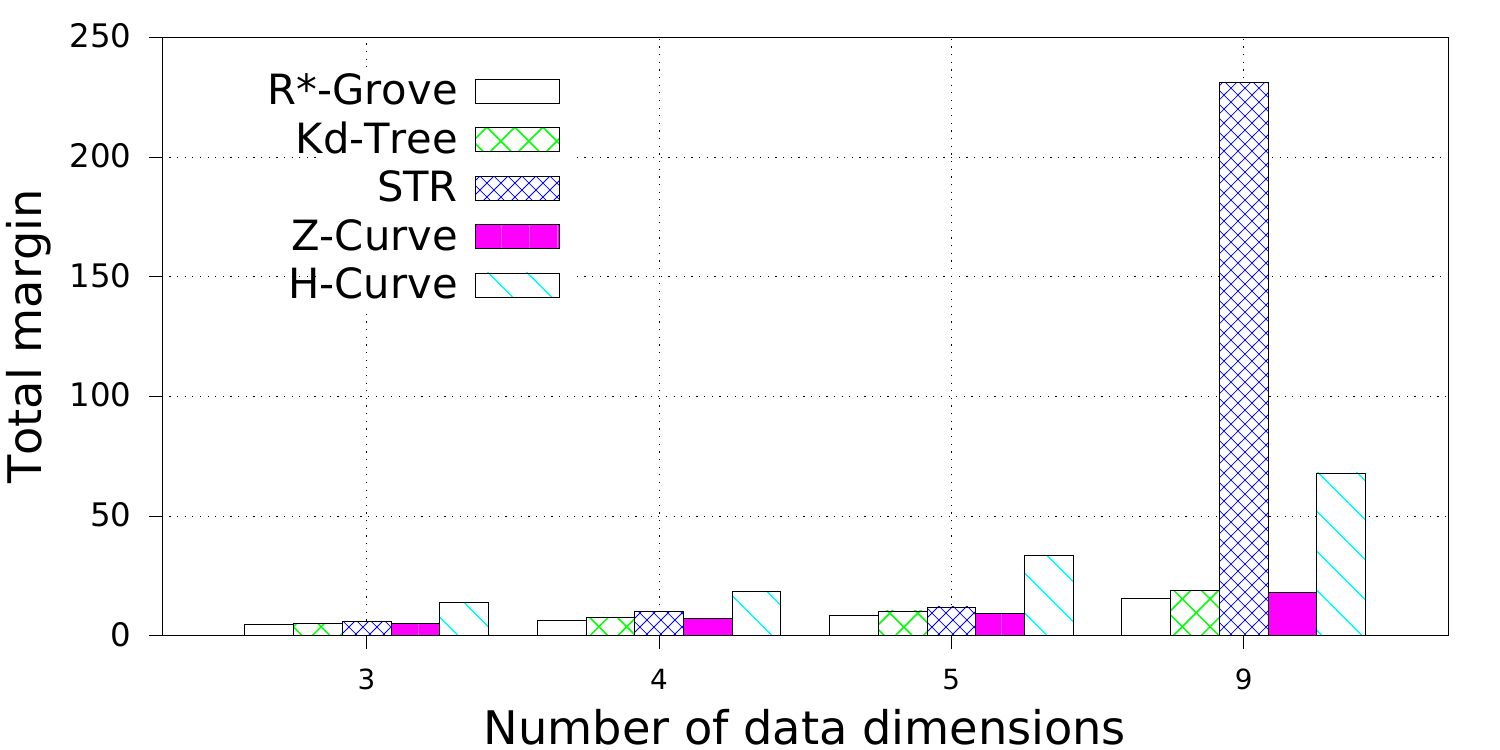} \\
(a)~Partitioning time & (b)~Total volume & (c) Total margin \\
\includegraphics[width=0.32\columnwidth]{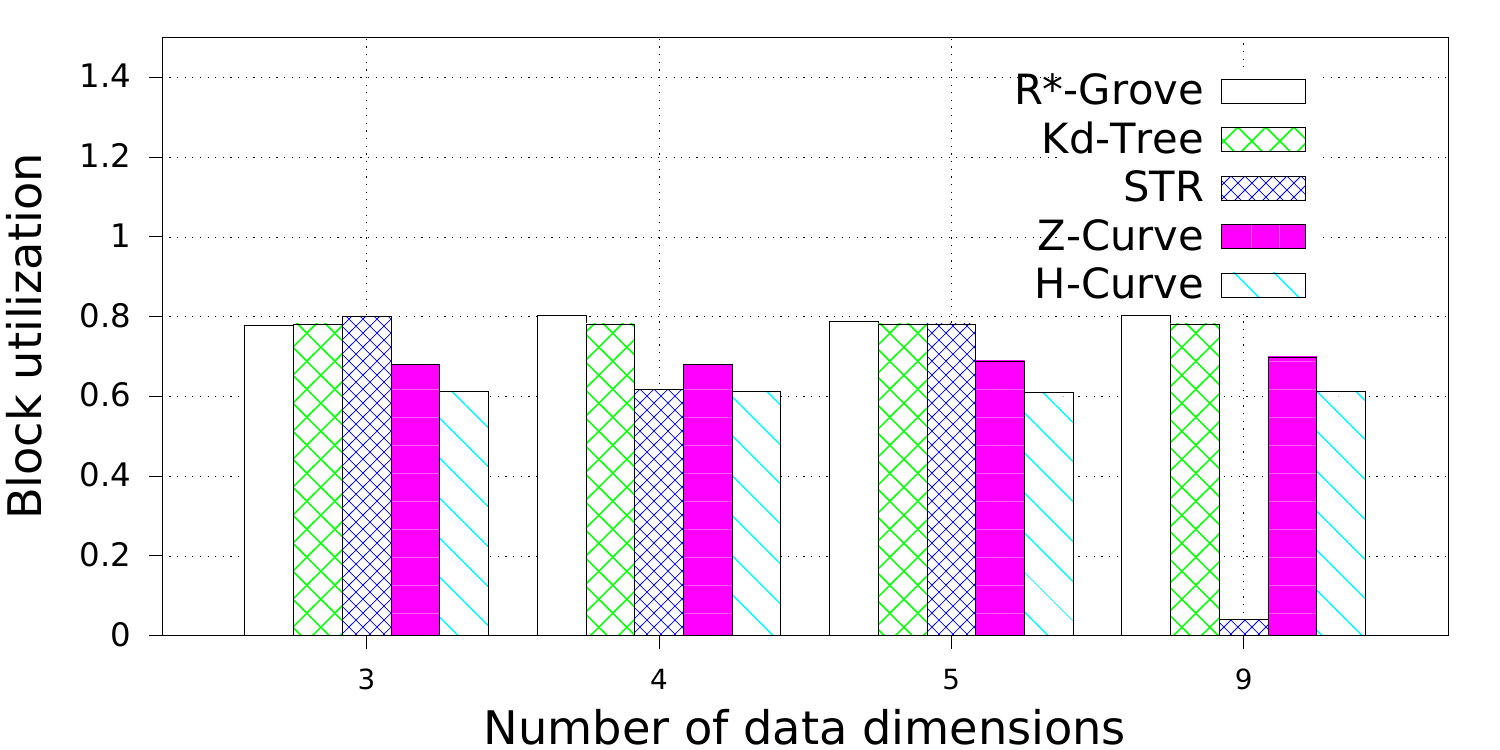}  & \includegraphics[width=0.32\columnwidth]{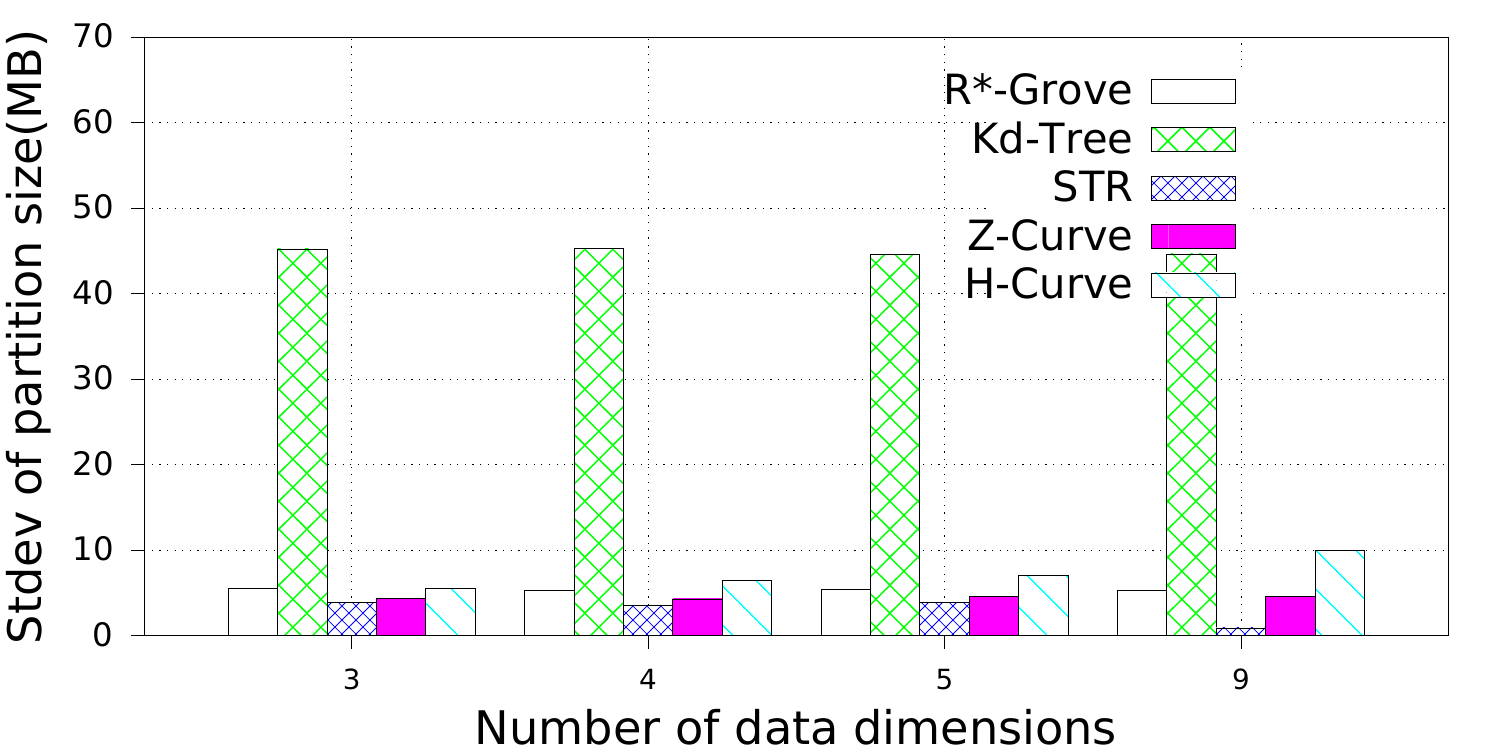} & \includegraphics[width=0.32\columnwidth]{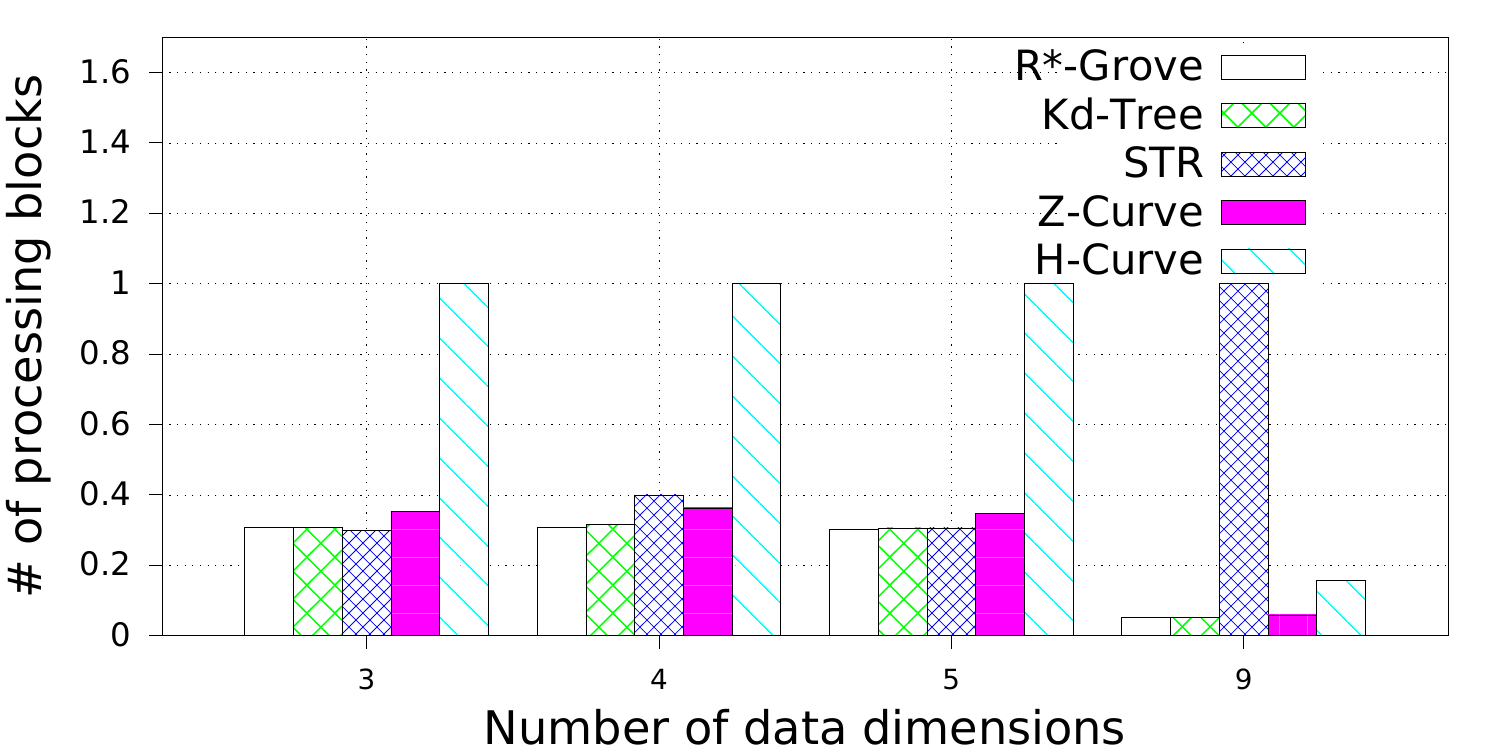} \\
(d)~Block utilization & (e)~Load balance & (f)~Range query performance
\end{tabular}
\end{minipage}
\caption{Indexing performance and partition quality of R*-Grove and other partitioning techniques on {\tt synthetic} multi-dimensional dataset.}
\label{fig:comparison_diagonal_points}
\end{figure}

In this experiment, we study the quality of R*-Grove on multi-dimensional datasets. Inspired by \cite{beckmann2008benchmark}, we use four {\tt synthetic} datasets with number of dimensions $3$, $4$, $5$, and $9$. We measure the running time and the quality of the five partitioning techniques: R*-Grove, STR, Kd-Tree, Z-Curve and H-Curve. Figure~\ref{fig:comparison_diagonal_points}(a) shows that R*-Grove is mostly the fastest technique to index the input dataset due to the best load balance among partitions. Figure~\ref{fig:comparison_diagonal_points}(b) shows that R*-Grove significantly reduces total area of partitions. Figure~\ref{fig:comparison_diagonal_points}(c) shows the total margin of all the techniques. While the total margin varies with the number of dimensions since they are different datasets, the techniques maintain the same order in terms of quality from best to worst, i.e., R*-Grove, Z-Curve, Kd-tree, STR and H-Curve, except the last group, where H-Curve is better than STR. This experiment indicates that R*-Grove could maintain its characteristics for multi-dimensional datasets. Figure~\ref{fig:comparison_diagonal_points}(d) and \ref{fig:comparison_diagonal_points}(e) report the block utilization and standard deviation of partition size, respectively. R*-Grove is the best technique that keeps both measures good. Figure~\ref{fig:comparison_diagonal_points}(f) depicts the normalized range query performance of different techniques, which affirms the advantages of R*-Grove. Notice that this is the only experiment where Z-Curve performs better than H-Curve. The reason is that the generated points are generated close to a diagonal line in the $d$-dimension. Since the Z-Curve just interleaves the bits of all dimensions, it will result in sorting these points along the diagonal line which results in a good partitioning. However, the way H-Curve rotates the space with each level will cause it to jump across the diagonal.

Additionally, STR becomes very bad as the number dimensions increases. This can be explained by the way STR computes the number of partitions given a sample data points. The existing STR implementation always creates a tree with a fixed node degree $n$ and $d$ levels where $d$ is the number of dimensions. This configuration results in $n^d$ leaf nodes or partitions. It computes the node degree $n$ as the smallest integer that satisfies $n^d\ge P$ where $P$ is the number of desired partitions. For example, for an input dataset of $100$~GB, $d=9$ dimensions, and a block size of $B=128$~MB, the number of desired partitions $P=100\cdot 1024/128=800$ partitions and $n=3$. This results in a total of $3^9=19683$ partitions. Obviously, as $d$ increases, the gap between the ideal number of partitions $P$ and the actual number of partitions $n^d$ increases which results in a very poor block utilization as shown in this experiment. Finally, Figure~\ref{fig:comparison_diagonal_points}(f) shows the average cost of a range query in terms of number of processed blocks, which indicates that R*-Grove is the winner when we want to speed up spatial query processing.

\begin{figure}[t]
\centering
\begin{minipage}{\columnwidth}
\begin{tabular}{ccc}
\includegraphics[width=0.32\columnwidth]{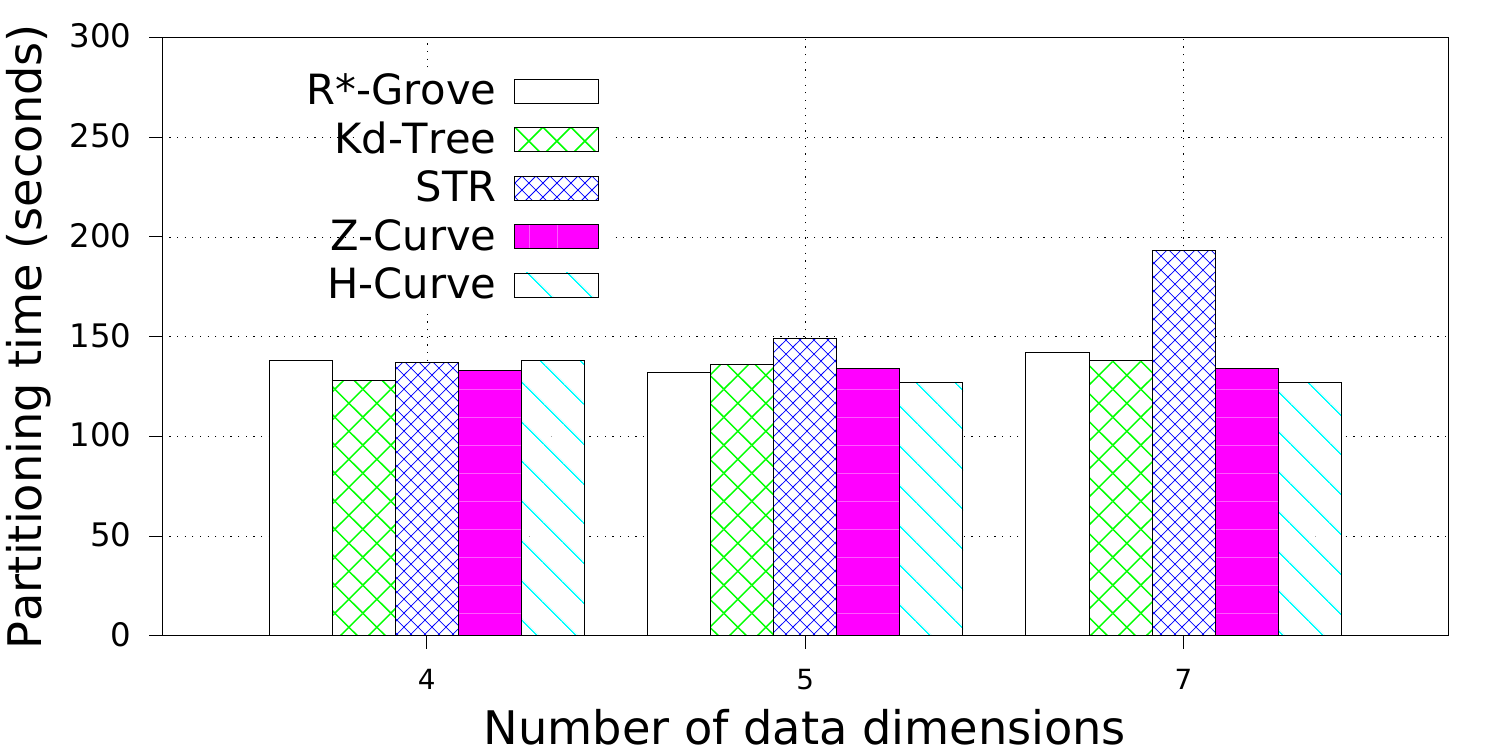}  & \includegraphics[width=0.32\columnwidth]{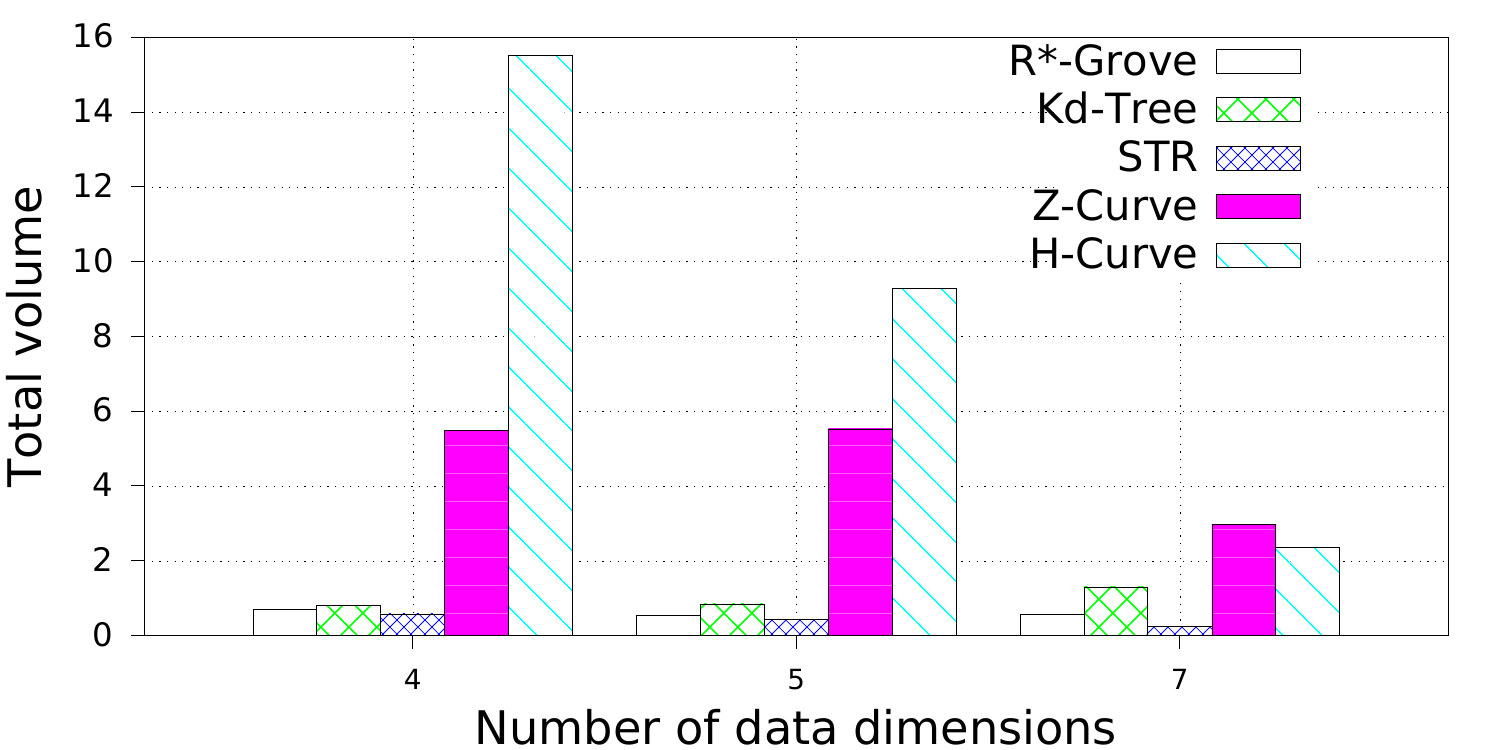} & \includegraphics[width=0.32\columnwidth]{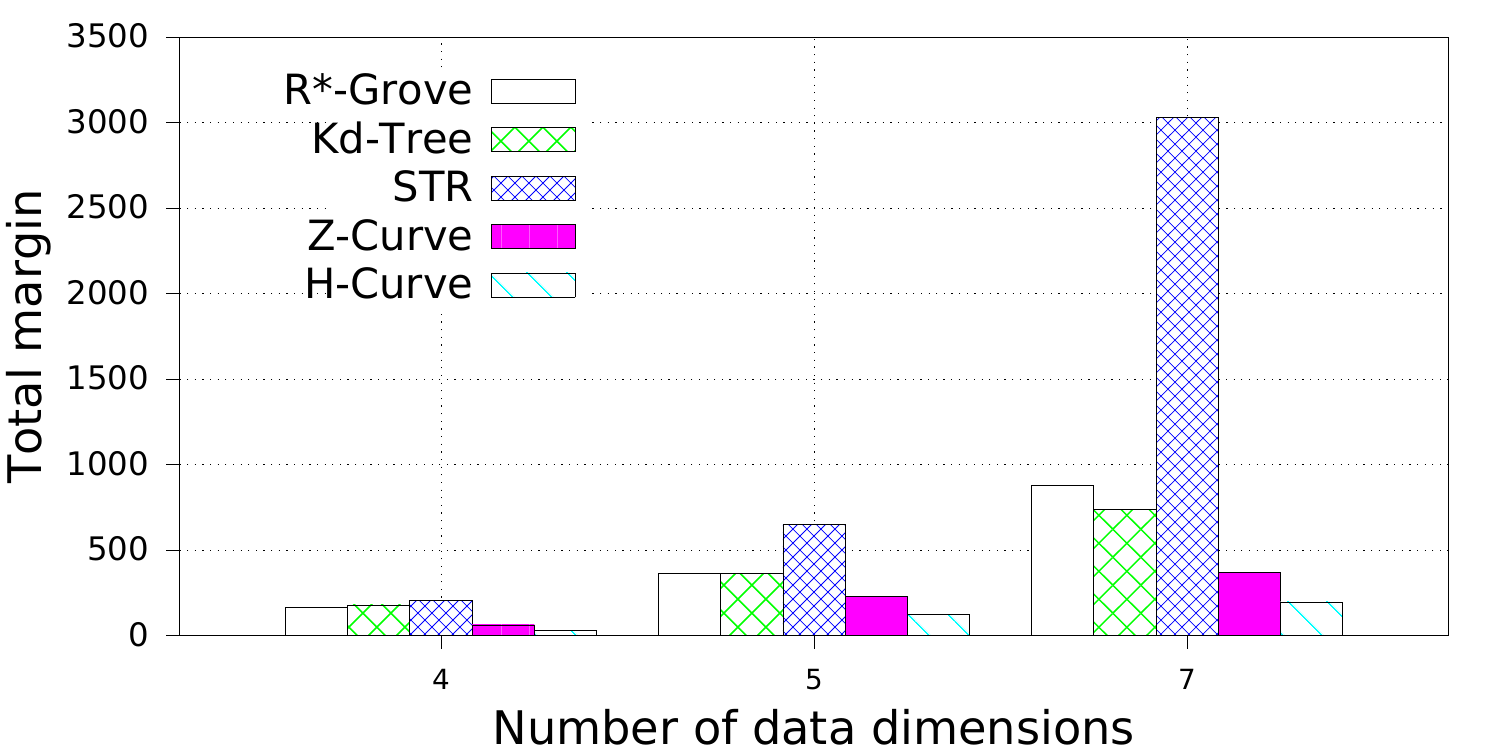} \\
(a)~Partitioning time & (b)~Total volume & (c) Total margin \\
\includegraphics[width=0.32\columnwidth]{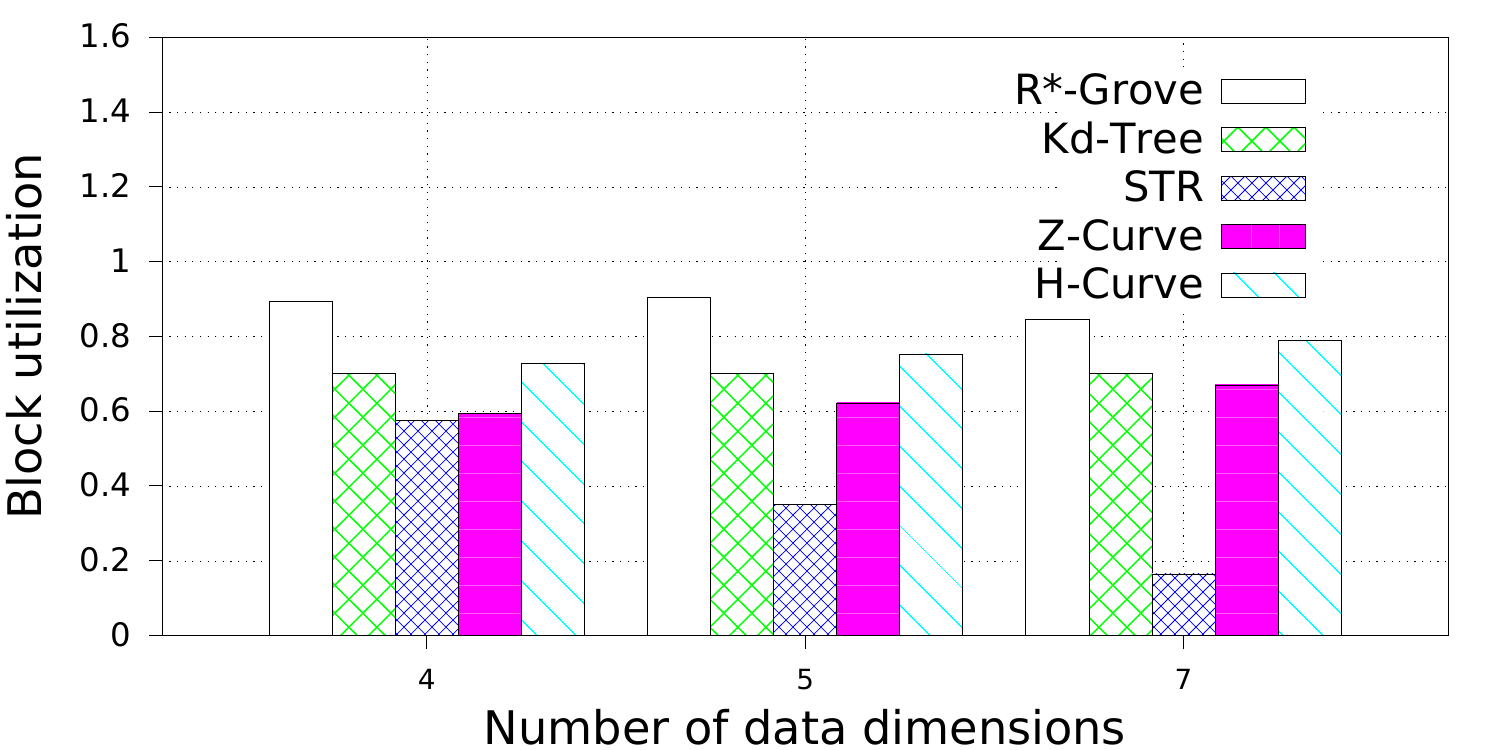}  & \includegraphics[width=0.32\columnwidth]{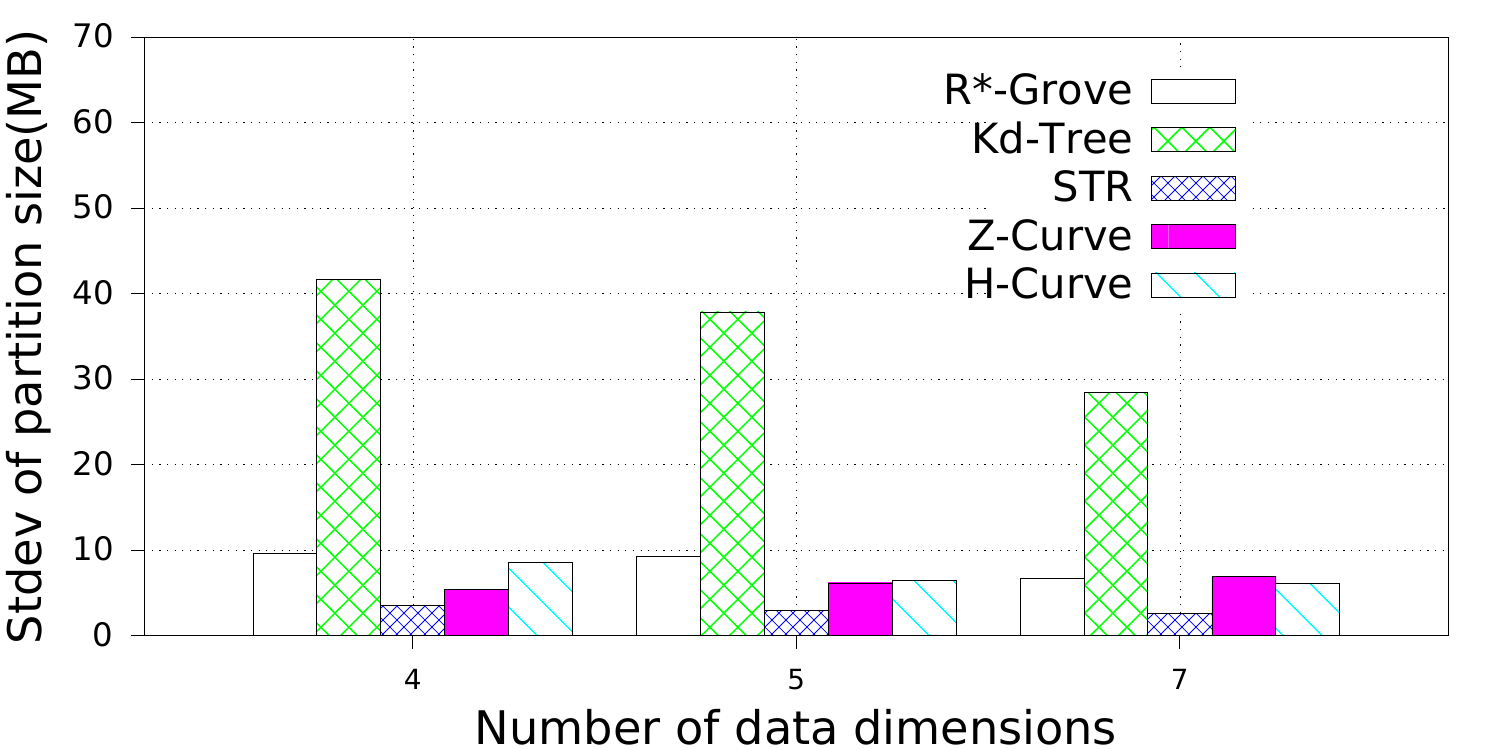} & \includegraphics[width=0.32\columnwidth]{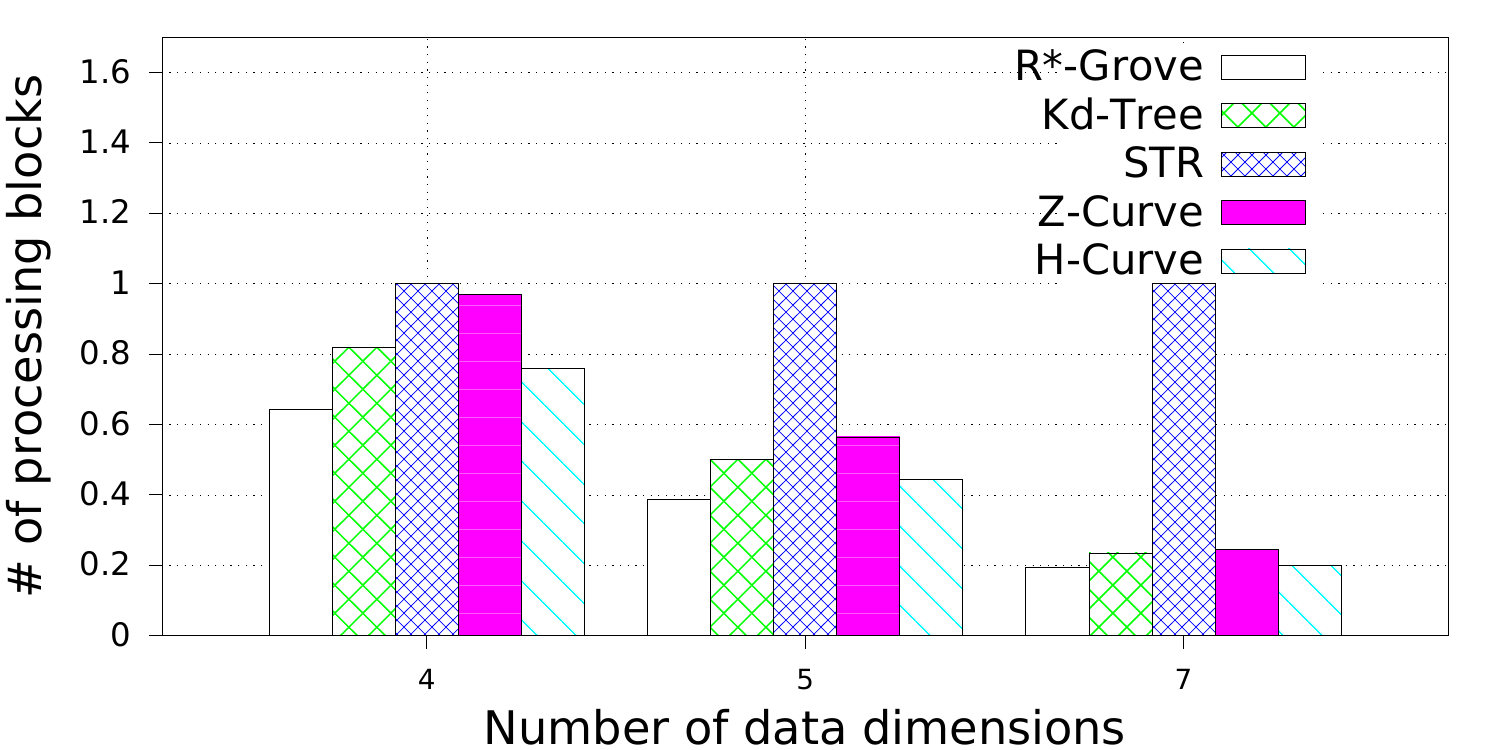} \\
(d)~Block utilization & (e)~Load balance & (f)~Range query performance
\end{tabular}
\end{minipage}
\caption{Indexing performance and partition quality of R*-Grove and other partitioning techniques on multi-dimensional {\tt NYC-Taxi} dataset.}
\label{fig:comparison_nyc_taxi}
\end{figure}

To further support our findings, we also execute similar experiment on {\tt NYC-Taxi} dataset, which contains up to seven dimensions as follows: $pickup\_latitude$, $pickup\_longitude$, $dropoff\_latitude$, $dropoff\_longitude$, $pickup\_datetime$, $trip\_time\_in\_secs$, $trip\_distance$. These attribute values are normalized in order to avoid the dominance of some columns. We decide to partition this dataset using multiple attributes which is picked in the aformentioned order with size $4,5$ and $7$. Figure~\ref{fig:comparison_nyc_taxi} shows that R*-Grove balances all the different quality metrics. Specially, Figure~\ref{fig:comparison_nyc_taxi}(f) indicates that R*-Grove is the winner when compared to other techniques in terms of spatial query performance. We also notice that H-Curve performs better than Z-Curve with this real dataset. We conclude that R*-Grove is a better option for indexing multi-dimensional spatial data since it outperforms or got an equivalent performance with other indexes in all metrics.

% -------------- Conclusion -------------
\section{Conclusion}
\label{sec:conclusion}
This paper proposes R*-Grove, a novel partitioning technique which can be widely used in many big spatial data processing systems. We highlighted three limitations in existing partitioning techniques such as STR, Kd-Tree, Z-Curve and Hilbert Curve. These limitations are the low quality of the partitions, the imbalance among partitions, and the failure to handle variable-size records. We showed that R*-Grove overcomes these three limitations to produce high quality partitions. We showed three case studies in which R*-Grove can be used to facilitate big spatial indexing, range query, and spatial join. An extensive experimental evaluation was carried out on big spatial datasets and showed that R*-Grove is scalable and speeds up all the operations in the case studies. We believe that R*-Grove promises to be a good replacement to existing big spatial data partitioning techniques in many systems. In the future, we will further study the proposed technique for in-memory and streaming applications to see how it behaves under these architectures.

\section*{Funding}
This work is supported in part by the National Science Foundation (NSF) under grants IIS-1838222 and CNS-1924694.

\section*{Data Availability Statement}
The datasets generated for this study are available on the UCR Spatio-temporal Active Repository (UCR-STAR, https://star.cs.ucr.edu/) or on request to the corresponding author. In particular, we used {\tt OSM2015/all\_nodes}, {\tt OSM2015/roads}, {\tt OSM2015/parks}, {\tt OSM2015/all\_objects}, {\tt NYCTaxi}. For the {\tt diagonal\ points} dataset, we generated them using the spatial data generator \cite{vu2019spatial} with following parameters: dataset size $|D|=80$ million points; number of dimensions $d=3,4,5,9$; the percentage (ratio) of the points that are exactly on the line $perc=0.05$; the size of the buffer around the line where additional points are scattered $buf=0.1$.

\bibliographystyle{abbrv}
\bibliography{frontiers_vector}

\end{document}

%% file: Figures/positions.pdf_tex
%% Creator: Inkscape inkscape 0.92.4, www.inkscape.org
%% PDF/EPS/PS + LaTeX output extension by Johan Engelen, 2010
%% Accompanies image file 'positions.pdf' (pdf, eps, ps)
%%
%% To include the image in your LaTeX document, write
%%   \input{<filename>.pdf_tex}
%%  instead of
%%   \includegraphics{<filename>.pdf}
%% To scale the image, write
%%   \def\svgwidth{<desired width>}
%%   \input{<filename>.pdf_tex}
%%  instead of
%%   \includegraphics[width=<desired width>]{<filename>.pdf}
%%
%% Images with a different path to the parent latex file can
%% be accessed with the `import' package (which may need to be
%% installed) using
%%   \usepackage{import}
%% in the preamble, and then including the image with
%%   \import{<path to file>}{<filename>.pdf_tex}
%% Alternatively, one can specify
%%   \graphicspath{{<path to file>/}}
%% 
%% For more information, please see info/svg-inkscape on CTAN:
%%   http://tug.ctan.org/tex-archive/info/svg-inkscape
%%
\begingroup%
  \makeatletter%
  \providecommand\color[2][]{%
    \errmessage{(Inkscape) Color is used for the text in Inkscape, but the package 'color.sty' is not loaded}%
    \renewcommand\color[2][]{}%
  }%
  \providecommand\transparent[1]{%
    \errmessage{(Inkscape) Transparency is used (non-zero) for the text in Inkscape, but the package 'transparent.sty' is not loaded}%
    \renewcommand\transparent[1]{}%
  }%
  \providecommand\rotatebox[2]{#2}%
  \newcommand*\fsize{\dimexpr\f@size pt\relax}%
  \newcommand*\lineheight[1]{\fontsize{\fsize}{#1\fsize}\selectfont}%
  \ifx\svgwidth\undefined%
    \setlength{\unitlength}{192.75590551bp}%
    \ifx\svgscale\undefined%
      \relax%
    \else%
      \setlength{\unitlength}{\unitlength * \real{\svgscale}}%
    \fi%
  \else%
    \setlength{\unitlength}{\svgwidth}%
  \fi%
  \global\let\svgwidth\undefined%
  \global\let\svgscale\undefined%
  \makeatother%
  \begin{picture}(1,0.32096706)%
    \lineheight{1}%
    \setlength\tabcolsep{0pt}%
    \put(0,0){\includegraphics[width=\unitlength,page=1]{positions.pdf}}%
    \put(0.38502758,0.01252394){\color[rgb]{0,0,0}\makebox(0,0)[lt]{\lineheight{1.25}\smash{\begin{tabular}[t]{l}$pos_4=\sum_{j\in[1,4]}{w_j}$\end{tabular}}}}%
    \put(0,0){\includegraphics[width=\unitlength,page=2]{positions.pdf}}%
    \put(0.07253103,0.27305997){\color[rgb]{0,0,0}\makebox(0,0)[t]{\lineheight{1.25}\smash{\begin{tabular}[t]{c}$w_1$\end{tabular}}}}%
    \put(0.19975093,0.27305997){\color[rgb]{0,0,0}\makebox(0,0)[t]{\lineheight{1.25}\smash{\begin{tabular}[t]{c}$w_2$\end{tabular}}}}%
    \put(0.28585785,0.27305997){\color[rgb]{0,0,0}\makebox(0,0)[t]{\lineheight{1.25}\smash{\begin{tabular}[t]{c}$w_3$\end{tabular}}}}%
    \put(0.39146039,0.27305997){\color[rgb]{0,0,0}\makebox(0,0)[t]{\lineheight{1.25}\smash{\begin{tabular}[t]{c}$w_4$\end{tabular}}}}%
    \put(0.55161216,0.27305997){\color[rgb]{0,0,0}\makebox(0,0)[t]{\lineheight{1.25}\smash{\begin{tabular}[t]{c}$w_5$\end{tabular}}}}%
    \put(0.72834057,0.27305997){\color[rgb]{0,0,0}\makebox(0,0)[t]{\lineheight{1.25}\smash{\begin{tabular}[t]{c}$w_6$\end{tabular}}}}%
    \put(0.89722872,0.27305997){\color[rgb]{0,0,0}\makebox(0,0)[t]{\lineheight{1.25}\smash{\begin{tabular}[t]{c}$w_7$\end{tabular}}}}%
  \end{picture}%
\endgroup%

%% file: Figures/validranges.pdf_tex
%% Creator: Inkscape 1.0 (4035a4f, 2020-05-01), www.inkscape.org
%% PDF/EPS/PS + LaTeX output extension by Johan Engelen, 2010
%% Accompanies image file 'validranges.pdf' (pdf, eps, ps)
%%
%% To include the image in your LaTeX document, write
%%   \input{<filename>.pdf_tex}
%%  instead of
%%   \includegraphics{<filename>.pdf}
%% To scale the image, write
%%   \def\svgwidth{<desired width>}
%%   \input{<filename>.pdf_tex}
%%  instead of
%%   \includegraphics[width=<desired width>]{<filename>.pdf}
%%
%% Images with a different path to the parent latex file can
%% be accessed with the `import' package (which may need to be
%% installed) using
%%   \usepackage{import}
%% in the preamble, and then including the image with
%%   \import{<path to file>}{<filename>.pdf_tex}
%% Alternatively, one can specify
%%   \graphicspath{{<path to file>/}}
%% 
%% For more information, please see info/svg-inkscape on CTAN:
%%   http://tug.ctan.org/tex-archive/info/svg-inkscape
%%
\begingroup%
  \makeatletter%
  \providecommand\color[2][]{%
    \errmessage{(Inkscape) Color is used for the text in Inkscape, but the package 'color.sty' is not loaded}%
    \renewcommand\color[2][]{}%
  }%
  \providecommand\transparent[1]{%
    \errmessage{(Inkscape) Transparency is used (non-zero) for the text in Inkscape, but the package 'transparent.sty' is not loaded}%
    \renewcommand\transparent[1]{}%
  }%
  \providecommand\rotatebox[2]{#2}%
  \newcommand*\fsize{\dimexpr\f@size pt\relax}%
  \newcommand*\lineheight[1]{\fontsize{\fsize}{#1\fsize}\selectfont}%
  \ifx\svgwidth\undefined%
    \setlength{\unitlength}{298.20283424bp}%
    \ifx\svgscale\undefined%
      \relax%
    \else%
      \setlength{\unitlength}{\unitlength * \real{\svgscale}}%
    \fi%
  \else%
    \setlength{\unitlength}{\svgwidth}%
  \fi%
  \global\let\svgwidth\undefined%
  \global\let\svgscale\undefined%
  \makeatother%
  \begin{picture}(1,0.53274786)%
    \lineheight{1}%
    \setlength\tabcolsep{0pt}%
    \put(0,0){\includegraphics[width=\unitlength,page=1]{validranges.pdf}}%
    \put(0.64673144,0.00837047){\color[rgb]{0,0,0}\makebox(0,0)[lt]{\lineheight{1.25}\smash{\begin{tabular}[t]{l}Valid ranges\end{tabular}}}}%
    \put(0,0){\includegraphics[width=\unitlength,page=2]{validranges.pdf}}%
    \put(0.64673144,0.32942807){\color[rgb]{0,0,0}\makebox(0,0)[lt]{\lineheight{1.25}\smash{\begin{tabular}[t]{l}Valid left ranges\end{tabular}}}}%
    \put(0.52456549,0.42193318){\color[rgb]{0,0,0}\makebox(0,0)[lt]{\lineheight{1.25}\smash{\begin{tabular}[t]{l}$[im,iM]$\end{tabular}}}}%
    \put(0,0){\includegraphics[width=\unitlength,page=3]{validranges.pdf}}%
    \put(0.07994319,0.25906068){\color[rgb]{0,0,0}\makebox(0,0)[t]{\lineheight{1.25}\smash{\begin{tabular}[t]{c}$m$\end{tabular}}}}%
    \put(0.15674256,0.26130625){\color[rgb]{0,0,0}\makebox(0,0)[t]{\lineheight{1.25}\smash{\begin{tabular}[t]{c}$2m$\end{tabular}}}}%
    \put(0.34088136,0.26130625){\color[rgb]{0,0,0}\makebox(0,0)[t]{\lineheight{1.25}\smash{\begin{tabular}[t]{c}$im$\end{tabular}}}}%
    \put(0.09455165,0.40130787){\color[rgb]{0,0,0}\makebox(0,0)[t]{\lineheight{1.25}\smash{\begin{tabular}[t]{c}$M$\end{tabular}}}}%
    \put(0.19045935,0.40130787){\color[rgb]{0,0,0}\makebox(0,0)[t]{\lineheight{1.25}\smash{\begin{tabular}[t]{c}$2M$\end{tabular}}}}%
    \put(0.38163959,0.40130787){\color[rgb]{0,0,0}\makebox(0,0)[t]{\lineheight{1.25}\smash{\begin{tabular}[t]{c}$iM$\end{tabular}}}}%
    \put(0,0){\includegraphics[width=\unitlength,page=4]{validranges.pdf}}%
    \put(0.64673144,0.13981174){\color[rgb]{0,0,0}\makebox(0,0)[lt]{\lineheight{1.25}\smash{\begin{tabular}[t]{l}Valid right ranges\end{tabular}}}}%
    \put(0.54106228,0.0604041){\color[rgb]{0,0,0}\makebox(0,0)[t]{\lineheight{1.25}\smash{\begin{tabular}[t]{c}$m$\end{tabular}}}}%
    \put(0.45399289,0.05756845){\color[rgb]{0,0,0}\makebox(0,0)[t]{\lineheight{1.25}\smash{\begin{tabular}[t]{c}$2m$\end{tabular}}}}%
    \put(0.28157945,0.07027147){\color[rgb]{0,0,0}\makebox(0,0)[t]{\lineheight{1.25}\smash{\begin{tabular}[t]{c}$jm$\end{tabular}}}}%
    \put(0.53026473,0.20900281){\color[rgb]{0,0,0}\makebox(0,0)[t]{\lineheight{1.25}\smash{\begin{tabular}[t]{c}$M$\end{tabular}}}}%
    \put(0.43689765,0.21027312){\color[rgb]{0,0,0}\makebox(0,0)[t]{\lineheight{1.25}\smash{\begin{tabular}[t]{c}$2M$\end{tabular}}}}%
    \put(0.24444713,0.20900281){\color[rgb]{0,0,0}\makebox(0,0)[t]{\lineheight{1.25}\smash{\begin{tabular}[t]{c}$jM$\end{tabular}}}}%
    \put(0,0){\includegraphics[width=\unitlength,page=5]{validranges.pdf}}%
    \put(0.31380794,0.50217408){\color[rgb]{0,0,0}\makebox(0,0)[t]{\lineheight{1.25}\smash{\begin{tabular}[t]{c}$W=\sum{w_i}$\end{tabular}}}}%
    \put(0,0){\includegraphics[width=\unitlength,page=6]{validranges.pdf}}%
  \end{picture}%
\endgroup%

%% file: Figures/correction.pdf_tex
%% Creator: Inkscape inkscape 0.92.4, www.inkscape.org
%% PDF/EPS/PS + LaTeX output extension by Johan Engelen, 2010
%% Accompanies image file 'correction.pdf' (pdf, eps, ps)
%%
%% To include the image in your LaTeX document, write
%%   \input{<filename>.pdf_tex}
%%  instead of
%%   \includegraphics{<filename>.pdf}
%% To scale the image, write
%%   \def\svgwidth{<desired width>}
%%   \input{<filename>.pdf_tex}
%%  instead of
%%   \includegraphics[width=<desired width>]{<filename>.pdf}
%%
%% Images with a different path to the parent latex file can
%% be accessed with the `import' package (which may need to be
%% installed) using
%%   \usepackage{import}
%% in the preamble, and then including the image with
%%   \import{<path to file>}{<filename>.pdf_tex}
%% Alternatively, one can specify
%%   \graphicspath{{<path to file>/}}
%% 
%% For more information, please see info/svg-inkscape on CTAN:
%%   http://tug.ctan.org/tex-archive/info/svg-inkscape
%%
\begingroup%
  \makeatletter%
  \providecommand\color[2][]{%
    \errmessage{(Inkscape) Color is used for the text in Inkscape, but the package 'color.sty' is not loaded}%
    \renewcommand\color[2][]{}%
  }%
  \providecommand\transparent[1]{%
    \errmessage{(Inkscape) Transparency is used (non-zero) for the text in Inkscape, but the package 'transparent.sty' is not loaded}%
    \renewcommand\transparent[1]{}%
  }%
  \providecommand\rotatebox[2]{#2}%
  \newcommand*\fsize{\dimexpr\f@size pt\relax}%
  \newcommand*\lineheight[1]{\fontsize{\fsize}{#1\fsize}\selectfont}%
  \ifx\svgwidth\undefined%
    \setlength{\unitlength}{191.32696918bp}%
    \ifx\svgscale\undefined%
      \relax%
    \else%
      \setlength{\unitlength}{\unitlength * \real{\svgscale}}%
    \fi%
  \else%
    \setlength{\unitlength}{\svgwidth}%
  \fi%
  \global\let\svgwidth\undefined%
  \global\let\svgscale\undefined%
  \makeatother%
  \begin{picture}(1,0.53229934)%
    \lineheight{1}%
    \setlength\tabcolsep{0pt}%
    \put(0,0){\includegraphics[width=\unitlength,page=1]{correction.pdf}}%
    \put(0.33271178,0.41924053){\color[rgb]{0,0,0}\makebox(0,0)[t]{\lineheight{1.25}\smash{\begin{tabular}[t]{c}$\Delta pos$\end{tabular}}}}%
    \put(0.9627687,0.27601522){\color[rgb]{0,0,0}\makebox(0,0)[rt]{\lineheight{1.25}\smash{\begin{tabular}[t]{r}Before\end{tabular}}}}%
    \put(0.96191675,0.05716835){\color[rgb]{0,0,0}\makebox(0,0)[rt]{\lineheight{1.25}\smash{\begin{tabular}[t]{r}After\end{tabular}}}}%
    \put(0.39561306,0.35637893){\color[rgb]{0,0,0}\makebox(0,0)[lt]{\lineheight{1.25}\smash{\begin{tabular}[t]{l}$p_1$\end{tabular}}}}%
    \put(0.59558242,0.36132867){\color[rgb]{0,0,0}\makebox(0,0)[lt]{\lineheight{1.25}\smash{\begin{tabular}[t]{l}$p_2$\end{tabular}}}}%
    \put(0.47080949,0.50190111){\color[rgb]{0,0,0}\makebox(0,0)[t]{\lineheight{1.25}\smash{\begin{tabular}[t]{c}$w_2$\end{tabular}}}}%
    \put(0.22777744,0.50190111){\color[rgb]{0,0,0}\makebox(0,0)[t]{\lineheight{1.25}\smash{\begin{tabular}[t]{c}$w_1$\end{tabular}}}}%
    \put(0.44111109,0.00522666){\color[rgb]{0,0,0}\makebox(0,0)[t]{\lineheight{1.25}\smash{\begin{tabular}[t]{c}$w_2'$\end{tabular}}}}%
    \put(0.20005893,0.00522666){\color[rgb]{0,0,0}\makebox(0,0)[t]{\lineheight{1.25}\smash{\begin{tabular}[t]{c}$w_1'$\end{tabular}}}}%
    \put(-0.06726265,0.2927727){\color[rgb]{0,0,0}\makebox(0,0)[lt]{\begin{minipage}{0.27223544\unitlength}\raggedleft Empty valid range\end{minipage}}}%
  \end{picture}%
\endgroup%

%% file: frontiers_vector.bbl
\begin{thebibliography}{10}

\bibitem{allobjects}
Openstreetmap all objects dataset, 2019.
\newblock http://star.cs.ucr.edu/\#dataset=OSM2015/all\_objects.

\bibitem{BKS+90}
N.~Beckmann, H.~Kriegel, R.~Schneider, and B.~Seeger.
\newblock The r*-tree: An efficient and robust access method for points and
  rectangles.
\newblock In {\em SIGMOD}, pages 322--331, Atlantic City, NJ, May 1990.

\bibitem{beckmann2008benchmark}
N.~Beckmann and B.~Seeger.
\newblock A benchmark for multidimensional index structures, 2008.

\bibitem{BS09}
N.~Beckmann and B.~Seeger.
\newblock A revised r*-tree in comparison with related index structures.
\newblock In {\em SIGMOD}, pages 799--812, Providence, RI, June 2009.

\bibitem{bentley1975multidimensional}
J.~L. Bentley.
\newblock Multidimensional binary search trees used for associative searching.
\newblock {\em Communications of the ACM}, 18(9):509--517, 1975.

\bibitem{CE17}
H.~Chasparis and A.~Eldawy.
\newblock Experimental evaluation of selectivity estimation on big spatial
  data.
\newblock In {\em Proceedings of the Fourth International {ACM} Workshop on
  Managing and Mining Enriched Geo-Spatial Data, Chicago, IL, USA, May 14,
  2017}, pages 8:1--8:6, 2017.

\bibitem{DS00}
J.~Dittrich and B.~Seeger.
\newblock Data redundancy and duplicate detection in spatial join processing.
\newblock In {\em Proceedings of the 16th International Conference on Data
  Engineering, San Diego, California, USA, February 28 - March 3, 2000}, pages
  535--546, 2000.

\bibitem{EAM15}
A.~Eldawy, L.~Alarabi, and M.~F. Mokbel.
\newblock Spatial partitioning techniques in spatial hadoop.
\newblock {\em {PVLDB}}, 8(12):1602--1605, 2015.

\bibitem{eldawy2015spatial}
A.~Eldawy et~al.
\newblock Spatial partitioning techniques in spatialhadoop.
\newblock {\em Proceedings of the VLDB Endowment}, 8(12):1602--1605, 2015.

\bibitem{ESE+17}
A.~Eldawy et~al.
\newblock {Sphinx: Empowering Impala for Efficient Execution of {SQL} Queries
  on Big Spatial Data}.
\newblock In {\em SSTD}, pages 65--83, Arlington, VA, Aug. 2017.

\bibitem{ELM+13}
A.~Eldawy, Y.~Li, M.~F. Mokbel, and R.~Janardan.
\newblock Cg{\_}hadoop: computational geometry in mapreduce.
\newblock In {\em SIGSPATIAL}, pages 284--293, Orlando, FL, Nov. 2013.

\bibitem{EM15}
A.~Eldawy and M.~F. Mokbel.
\newblock Spatialhadoop: {A} mapreduce framework for spatial data.
\newblock In {\em ICDE}, pages 1352--1363, Seoul, South Korea, Apr. 2015.

\bibitem{EM16}
A.~Eldawy and M.~F. Mokbel.
\newblock {The Era of Big Spatial Data: A Survey}.
\newblock {\em {Foundations and Trends in Databases}}, 6(3-4):163--273, 2016.

\bibitem{EMA+15}
A.~Eldawy, M.~F. Mokbel, S.~Al{-}Harthi, A.~Alzaidy, K.~Tarek, and S.~Ghani.
\newblock {SHAHED:} {A} mapreduce-based system for querying and visualizing
  spatio-temporal satellite data.
\newblock In {\em ICDE}, pages 1585--1596, Seoul, South Korea, Apr. 2015.

\bibitem{EMJ16}
A.~Eldawy, M.~F. Mokbel, and C.~Jonathan.
\newblock Hadoopviz: {A} mapreduce framework for extensible visualization of
  big spatial data.
\newblock In {\em ICDE}, pages 601--612, Helsinki, Finland, May 2016.

\bibitem{fox2013spatio}
A.~D. Fox, C.~N. Eichelberger, J.~N. Hughes, and S.~Lyon.
\newblock Spatio-temporal indexing in non-relational distributed databases.
\newblock In {\em Big Data}, pages 291--299, Santa Clara, CA, Oct. 2013.

\bibitem{GEJ19}
S.~Ghosh, A.~Eldawy, and S.~Jais.
\newblock Aid: An adaptive image data index for interactive multilevel
  visualization.
\newblock In {\em ICDE}, page~4, Macau, China, Apr. 2019.

\bibitem{G07}
M.~F. Goodchild.
\newblock Citizens as voluntary sensors: Spatial data infrastructure in the
  world of web 2.0.
\newblock {\em {IJSDIR}}, 2:24--32, 2007.

\bibitem{G84}
A.~Guttman.
\newblock R-trees: {A} dynamic index structure for spatial searching.
\newblock In {\em SIGMOD}, pages 47--57, Boston, MA, June 1984.

\bibitem{HBC+16}
N.~Henke et~al.
\newblock {The Age of Analytics: Competing in a Data-driven World}.
\newblock Technical report, McKinsey Global Institute, Dec. 2016.

\bibitem{HS94}
E.~G. Hoel and H.~Samet.
\newblock Performance of data-parallel spatial operations.
\newblock In {\em VLDB'94, Proceedings of 20th International Conference on Very
  Large Data Bases, September 12-15, 1994, Santiago de Chile, Chile}, pages
  156--167, 1994.

\bibitem{jacox2007spatial}
E.~H. Jacox and H.~Samet.
\newblock Spatial join techniques.
\newblock {\em ACM Transactions on Database Systems (TODS)}, 32(1):7, 2007.

\bibitem{KF94}
I.~Kamel and C.~Faloutsos.
\newblock Hilbert r-tree: An improved r-tree using fractals.
\newblock In {\em VLDB}, pages 500--509, Santiago de Chile, Chile, Sept. 1994.

\bibitem{lee1977worst}
D.-T. Lee and C.~Wong.
\newblock Worst-case analysis for region and partial region searches in
  multidimensional binary search trees and balanced quad trees.
\newblock {\em Acta Informatica}, 9(1):23--29, 1977.

\bibitem{lee2003omt}
T.~Lee et~al.
\newblock Omt: Overlap minimizing top-down bulk loading algorithm for r-tree.
\newblock In {\em CAISE Short paper proceedings}, volume~74, pages 69--72,
  2003.

\bibitem{leutenegger1997str}
S.~T. Leutenegger et~al.
\newblock Str: A simple and efficient algorithm for r-tree packing.
\newblock In {\em ICDE}, pages 497--506. IEEE, 1997.

\bibitem{LEX+19}
Y.~Li, A.~Eldawy, J.~Xue, N.~Knorozova, M.~F. Mokbel, and R.~Janardan.
\newblock {Scalable Computational Geometry in MapReduce}.
\newblock {\em The VLDB Journal}, Jan 2019.

\bibitem{LR94}
M.~Lo and C.~V. Ravishankar.
\newblock Spatial joins using seeded trees.
\newblock In {\em SIGMOD}, pages 209--220, Minneapolis, MN, May 1994.

\bibitem{LCC+14}
P.~Lu et~al.
\newblock {ScalaGiST: Scalable Generalized Search Trees for MapReduce Systems}.
\newblock {\em PVLDB}, 7(14):1797--1808, 2014.

\bibitem{MAA+14}
A.~Magdy, L.~Alarabi, S.~Al{-}Harthi, M.~Musleh, T.~M. Ghanem, S.~Ghani, and
  M.~F. Mokbel.
\newblock Taghreed: a system for querying, analyzing, and visualizing geotagged
  microblogs.
\newblock In {\em SIGSPATIAL}, pages 163--172, Dallas/Fort Worth, TX, Nov.
  2014.

\bibitem{nishimura2013mathcal}
S.~Nishimura, S.~Das, D.~Agrawal, and A.~{El Abbadi}.
\newblock $\mathcal{MD}$-hbase: design and implementation of an elastic data
  infrastructure for cloud-scale location services.
\newblock {\em Distributed and Parallel Databases}, 31(2):289--319, 2013.

\bibitem{SM17}
I.~Sabek and M.~F. Mokbel.
\newblock On spatial joins in mapreduce.
\newblock In {\em SIGSPATIAL}, pages 21:1--21:10, Redondo Beach, CA, Nov. 2017.

\bibitem{S84}
H.~Samet.
\newblock The quadtree and related hierarchical data structures.
\newblock {\em {ACM} Computing Surveys}, 16(2):187--260, 1984.

\bibitem{siddique2019comparing}
A.~B. Siddique, A.~Eldawy, and V.~Hristidis.
\newblock Comparing synopsis techniques for approximate spatial data analysis.
\newblock {\em Proceedings of the VLDB Endowment}, 12(11):1583--1596, 2019.

\bibitem{TYM+16}
M.~Tang, Y.~Yu, Q.~M. Malluhi, M.~Ouzzani, and W.~G. Aref.
\newblock {LocationSpark: {A} Distributed In-Memory Data Management System for
  Big Spatial Data}.
\newblock {\em PVLDB}, 9(13):1565--1568, 2016.

\bibitem{ucrstar}
The ucr spatio-temporal active repository (ucr-star), 2019.
\newblock https://star.cs.ucr.edu/.

\bibitem{VAW14}
H.~Vo, A.~Aji, and F.~Wang.
\newblock {SATO:} a spatial data partitioning framework for scalable query
  processing.
\newblock In {\em SIGSPATIAL}, pages 545--548, Dallas/Fort Worth, TX, Nov.
  2014.

\bibitem{VE18}
T.~Vu and A.~Eldawy.
\newblock {R-Grove: growing a family of R-trees in the big-data forest}.
\newblock In {\em SIGSPATIAL}, pages 532--535, Seattle, WA, Nov. 2018.

\bibitem{vu2019spatial}
T.~Vu, S.~Migliorini, A.~Eldawy, and A.~Belussi.
\newblock Spatial data generators.
\newblock In {\em 1st ACM SIGSPATIAL International Workshop on Spatial Gems
  (SpatialGems 2019)}, page~7, 2019.

\bibitem{WPA+14}
R.~T. Whitman et~al.
\newblock Spatial indexing and analytics on hadoop.
\newblock In {\em SIGSPATIAL}, pages 73--82, Dallas/Fort Worth, TX, Nov. 2014.

\bibitem{XLY+16}
D.~Xie, F.~Li, B.~Yao, G.~Li, L.~Zhou, and M.~Guo.
\newblock Simba: Efficient in-memory spatial analytics.
\newblock In {\em SIGMOD}, pages 1071--1085, San Francisco, CA, July 2016.

\bibitem{YWS15}
J.~Yu, J.~Wu, and M.~Sarwat.
\newblock Geospark: a cluster computing framework for processing large-scale
  spatial data.
\newblock In {\em SIGSPATIAL}, pages 70:1--70:4, Bellevue, WA, Nov. 2015.

\bibitem{yu2015geospark}
J.~Yu, J.~Wu, and M.~Sarwat.
\newblock Geospark: A cluster computing framework for processing large-scale
  spatial data.
\newblock In {\em Proceedings of the 23rd SIGSPATIAL International Conference
  on Advances in Geographic Information Systems}, page~70. ACM, 2015.

\bibitem{Zhou1998}
X.~Zhou, D.~J. Abel, and D.~Truffet.
\newblock Data partitioning for parallel spatial join processing.
\newblock {\em GeoInformatica}, 2(2):175--204, Jun 1998.

\end{thebibliography}
